\documentclass[3p, authoryear]{elsarticle}
\usepackage{amsmath}
\usepackage{amssymb}
\usepackage{multirow}
\usepackage{multicol}
\usepackage{dblfloatfix}
\usepackage{setspace}
\usepackage{lscape}
\usepackage{changepage}
\usepackage{psfrag}
\usepackage{color}
\usepackage{rotating}
\usepackage{tablefootnote}
\usepackage{arydshln}
\usepackage{gensymb}
\usepackage{url}
\usepackage{subfigure}
\usepackage{booktabs}
\usepackage{caption}

\makeatletter
\def\maxwidth{ %
  \ifdim\Gin@nat@width>\linewidth
    \linewidth
  \else
    \Gin@nat@width
  \fi
}
\makeatother

\definecolor{fgcolor}{rgb}{0.345, 0.345, 0.345}
\newcommand{\hlnum}[1]{\textcolor[rgb]{0.686,0.059,0.569}{#1}}%
\newcommand{\hlstr}[1]{\textcolor[rgb]{0.192,0.494,0.8}{#1}}%
\newcommand{\hlcom}[1]{\textcolor[rgb]{0.678,0.584,0.686}{\textit{#1}}}%
\newcommand{\hlopt}[1]{\textcolor[rgb]{0,0,0}{#1}}%
\newcommand{\hlstd}[1]{\textcolor[rgb]{0.345,0.345,0.345}{#1}}%
\newcommand{\hlkwa}[1]{\textcolor[rgb]{0.161,0.373,0.58}{\textbf{#1}}}%
\newcommand{\hlkwb}[1]{\textcolor[rgb]{0.69,0.353,0.396}{#1}}%
\newcommand{\hlkwc}[1]{\textcolor[rgb]{0.333,0.667,0.333}{#1}}%
\newcommand{\hlkwd}[1]{\textcolor[rgb]{0.737,0.353,0.396}{\textbf{#1}}}%

\usepackage{framed}
\makeatletter
\newenvironment{kframe}{%
 \def\at@end@of@kframe{}%
 \ifinner\ifhmode%
  \def\at@end@of@kframe{\end{minipage}}%
  \begin{minipage}{\columnwidth}%
 \fi\fi%
 \def\FrameCommand##1{\hskip\@totalleftmargin \hskip-\fboxsep
 \colorbox{shadecolor}{##1}\hskip-\fboxsep
     \hskip-\linewidth \hskip-\@totalleftmargin \hskip\columnwidth}%
 \MakeFramed {\advance\hsize-\width
   \@totalleftmargin\z@ \linewidth\hsize
   \@setminipage}}%
 {\par\unskip\endMakeFramed%
 \at@end@of@kframe}
\makeatother

\definecolor{shadecolor}{rgb}{.97, .97, .97}
\definecolor{messagecolor}{rgb}{0, 0, 0}
\definecolor{warningcolor}{rgb}{1, 0, 1}
\definecolor{errorcolor}{rgb}{1, 0, 0}
\newenvironment{knitrout}{}{} 
\usepackage{alltt}
\usepackage{Rd}

\newcommand{\by}{ {\boldsymbol y} }

\newcommand{\balpha}{ {\boldsymbol \alpha} }

\newcommand{\bgamma}{ {\boldsymbol \gamma} }

\newcommand{\U}{\mathcal{U}}
\newcommand{\given}{\,|\,}

\usepackage{xspace}

\newcommand{\beginsupplement}{%
        \setcounter{table}{0}
        \renewcommand{\thetable}{S\arabic{table}}%
        \setcounter{figure}{0}
        \renewcommand{\thefigure}{S\arabic{figure}}%
     }

\title{A Bayesian hierarchical model to estimate land surface phenology parameters with harmonized Landsat 8 and Sentinel-2 images}

\author[umn]{Chad Babcock\corref{cor1}}
\author[msu]{Andrew O. Finley}
\author[umn-soil]{Nathaniel Looker}

\cortext[cor1]{Corresponding author}
\address[umn]{University of Minnesota, Department of Forest Resources, St.~Paul, MN, 55108, USA}
\address[msu]{Michigan State University, Forestry Department, East Lansing, MI, 48824, USA}
\address[umn-soil]{University of Minnesota, Department of Soil, Water and Climate, St.~Paul, MN, 55108, USA}

\begin{document}
\begin{frontmatter}

\begin{abstract}
We develop a Bayesian Land Surface Phenology (LSP) model and examine its performance using Enhanced Vegetation Index (EVI) observations derived from the Harmonized Landsat Sentinel-2 (HLS) dataset. Building on previous work, we propose a double logistic function that, once couched within a Bayesian model, yields posterior distributions for all LSP parameters. We assess the efficacy of the Normal, Truncated Normal, and Beta likelihoods to deliver robust LSP parameter estimates. Two case studies are presented and used to explore aspects of the proposed model. The first, conducted over forested pixels within a HLS tile, explores choice of likelihood and space-time varying HLS data availability for long-term average LSP parameter point and uncertainty estimation. The second, conducted on a small area of interest within the HLS tile on an annual time-step, further examines the impact of sample size and choice of likelihood on LSP parameter estimates. Results indicate that while the Truncated Normal and Beta likelihoods are theoretically preferable when the vegetation index is bounded, all three likelihoods performed similarly when the number of index observations is sufficiently large and values are not near the index bounds. Both case studies demonstrate how pixel-level LSP parameter posterior distributions can be used to propagate uncertainty through subsequent analysis. As a companion to this article, we provide an open-source \R package \pkg{rsBayes} and supplementary data and code used to reproduce the analysis results. The proposed model specification and software implementation delivers computationally efficient, statistically robust, and inferentially rich LSP parameter posterior distributions at the pixel-level across massive raster time series datasets. 
\end{abstract}


\end{frontmatter}

\section{Introduction}
From spring budbreak through winter dormancy, temperate forest vegetation changes seasonally. The timing and magnitude of the phenological events that indicate transitions in vegetation behavior are primarily driven by temperature and water availability \citep{aber2001}. The vegetation phenology cycle shifts with changes in climatic variables, such as seasonal average temperatures and precipitation \citep{jeong2011}. This makes tracking changes in vegetation phenology important for understanding how climate change will impact terrestrial ecosystems \citep{parmesan2003}. Improving climate model projections and monitoring ecosystem response to climate change require spatially and temporally accurate information about vegetation phenology \citep{piao2019}. This need continues to motivate research activities aimed at collecting data and developing methods to provide high-quality spatio-temporal phenology information \citep{morisette2009}.

Monitoring seasonal changes in vegetation using using imagery, i.e., land surface phenology (LSP), has a long history in remote sensing \citep{zeng2020}. Exploring the use of early satellite remote sensing systems, such as the Advanced Very High Resolution Radiometer (AVHRR) and Landsat Mulitspectral Scanner (MSS), to estimate LSP parameters started in the 1970s \citep{rea1976,goward1985,taylor1985}. These early studies showed images from satellites could be used to effectively estimate many phenological parameters for both natural and cultivated landscapes \citep{henderson1984,spanner1990}; however, low spatial and temporal resolution of early remote sensing datasets limited LSP parameter mapping at fine scales. 

The launch of the Moderate Resolution Imaging Spectroradiometer (MODIS) in 2000 ushered in a new wave of LSP research. These studies aimed to exploit the high temporal resolution reflectance data products MODIS continues to offer \citep{ahl2006, zhang2003, ganguly2010}. Currently, the United States National Aeronautics and Space Administration (NASA) hosts the MODIS Land Cover Dynamics product (MCD12Q2 C6) that provides annual LSP parameter estimates at a 500 m spatial resolution for the globe from 2001 to present \citep{sullamenashe2019}. In 2011, the Visible Infrared Imaging Radiometer Suite (VIIRS) was launched. This sensor system was designed as a replacement to the aging MODIS Aqua and Terra satellites. NASA also hosts the VIIRS Global Land Surface Phenology Product (VNP22Q2) that provides annual 500 m resolution LSP parameters from 2013 through 2018 \citep{zhang2018}.

Over the past decade, remote sensing technology has advanced tremendously and, with it, the ability to collect imagery useful for higher spatial resolution LSP parameter estimation \citep{melaas2013,an2018}. NASA's Landsat 8 and the European Space Agency's (ESA) Sentinel-2 A and B satellites now collect and make publicly available reflectance data for the entire globe at unprecedented spatial, temporal, and radiometric resolutions. Combined, Landsat 8 and Sentinel-2 provide global coverage imagery at a nearly 3 day repeat interval \citep{li2017}. The Harmonized Landsat Sentinel-2 (HLS) project is an effort by NASA to bring Landsat 8 Operational Land Imager (OLI) and Sentinel-2 Multispectral Instrument (MSI) reflectance data together \citep{claverie2018}. The HLS product is produced using processing steps designed so researchers can combine OLI and MSI surface reflectance images seamlessly into a single time series for analysis. With the HLS dataset, the remote sensing and geospatial analysis community has access to analysis-ready imagery that is temporally dense enough to potentially estimate LSP parameters at 30 m resolution annually \citep{bolton2020, kowalski2020}.

Along with improvements to remote sensing data collection, there have been methodological advancements for translating reflectance readings into meaningful LSP parameters. \citet{vorobiova2017} and \citet{zeng2020} provide an extensive survey of approaches used to extract LSP parameters from vegetation index (VI) time series. We focus our attention on the double logistic curve fitting technique \citep{zhang2003,elmore2012,melaas2013}. The approach generally involves fitting separate spring and autumn logistic functions to a VI time series for a pixel using a nonlinear model fitting algorithm. The spring function models VI behavior from winter dormancy to the midsummer growing season, and the autumn logistic function models VI behavior from midsummer back to dormancy. To determine the day of year (DOY) transition date from spring to autumn functions, \citet{melaas2013} calculated slopes for moving windows across the time series. The earliest (after DOY = 90) detection of a negative slope in a moving window was set as the transition DOY for the pixel. One of the key strengths to the phenology curve fitting approach is the spring and autumn function coefficients can be meaningfully interpreted as LSP parameters \citep{melaas2013, elmore2012} . Many other approaches require further manipulation of the fitted curve to derive ecologically relevant metrics (e.g., derivatives to find inflection points, minimums, and maximums). The logistic function approach is also well constrained to ensure sensible phenology curve fits. Many spline fitting algorithms have a tendency to over-fit noisy data, leading to multiple peaks and valleys that are likely artifacts of data noise rather than true phenological signal. 

Building upon \citet{melaas2013}, \citet{senf2017}  placed the spring logistic function in a Bayesian model to estimate the LSP parameters and demonstrate the usefulness of this statistical paradigm for quantifying parameter uncertainty.  \citet{senf2017}, however, left for future research the derivation of a single Bayesian LSP model capable of fitting both spring and autumn functions and development of model fitting algorithms able to feasibly deliver Bayesian inference on more than a few hundred pixels. Our work aims to address both of these points.

The overarching objective of this paper is to advance model development work detailed in \citet{melaas2013} and \citet{senf2017}. Specifically, we propose an approach to simultaneously estimate spring and autumn logistic function parameters for a VI time series within a single Bayesian model. Casting these functions in a single Bayesian model allows the entire phenology curve to be estimated with statistical rigor and yields full uncertainty quantification for LSP parameters. Additionally, we developed an open-source \R package \pkg{rsBayes} \citep{rsBayes} that provides functions for efficiently sampling from the proposed LSP model parameters' posterior distributions using a Markov chain Monte Carlo (MCMC) algorithm. The proposed LSP model's ability to deliver robust LSP parameter estimates is assessed using two case studies. The case studies consider an Enhanced Vegetation Index (EVI) time series for cloud-free forested pixels within a HLS tile. In the first case study, we pool EVI time series across years (2013-2019) to estimate long-term average LSP curves as was done in \citet{fisher2006} and \citet{elmore2012}. We examine how EVI observation sample size and data variability contribute to LSP parameter uncertainty at the pixel level. In the second case study, we fit the proposed LSP model using annual EVI observations  to assess reliability of LSP parameter estimates given differing sample sizes. Using an abbreviated forest change detection analysis, this case study also demonstrates how the proposed error propagation approach is useful for estimating the posterior distribution of new parameters computed from the LSP model's parameters.

Supplementary material includes the data and software needed to reproduce our results and additional \R functions that facilitate efficient fitting of the proposed model over massive pixel arrays (e.g., entire Sentinel-2 tiles) for multi-core processor computer systems. We conclude by discussing potential ways to further improve the proposed Bayesian LSP approach through model refinement and increased posterior sampling efficiency.

\section{Model development}\label{sec:model}
\subsection{Land surface phenology function}

\begin{figure}[!h]
    \begin{center}
    \includegraphics[width=16cm]{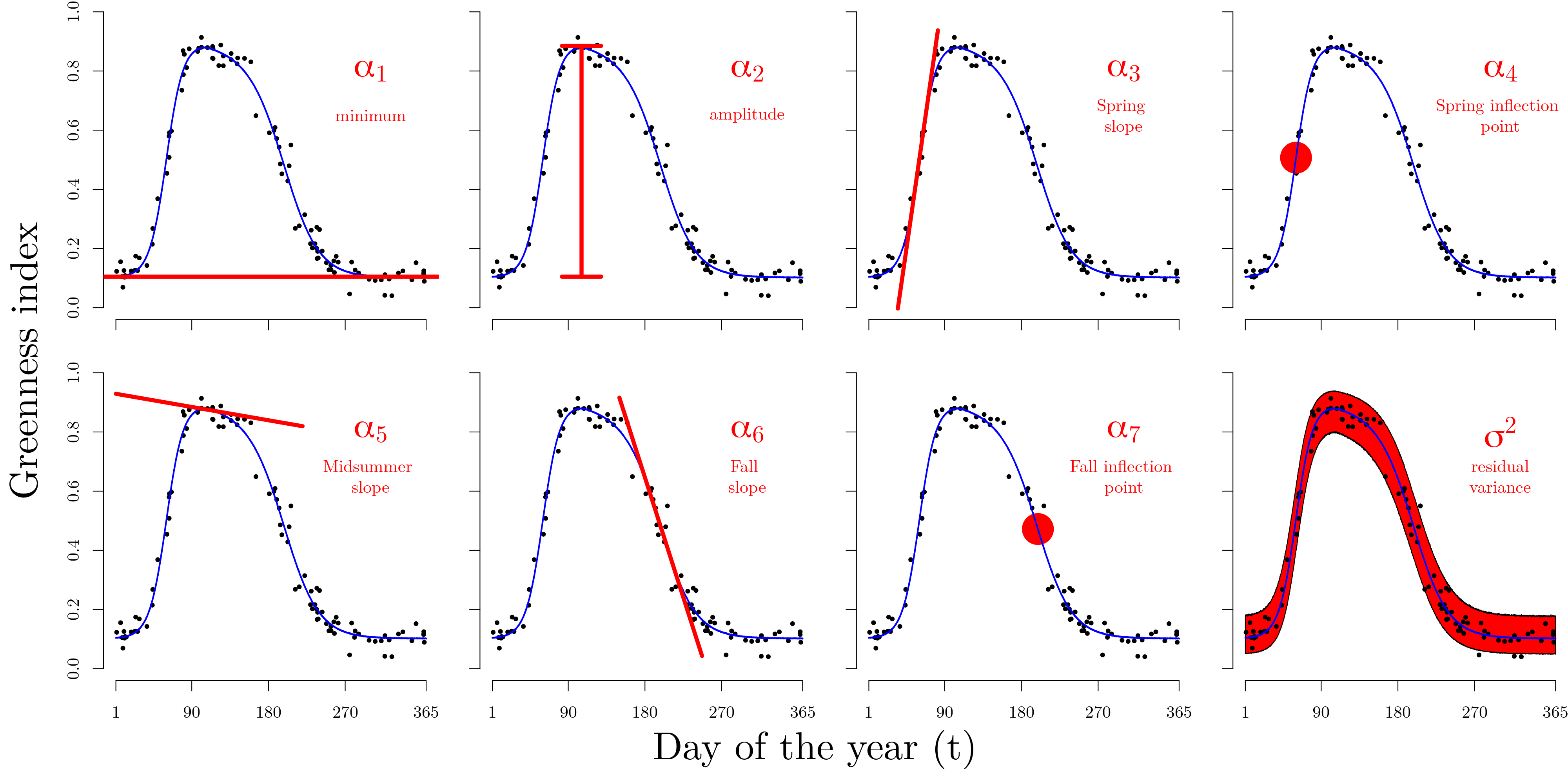}
    \caption{Graphic depictions of proposed LSP function (\ref{eqn:G}) and Bayesian LSP model (\ref{eqn:post}) parameters. $\alpha_1$ is the LSP curve minimum. $\alpha_2$ is the LSP curve amplitude. $\alpha_3$ controls the spring side slope of the LSP curve. $\alpha_4$ is the spring inflection point. $\alpha_5$ controls the midsummer slope. $\alpha_6$ controls the fall side slope of the LSP curve. $\alpha_7$ is the fall inflection point. $\sigma_2$ models residual variance.}\label{fig:model-graphic}
    \end{center}
\end{figure}

We develop a LSP function by building on the approaches outlined in \citet{melaas2013} and \citet{senf2017}. Here, for a given spatial location, e.g., pixel, we denote $y(t)$ as the observed VI at time $t$. As in \citet{melaas2013}, the first order behavior of $y$ over a year, i.e., $t = 1, 2, \ldots, T$ with $T=365$, can be modeled using two logistic functions. The spring function is
\begin{equation}\label{eqn:S}
    S(t) = \alpha_1 + \frac{\alpha_2-\alpha_5t}{1 + e^{-\alpha_3(t-\alpha_4)}},
\end{equation}
where $\alpha_1$ is seasonal minimum greenness, $\alpha_2$ is the seasonal amplitude, $\alpha_3$ is the green-up rate, $\alpha_4$ is the green-up inflection point, and $\alpha_5$ controls the mid-growing season greenness trajectory. The autumn function is defined as
\begin{equation}\label{eqn:A}
    A(t) = \alpha_1 + \frac{\alpha_2-\alpha_5t}{1 + e^{-\alpha_6(\alpha_7-t)}},
\end{equation}
where $\alpha_6$ is the green-down rate and $\alpha_7$ is the autumn inflection point.
We propose the following step function to seamlessly combine combine (\ref{eqn:S}) and (\ref{eqn:A}),
\begin{equation}\label{eqn:G}
    G(t) = \begin{cases}
            S(t) & 1 \le t \le \delta\\
            A(t) & \delta < t \le 365 
           \end{cases},
\end{equation}
where $\delta$ is the time $t$ where $S(t) = A(t)$ and is itself a random variable equal to $(\alpha_3\alpha_4 + \alpha_6\alpha_7)/(\alpha_3 + \alpha_6)$. This step function provides a smooth transition between the spring and autumn components and facilitates full uncertainty quantification for $\delta$ and those parameters that describe the transition between mid-season phenology stages, e.g., mid-growing season greenness trajectory $\alpha_5$. The function parameters are illustrated in Figure~\ref{fig:model-graphic}.

\subsection{Bayesian model}\label{sec:posterior}
VIs derived from remotely sensed data typically have bounded support, e.g., index values can range from -1 to 1, or perhaps 0 to 1. Most functions proposed to approximate the seasonal trend in greenness are explicitly designed to accommodate this bounded support, e.g., leading to the logistic forms of $S(t)$ and $A(t)$ in \citet{melaas2013} and our proposed $G(t)$. In such settings, the posited statistical model, within which $G(t)$ is couched, should also only be defined over the support of the VI. It is generally not desirable to use a statistical model that does not match the support of the response variable, as doing so can cause issues with parameter inference, uncertainty quantification, and subsequent prediction. In the current setting, such a mismatch occurs when a Normal likelihood is used to model a bounded VI as a response variable (this is the common approach used for model-based VI function fitting in the studies noted above). In this case, issues could arise because the Normal likelihood assumes the response variable and resulting model residuals can take any value on the whole real line---when in fact both the response and residuals are bounded. From a practical standpoint, applying a Normal likelihood here can inflate the estimated variance and generate fitted values and predictions outside the response variable's support \citep{greene2008}. A standard fix in these cases is to transform the response variable so that its support becomes the real line. The key drawback to this fix, however, is loss of model parameter interpretation with respect to the original response variable. Interpretation of these model parameters are key to many of the phenological characteristics of interest, so we would like to avoid transformation if possible.

We define a general Bayesian model that accommodates (\ref{eqn:G}) for a variety of likelihoods with bounded and unbound support. The joint posterior distribution of the model's parameters $\balpha = (\alpha_1, \alpha_2, \ldots, \alpha_7)^\top$ and likelihood variance $\sigma^2$ is
\begin{equation}\label{eqn:post}
    p(\boldsymbol{\alpha},\sigma^2|\boldsymbol{y}) \propto \prod_{t=1}^T\mathcal{L}(y(t)\given G(t;\balpha),\sigma^2, \;\cdot\;)\times \prod_{j=1}^7 U(\alpha_j\given a_{\alpha_j},b_{\alpha_j})\times IG(\sigma^2\given a_{\sigma},b_{\sigma}),
\end{equation}
where $\by = (y(1), y(2), \ldots, y(T))^\top$ is the vector of observed VI values, $\mathcal{L}$ is the posited likelihood, $U(\cdot\given a_{\cdot}, b_{\cdot})$ is a Uniform prior distribution with lower and upper bounds $a$ and $b$, respectively, and $IG$ is an inverse-Gamma prior distribution with shape and scale hyperparameters $a_{\sigma}$ and $b_{\sigma}$, respectively.

\subsubsection{Candidate likelihoods}\label{sec:likelihoods}
In the subsequent analyses we apply (\ref{eqn:post}) using a Normal, Truncated Normal, and Beta likelihood. We denote the Normal likelihood as $\mathcal{L}_N(y(t)\given G(t;\balpha),\sigma^2)$. For values of $y(t)$ from $a$ to $b$, the Truncated Normal is denoted $\mathcal{L}_{TN}(y(t)\given G(t;\balpha),\sigma^2, a, b)$. For values of $y(t)$ outside the interval $\left[a,b\right]$ the Truncated Normal likelihood equals 0, see, e.g., \cite{laxman1989} and \cite{greene2008} for more details. Lastly, we denote the Beta likelihood defined for values of $y(t)$ in the interval $(0,1)$ as $\mathcal{L}_{B}(y(t)\given G(t;\balpha),\sigma^2)$. \cite{Smithson06} offer a mild transformation that allows the response variable to assume values equal to 0 and 1, i.e., to accommodate the inclusive interval $\left[a,b\right]$ (although such a transformation is likely not needed in practice because values of exactly 0 and 1 are rare). Following \cite{ferrari2004}, we use the variate mean and a precision parameterization of the Beta likelihood where the precision is defined as $1/\sigma^2$.

\subsubsection{Prior distributions}\label{sec:priors}

Given the somewhat complex nature of the non-linear function (\ref{eqn:G}), some care is needed when selecting prior distributions to ensure: $i$) identifiability of model parameters; $ii$) sufficient flexibility to accommodate a variety of phenology curves; $iii$) meaningful quantification of uncertainty in parameter estimates. Selection of several prior distributions depends on the theoretical support of the VI, which we define as $\bgamma = (\gamma_1, \gamma_2)$ where the elements hold the lower and upper bounds, respectively. For example, EVI typically ranges from $\gamma_1=0$ to $\gamma_2=1$ in forested settings. As noted in (\ref{eqn:post}), we use a Uniform prior distribution for each element in $\balpha$. The seasonal minimum $\alpha_1$ has prior support from $\gamma_1$ to $\gamma_2$. We constrain the seasonal amplitude $\alpha_2$ with support from 0 to $\gamma_2-\alpha_1$. This prior upper-bound on $\alpha_2$ ensures $\alpha_1+\alpha_2$ does not exceed the VI's theoretical maximum $\gamma_2$. Constraining $\alpha_3$ and $\alpha_6$ to support between 0 and 1 keeps estimated green-up and green-down rates within biologically sensible ranges. We ensure the spring inflection point $\alpha_4$ occurs before the autumn inflection point $\alpha_7$, by setting its prior support from 0 to $\alpha_7$. We make explicit the assumption that the VIs are observed over a year by setting the support for $\alpha_7$ from 1 to 365 (other ranges could be used). 

For the midsummer slope parameter $\alpha_5$, one might constrain it to be between 0 and a small positive value, say 0.01, to capture the common slight decrease in midsummer greenness (note, given the parameterization of (\ref{eqn:S}) and (\ref{eqn:A}) a positive $\alpha_5$ equates to a negative slope). Additional flexibility for slight midsummer greening can be accommodated by setting $\alpha_5$'s prior to be bounded by values centered on 0, e.g., from -0.01 to 0.01. This would work as long as the sum of $\alpha_1$, $\alpha_2$, and additional greening captured by a negative $\alpha_5$ does not exceed $\gamma_2$. Of course this minor wrinkle could be solved by redefining the upper bound on $\alpha_2$, but we have not seen the need for this in practice, because rarely does $\alpha_1+\alpha_2$ come close to $\gamma_2$ and there is rarely substantial midsummer greening. Finally, we prescribe $\sigma^2$ an inverse-Gamma $IG(a_\sigma, b_\sigma)$ with shape $a_\sigma$ and scale $b_\sigma$. Given an $a_\sigma=2$ this prior's mean is equal to $b_\sigma$ and variance is infinite. A summary of the default prior distributions and hyperparameters used in the analyses presented in Section~\ref{sec:case_study} are given in (\ref{eqn:priors}).
\begin{equation}\label{eqn:priors}
  \begin{split}
    \alpha_1 &\sim \U(\gamma_1,\gamma_2)\\
    \alpha_2 &\sim \U(0,\gamma_2-\alpha_1)\\
  \end{split}
\quad\quad
  \begin{split}
    \alpha_3 &\sim \U(0,1)\\
    \alpha_4 &\sim \U(1,\alpha_7)
  \end{split}
  \quad\quad
    \begin{split}
    \alpha_5 &\sim \U(-0.01,0.01)\\
    \alpha_6 &\sim \U(0,1)\\
  \end{split}
\quad\quad
  \begin{split}
    \alpha_7 &\sim \U(1,365)\\
    \sigma^2&\sim IG(2,\cdot)
  \end{split}
\end{equation}

\subsection{Implementation and inference}\label{sec:implementation}
Posterior inference about $\balpha$ and $\sigma^2$ is based on MCMC samples that are generated using a Metropolis sampling algorithm \citep{gelman2013}. With an aim to deliver inference for massive datasets comprising 10s of millions of pixels, we developed an MCMC algorithm and associated software tailored specifically to efficiently evaluate (\ref{eqn:post}) with the flexibility to accommodate different candidate likelihoods and prior specifications. This work was implemented in \pkg{rsBayes}, an open source \R package that provides a function to generate the desired MCMC samples and a set of supporting utility functions for efficiently summarizing model parameters and other derived quantities of interest \citep{rsBayes}.

A key advantage to using a Bayesian model is that we can sample the posterior distribution for any function of parameters in $\balpha$ and $\sigma^2$ at each pixel \citep{gelman2013}. Two posterior distributions of immediate interest are the model fitted and predicted values. A single sample from these, and other, derived posterior distributions is generated as output from the function of inferential interest given input of one post-convergence joint posterior sample from (\ref{eqn:post}). For example, to generate $M$ samples from the posterior distribution of fitted values for observed time point $t$ we use $\hat{y}(t)^{l} = G(t; \balpha^{l})$ for $l = 1, 2, \ldots, M$ where $\balpha^{l}$ is the $l$-th post-convergence sample of $\balpha$. Similarly, to generate $M$ samples from the posterior predictive distribution of a new time point $t_0$, we draw $\tilde{y}(t_0)^{l}$ as a random variate from the distribution corresponding to the likelihood used in (\ref{eqn:post}), i.e., draws from a Normal, Truncated Normal, or Beta distribution with mean $G(t; \balpha^{l})$ and variance $\sigma^{2,l}$.

Posterior distributions for other phenological characteristics of interest are generated in a similar fashion. For example, given $M$ post-convergence MCMC samples of $\balpha$, the corresponding $M$ samples of season length, say $\lambda = \alpha_7 - \alpha_4$, are obtained one-for-one $\lambda^{l} = \alpha_7^{l}-\alpha_4^{l}$ for $l = 1, 2, \ldots, M$. This inferential approach extends to more complex functions involving $\balpha$. For example, say we would like to estimate the posterior distribution for the area under the phenology curve (AUC) over some time interval from $t_1$ to $t_2$ and denote this area as $\Phi$. Again, the machinery is the same, we generate $M$ samples from $\Phi$'s posterior by evaluating $\Phi^{l} = \int_{t_1}^{t_2} G(t; \balpha^{l})dt$ for each of $l = 1, 2, \ldots, M$ posterior samples from $\balpha$. Given the form of the integrand function $G$ we could employ any number of numerical quadrature methods to estimate the area. This mechanism for estimating posterior distributions for parameters derived from $\balpha$ offers some exciting opportunities to explore uncertainty/sensitivity of complex quantities, e.g., phenology curve characteristics derived via differential equations \citep{zhang2003} or output from ecosystem process models that ingest phenology information as MCMC samples.

Given $M$ samples from the posterior of $\balpha$, $\sigma^2$, and functions of these parameters, point and interval summaries are easily computed (see, e.g., \cite{gelman2013} for details). In the subsequent analyses we use the median, standard deviation, and 95\% credible interval computed using $M$ post-convergence samples as point and associated uncertainty estimates for a given pixel. Importantly, these posterior summaries can be generated for each pixel and used to map parameter and derived quantities of interest with associated uncertainty.

\section{Analyses of forest phenology inferred from Harmonized Landsat Sentinel-2 (HLS) data}\label{sec:case_study}
The proposed modeling approach efficacy was assessed by estimating and analyzing key LSP characteristics using multi-year and multi-pixel reflectance data from the HLS tile described in Section~\ref{sec:data}. The subsequent analyses consider the performance of the proposed phenology function (\ref{eqn:G}) and estimation of its parameters using the Bayesian model (\ref{eqn:post}). Posterior summaries for model parameters, along with derived quantities such as the maximum greenness and area under the phenology curve (AUC) were collected using the \code{pheno} function in the \pkg{rsBayes} \R package. 

Two case studies are detailed in the subsequent sections. Section~\ref{sec:tile-analysis} offers an analysis that yields pixel-level posterior distributions for the proposed phenology function and error model variance parameters. Here, the phenology function is estimated using composite EVI data, i.e., combining EVI observations over available years to estimate long-term average LSP parameters (2013-2019). Point and dispersion summaries for the pixel-level posterior distributions are then presented as raster data layers and compared with 2016 National Land Cover Database \cite[NLCD;][]{nlcd} forest categories. Section~\ref{sec:quabbin-analysis} offers an analysis of a small area within the larger tile. This analysis assesses the robustness of phenology function parameters estimated using a small number of EVI observations collected on an annual time-step. The analysis also explores the usefulness of these data and inferential approach to detect interannual phenology change.

A data subset and \R code needed to reproduce key portions of the subsequent analyses is provided as Supplementary Material.

\subsection{Data}\label{sec:data}
\begin{figure}[!ht]
  \centering
    \subfigure[Study area map.]{\includegraphics[height=.52\textheight, trim = 4cm 0 4cm  0, clip]{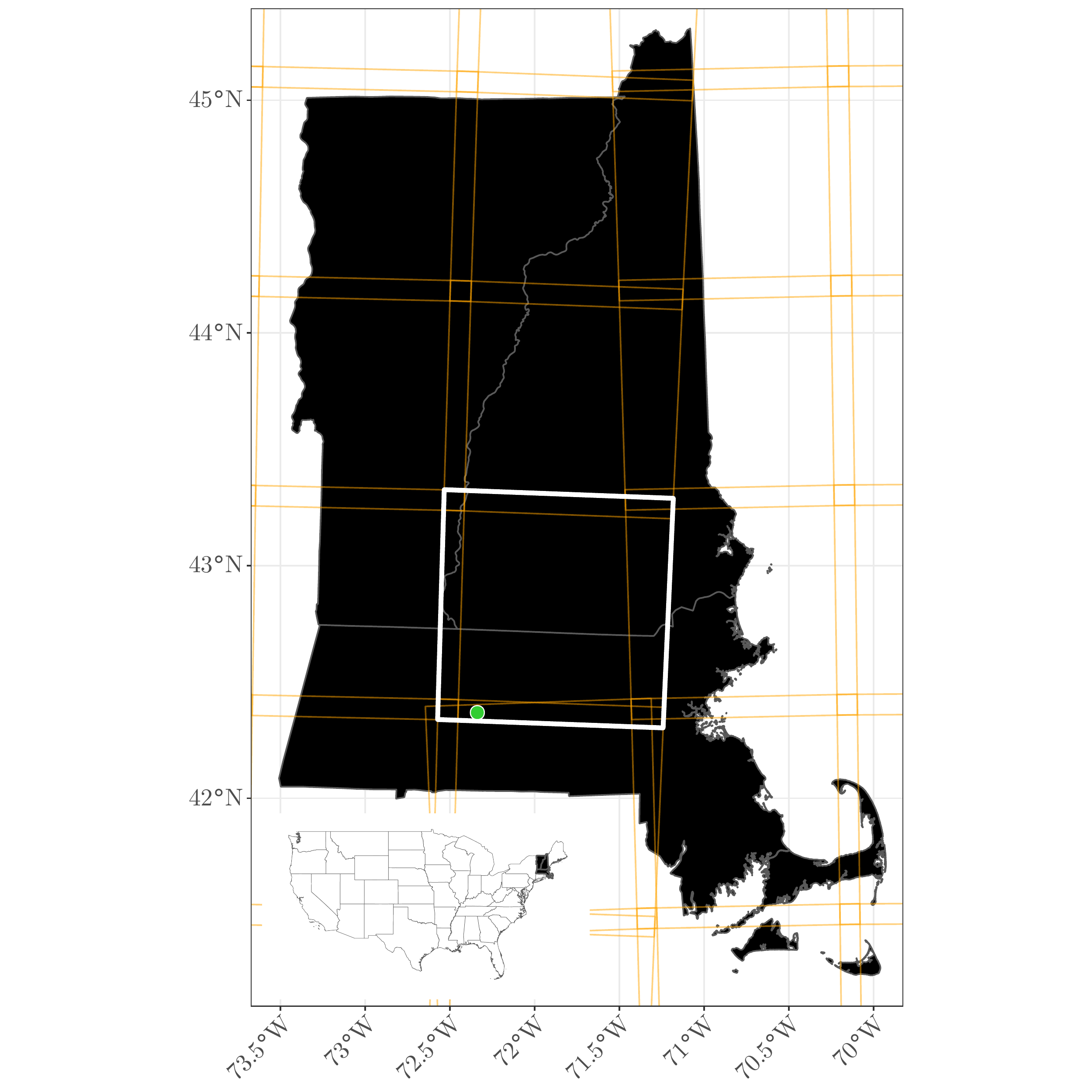}\label{fig:spatial_coverage}}
    \subfigure[Available L30 (blue) and S30 (red) HLS images.]{\includegraphics[height=.52\textheight]{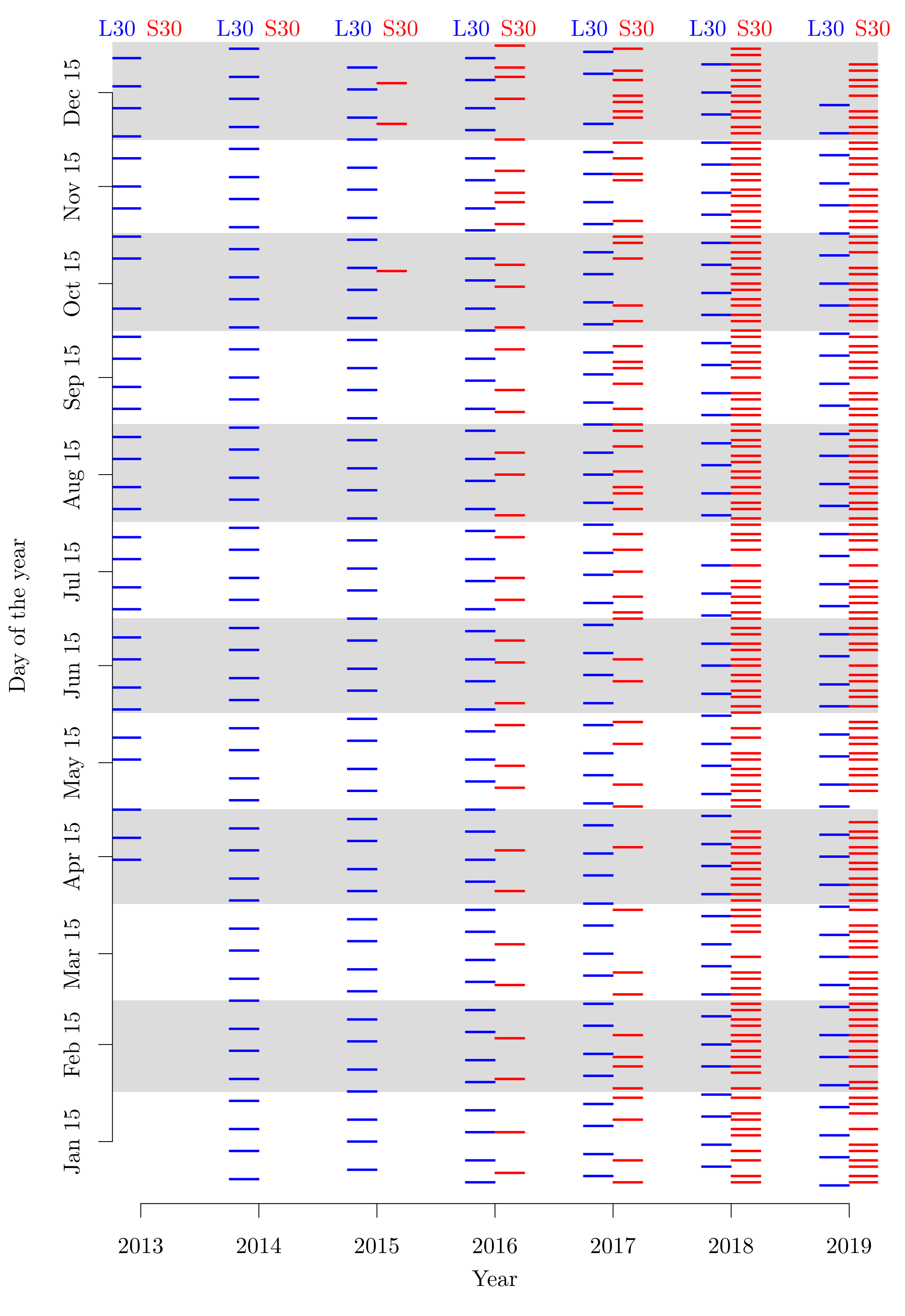}\label{fig:time_coverage}}\\
    \caption{\subref{fig:spatial_coverage} White box delineates the Harmonized Landsat Sentinel-2 (HLS) study tile 18TYN analyzed in Section~\ref{sec:tile-analysis}. The green point identifies the location of analysis in Section~\ref{sec:quabbin-analysis}. \subref{fig:time_coverage} Availability of Landsat and Sentinel-2 data in the study tile delineated in Figure~\ref{fig:spatial_coverage} over time.}\label{fig:data}
\end{figure}

\begin{figure}[!ht]
    \begin{center}
    \subfigure[Number of cloud-free HLS EVI observations.]{\includegraphics[width = .45\textwidth]{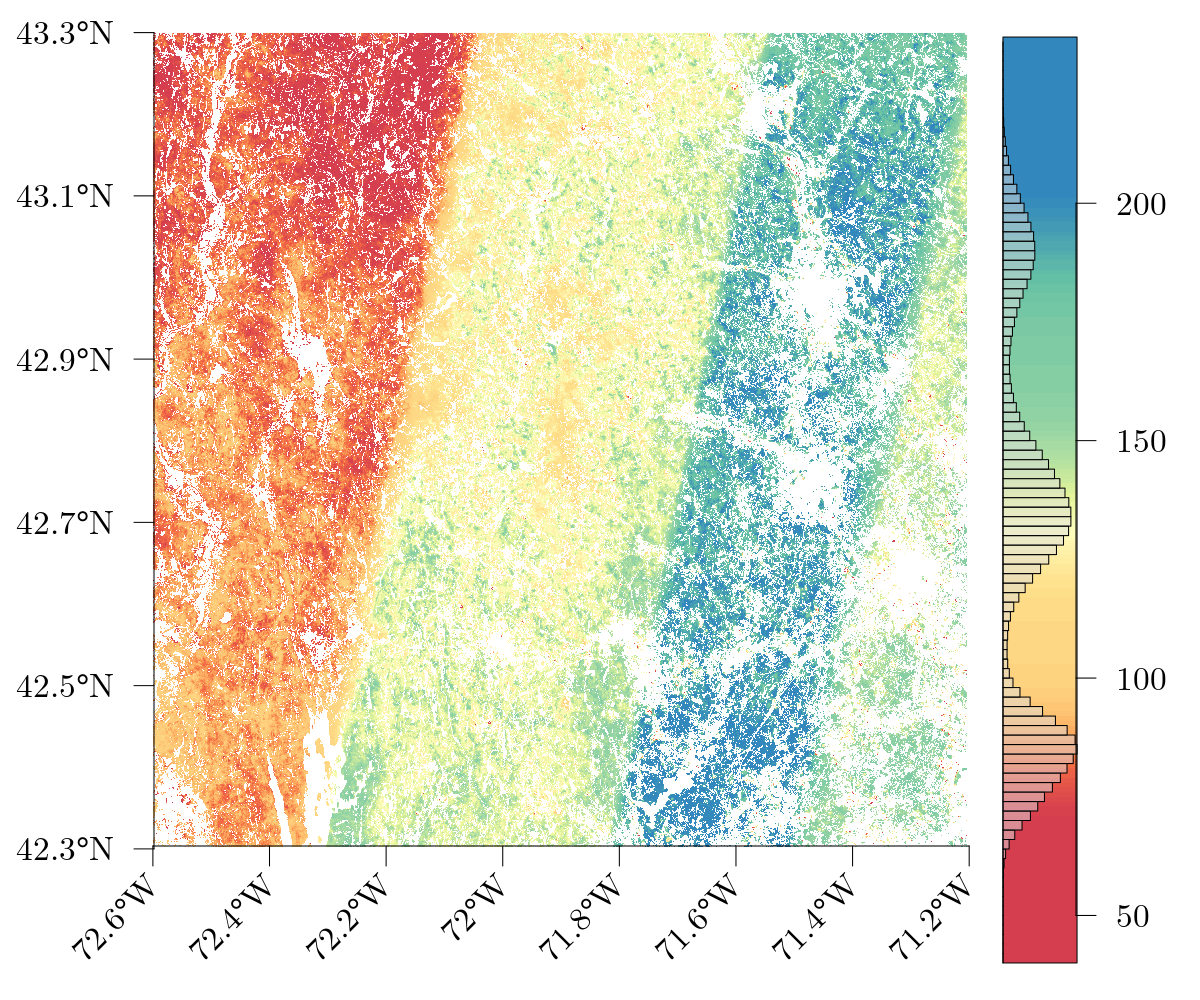}\label{fig:nobs}}
    \subfigure[Digital elevation model.]{\includegraphics[width = .45\textwidth]{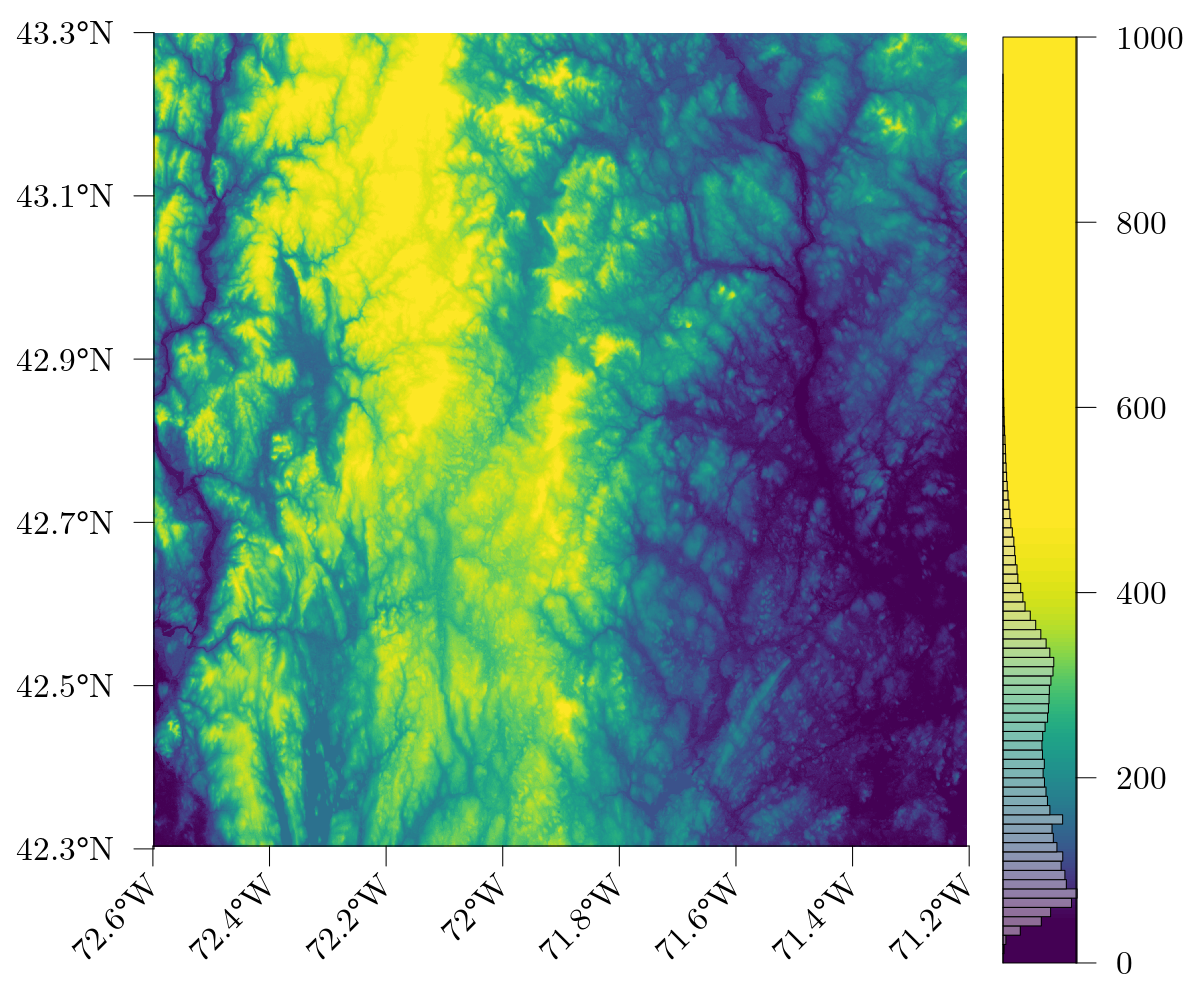}\label{fig:dem}}
    \end{center}    
    \caption{\subref{fig:nobs} Number of EVI observations at each 30 m pixel in the study tile. The histogram in the scale bar shows frequency of the given number of observations within the color range. \subref{fig:dem} Digital elevation model derived from Shuttle Radar Topography Mission (SRTM) data for the study tile. Elevation units are meters.}\label{fig:time_coverage}
\end{figure}

The motivating dataset comprised all 30 m HLS reflectance data (the L30 and S30 products) from January 1, 2013 to December 31, 2019, for tile ID 18TYN (available via \url{https://hls.gsfc.nasa.gov/data}). The tile, delineated by the white box in Figure~\ref{fig:spatial_coverage}, is positioned in the northeastern United States. The HLS imagery temporal resolution is given in Figure~\ref{fig:time_coverage} and shows sparse coverage from 2013 to 2016 when only Landsat 8 data are available. Then, with the addition of Sentinel-2 data, temporal resolution increases from 2016 through 2019. Within the tile, the availability of EVI observations used in the subsequent analyses varies based on satellite orbital path and cloud cover as shown in Figure~\ref{fig:nobs}. A total of 8,698,531 forested pixels, as specified by 2016 NLCD codes 41, 42, and 43, collectively comprising 1,101,352,942 EVI observations (i.e., sum of the pixel values in Figure~\ref{fig:nobs}) were retained for subsequent analyses. 

\begin{figure}[!ht]
    \centering
     \subfigure[]{\includegraphics[width = .45\textwidth]{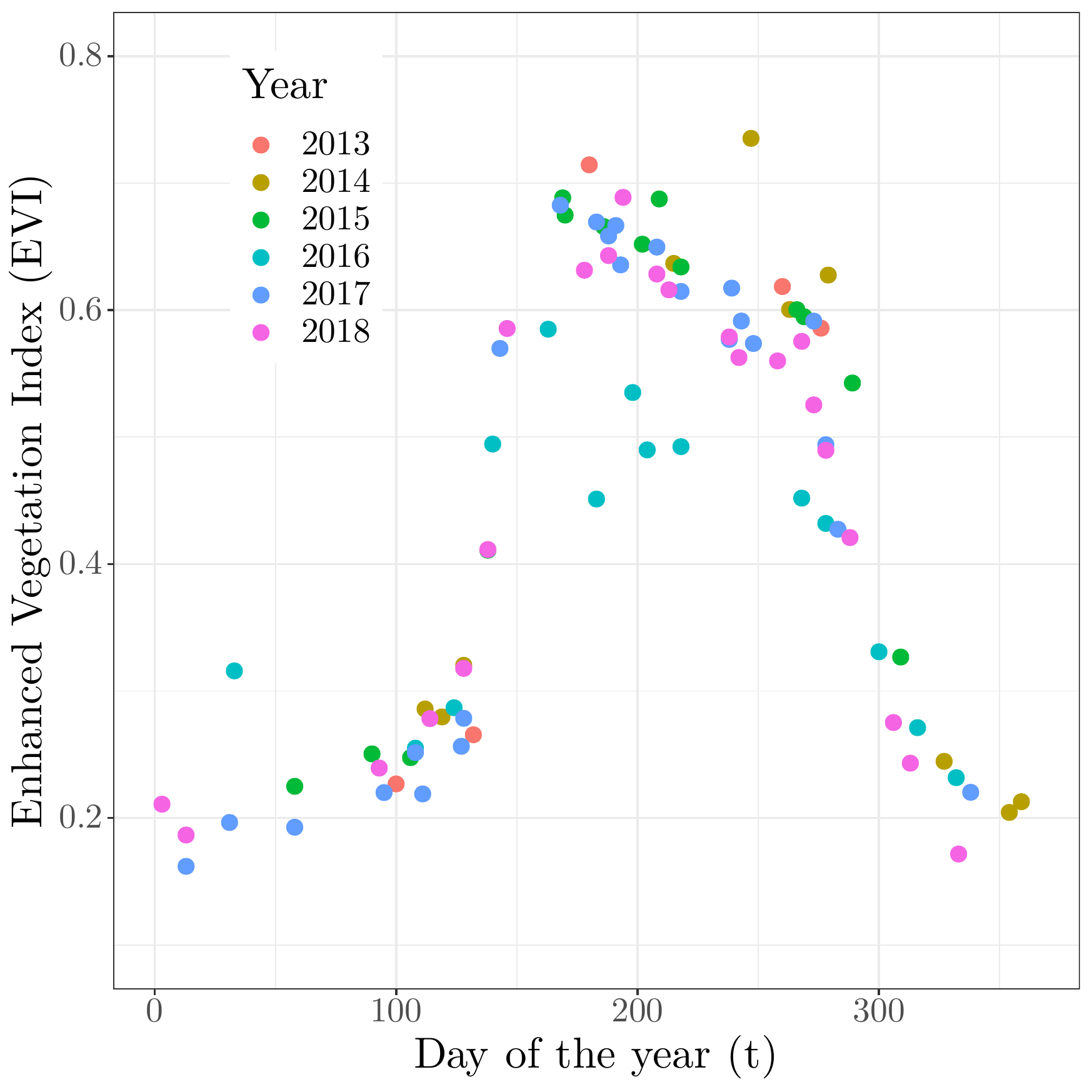}\label{fig:year_example_pixel_data}}
     \subfigure[]{\includegraphics[width = .45\textwidth]{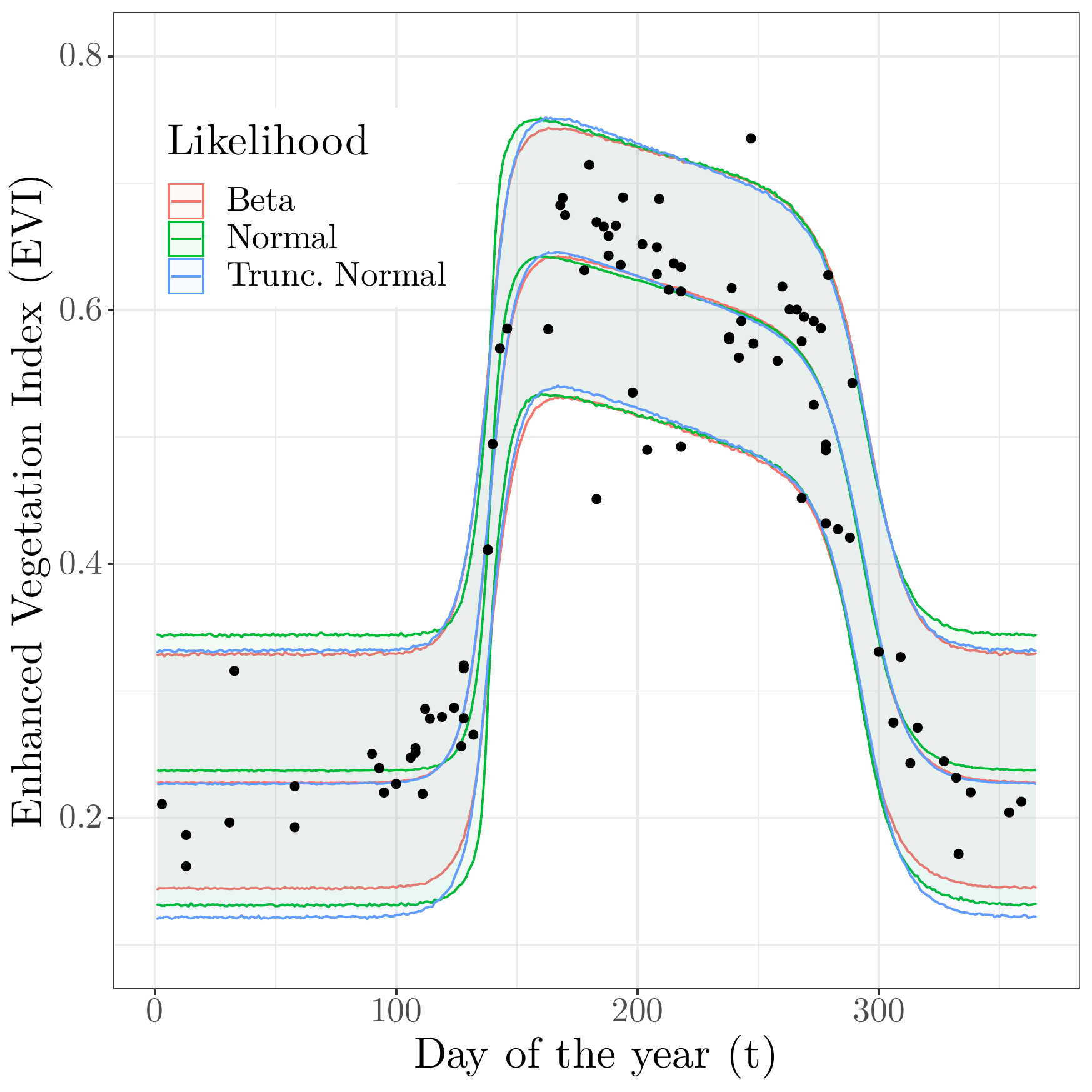}\label{fig:pred_example_pixel_data}}
    \caption{Example of composite Harmonized Landsat Sentinel-2 Enhanced Vegetation Index (EVI) time series for a continuously forested pixel. \subref{fig:year_example_pixel_data} EVI values by year. \subref{fig:pred_example_pixel_data} Posterior predictive distribution median and 95\% credible interval band for each of the three candidate likelihoods.}\label{fig:example_pixel_data}
\end{figure}

\begin{figure}[!ht]
    \begin{center}
    \includegraphics[width = .65\textwidth]{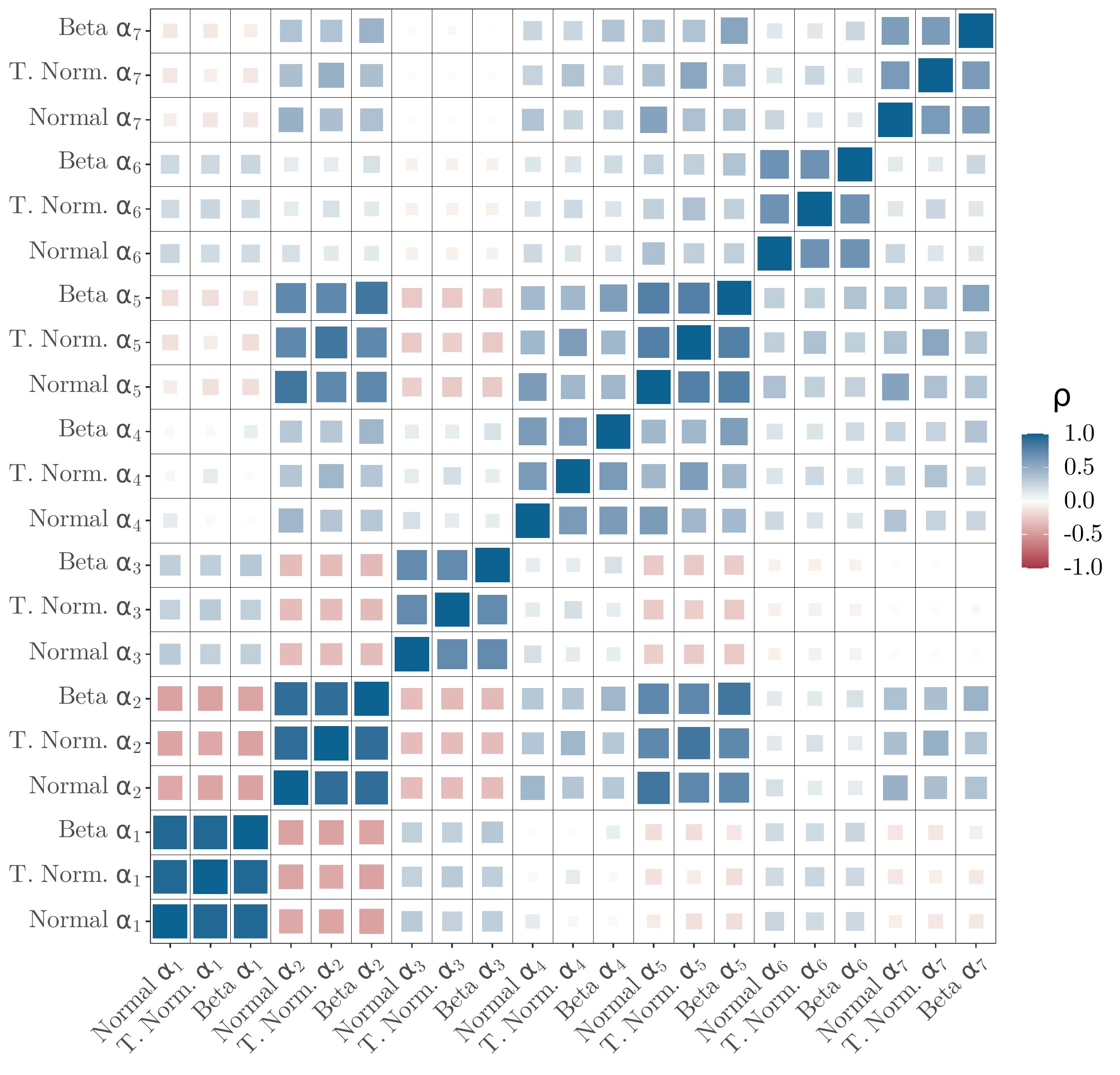}
    \end{center}    
    \caption{Person correlation ($\rho$) between phenology function (\ref{eqn:G}) parameters estimated using (\ref{eqn:post}) with the Normal, Truncated Normal (T. Norm.), and Beta likelihoods for all pixels in the study tile.}\label{fig:likelihood_corr}
\end{figure}

\subsection{Tile-wide composite long-term average LSP estimates and comparison with NLCD}\label{sec:tile-analysis}

The proposed phenology function (\ref{eqn:post}) was fit using a Normal, Truncated Normal, and Beta likelihood to each pixel's composite EVI observations within the study tile. Pooling the EVI data to produce a composite phenology curve is common practice when there is a paucity of observations within any given year \citep{fisher2006}. Figure~\ref{fig:example_pixel_data} illustrates composite EVI data for a typical pixel that remained forested over the seven year study period. Depending on the study objectives, pooling EVI observations across years might not be ideal, especially when there is large interannual variation in EVI values resulting from forest disturbance or other factors. Such settings have motivated work by \cite{senf2017} who propose partial pooling by introducing a year random effect on select parameters in (\ref{eqn:S}), (\ref{eqn:A}), and similar LSP functions. While the addition of such random effects is beyond the scope of this study, it is a next logical step in the development and discussed further in Section~\ref{sec:discussion-summary}. 

As noted in Section~\ref{sec:model}, there is sound theoretical motivation to use a likelihood that matches the bounded support of the VI response variable, i.e., we would favor use of the Truncated Normal or Beta distribution to estimate the phenology function's parameters given EVI. However, for the composite EVI data considered here, the three likelihoods delivered highly comparable parameter estimates and fits to the observed data. As an example from one pixel, Figure~\ref{fig:pred_example_pixel_data} shows the posterior predictive distribution median and 95\% credible interval bound for each likelihood. More broadly, Figure~\ref{fig:likelihood_corr} illustrates the Pearson correlation between parameters estimated (posterior median) using the three likelihoods for all pixels in the study tile. It is evident from this figure that the choice of likelihood has negligible effect on the resulting point estimate for model parameters. Similarly, although not shown, the choice of likelihood has little effect on any given parameter's estimated posterior distribution shape and spread. As a result, there is essentially no difference between the predicted values generated at the observed time points among the three likelihoods---all three likelihoods yielded the same root mean squared error of 0.05 and nominal 95\% posterior predictive distribution credible interval of 95.3. Despite the theoretical argument for using a bounded likelihood, these empirical findings are not surprising because EVI observations rarely have values close to the lower bound of 0 and upper bound of 1 in this dataset. If one considered data that frequently came up against the bounds, then inferential issues could arise if a Normal error model was assumed. Given the nearly indistinguishable results among the likelihoods, the remainder of the results presented in this section will be from the Beta regression model.

\begin{figure}[!ht]
\centering
\subfigure[$\alpha_1$ median]{\includegraphics[width = .45\textwidth]{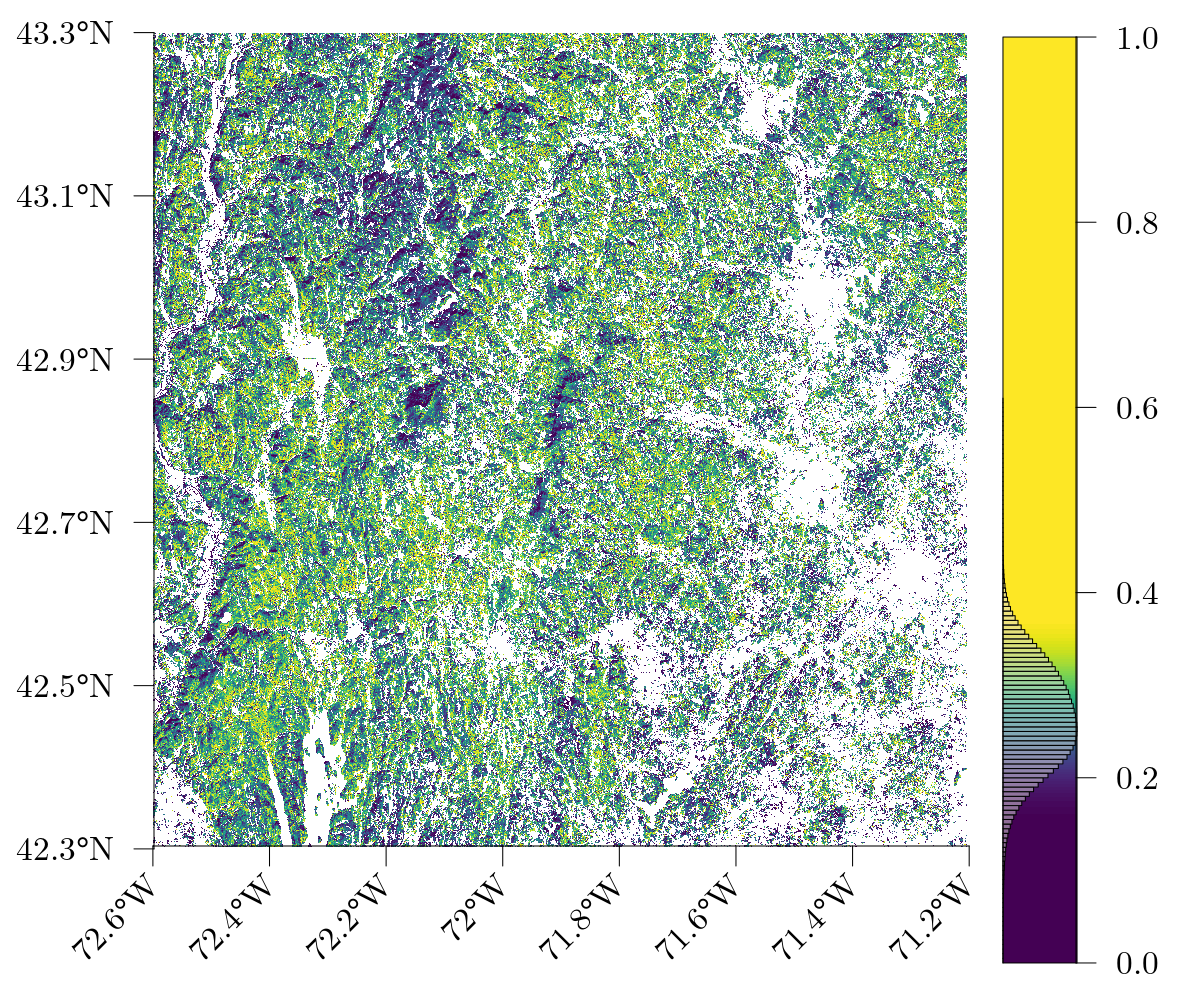}\label{fig:a1median}}
\subfigure[$\alpha_1$ standard deviation]{\includegraphics[width = .45\textwidth]{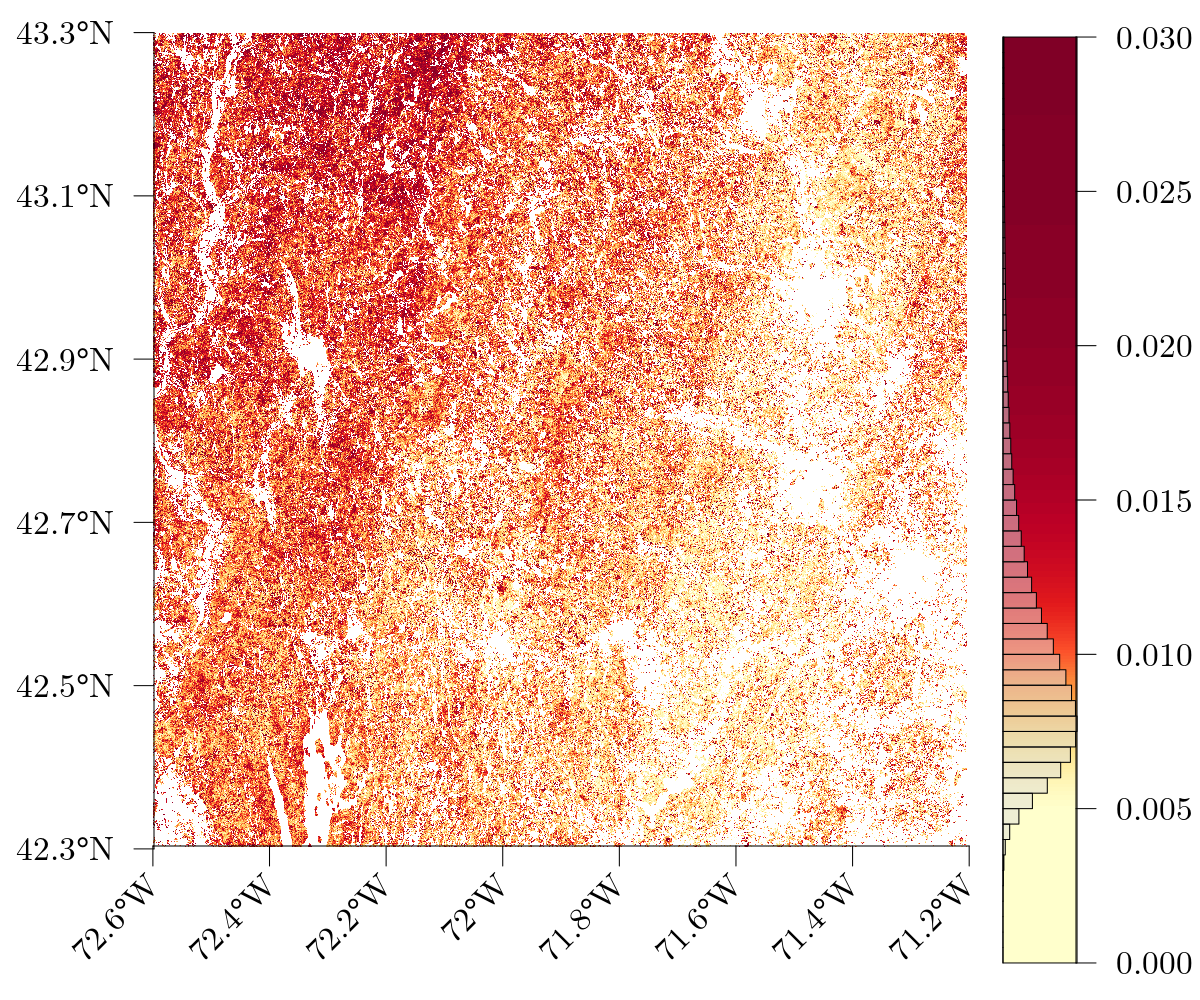}\label{fig:a1sd}}\\
\subfigure[$\alpha_2$ median]{\includegraphics[width = .45\textwidth]{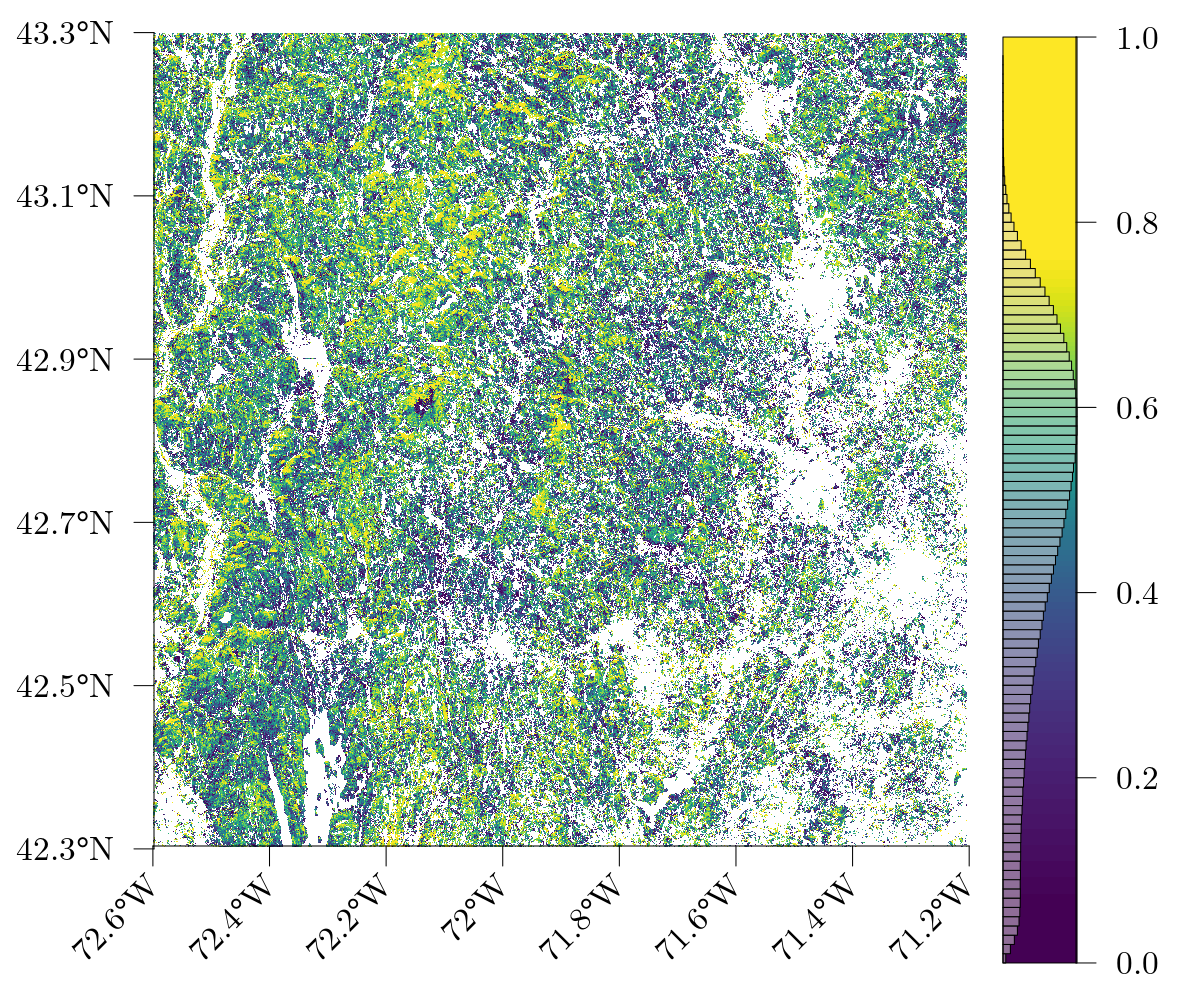}\label{fig:a2median}}
\subfigure[$\alpha_2$ standard deviation]{\includegraphics[width = .45\textwidth]{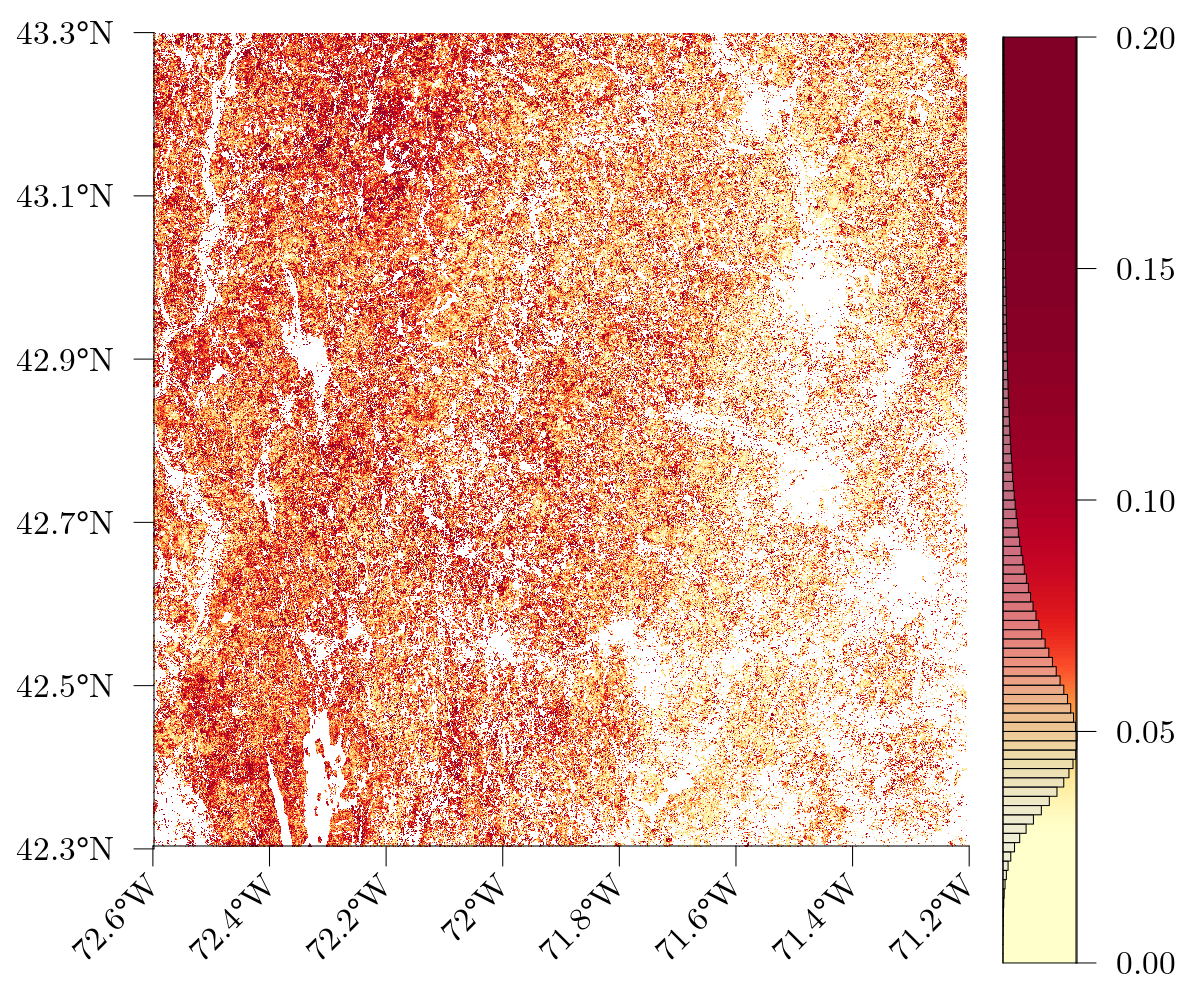}\label{fig:a2sd}}\\
\caption{Maps showing spatial distribution of long-term average (2013-2019) minimum $\alpha_1$ and amplitude $\alpha_2$ and associated uncertainties estimated using the Bayesian LSP model.}\label{fig:alpha1-2-map}
\end{figure}

Figures~\ref{fig:alpha1-2-map} and \ref{fig:alpha4-7-map} show composite (long-term average) $\alpha_1$, $\alpha_2$, $\alpha_4$ and $\alpha_7$ parameter estimates (posterior medians) and associated uncertainties (posterior standard deviations) generated using the Bayesian LSP model (with Beta likelihood) for the study tile. Corresponding maps for the remaining LSP parameters are provided as Supplementary Material. We see from the scale bar histogram in Figure~\ref{fig:a1median} that $\alpha_1$ (LSP curve minimum) tends to be between 0.2 and 0.3 with few estimates falling outside the range of 0.1 and 0.5. As seen in the scale bar histogram for $\alpha_2$ (LSP curve amplitude) in Figure~\ref{fig:a2median}, the range of estimated LSP amplitudes spans from nearly 0 to 0.85 with an apparent average around 0.6. Shown in Figure~\ref{fig:likelihood_corr}, there is a moderate negative correlation between the $\alpha_1$ and $\alpha_2$ parameter estimates. Given the interpretation that $\alpha_1$ is the LSP curve minimum and $\alpha_2$ is its amplitude, we can reasonably expect forested pixels with lower minimum greenness to tend to have higher greenness amplitudes. This phenomenon may be related to the relative proportion of evergreen and deciduous vegetation in a pixel. 
\begin{figure}[!ht]
\centering
\subfigure[$\alpha_1$]{\includegraphics[width = .32\textwidth]{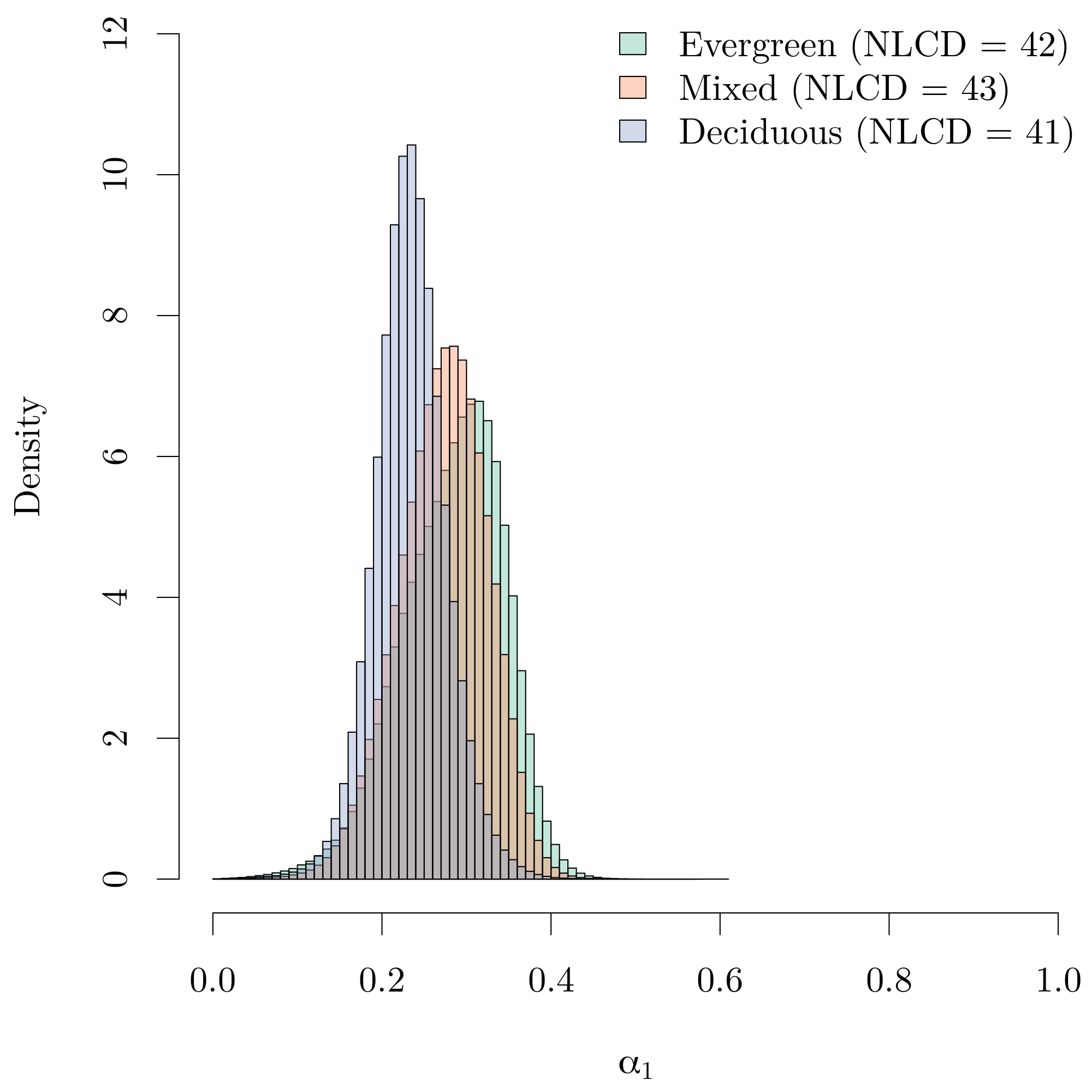}\label{fig:a1nlcd}}
\subfigure[$\alpha_2$]{\includegraphics[width = .32\textwidth]{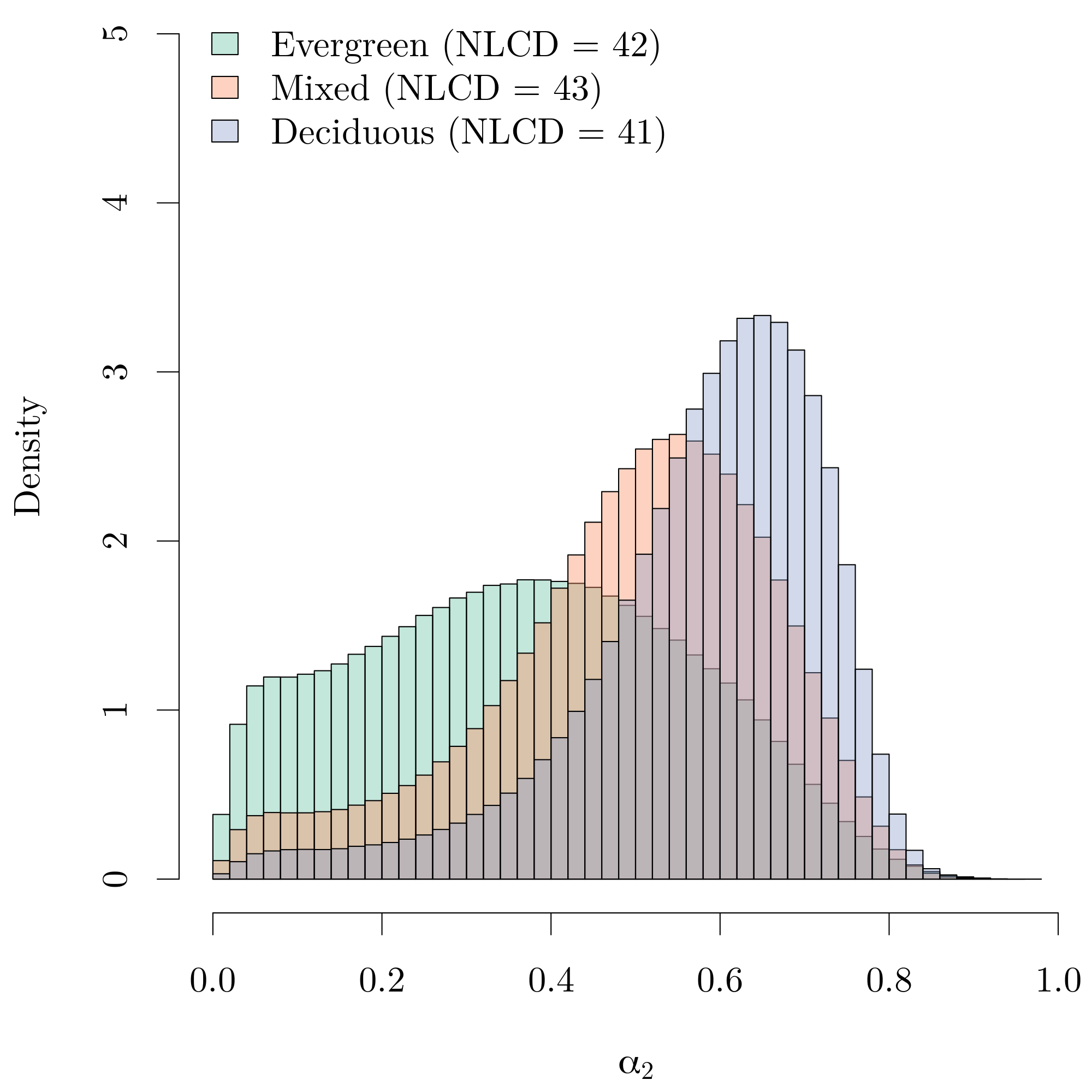}\label{fig:a2nlcd}}
\subfigure[$\alpha_1 + \alpha_2$]{\includegraphics[width = .32\textwidth]{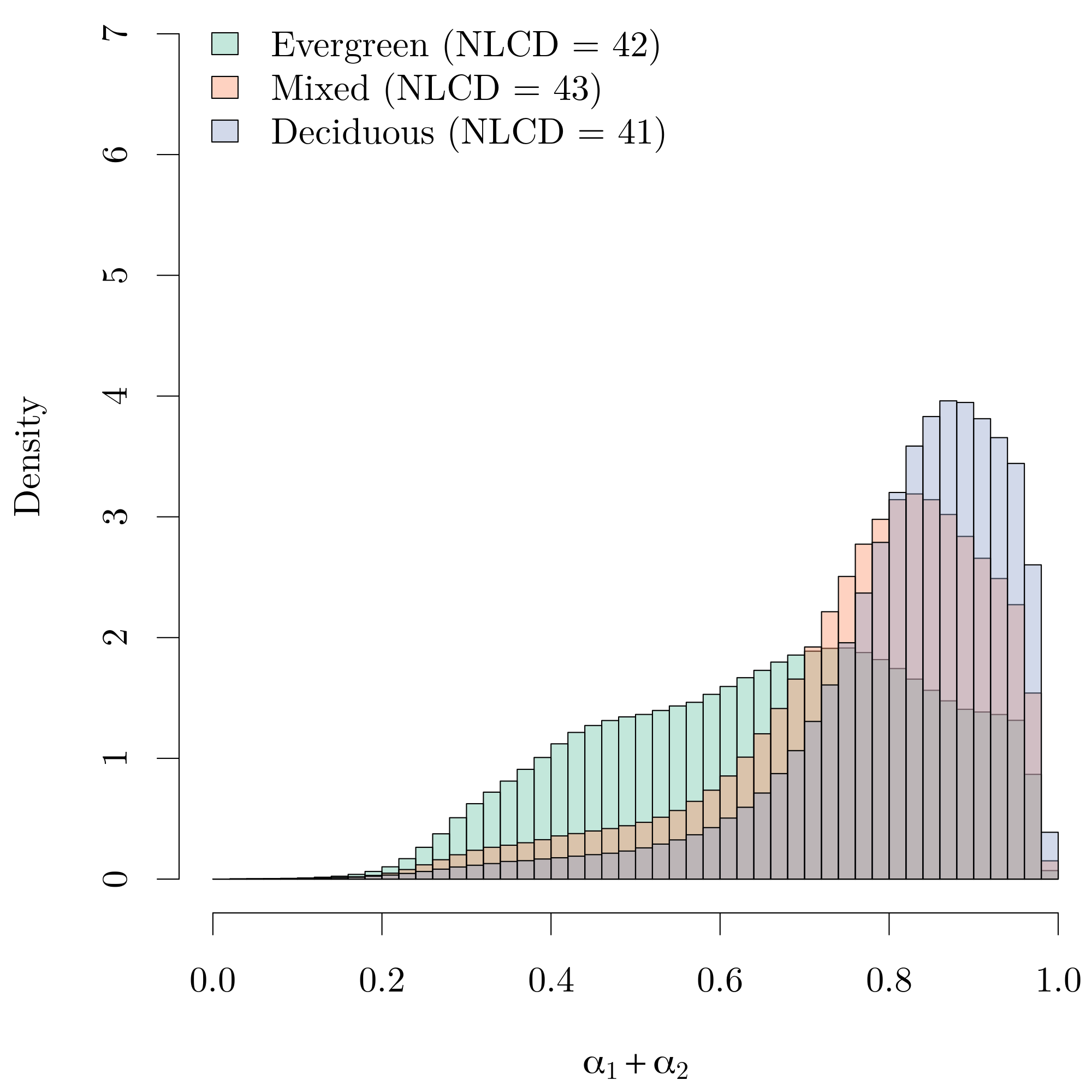}\label{a1a2nlcd}}
\caption{Histograms of composite (long-term average) $\alpha_1$, $\alpha_2$ and $\alpha_1+\alpha_2$ estimates (medians) by National Land Cover Database (NLCD) forest code.}\label{fig:histograms}
\end{figure}

Figure~\ref{fig:histograms} shows histograms for $\alpha_1$, $\alpha_2$ and $\alpha_1+\alpha_2$ (LSP curve maximum) broken out by NLCD forest type code. Figure~\ref{fig:a1nlcd} shows a tendency for deciduous-class pixels to have lower seasonal minimums than mixed- and evergreen-class pixels. We also see in Figure~\ref{fig:a2nlcd} that deciduous-class pixels tend to have higher LSP curve amplitudes than the mixed and evergreen classes. This is a sensible result. We expect deciduous forests to have lower seasonal minimums than evergreen forests due to leaf shedding. Mixed and evergreen forests will tend to have higher EVI values in the dormant season because conifers retain needles year round. We see in Figure~\ref{a1a2nlcd} that evergreen-class pixels tend to have lower LSP curve maximum estimates than mixed- and deciduous-class pixels. It is well-known that coniferous forests tend have lower summer EVI values than deciduous forests. It is encouraging to see this trend present in our LSP parameter estimates and indicates the moderate resolution LSP products using HLS data and our proposed modeling approach yield potentially insightful findings about forest phenology.

It is noticeable from Figures~\ref{fig:a1sd} and \ref{fig:a2sd} that $\alpha_1$ and $\alpha_2$ uncertainties tend to vary with EVI observation sample size (comparing to Figure~\ref{fig:nobs}). We see a tendency for higher posterior standard deviations in the north-west edge of the uncertainty maps. This is the region in the HLS tile with the fewest EVI observations per pixel, due to less sensor track swath overlap. It is even possible to make out borders where the amount of sensor path overlap changes across the tile by examining the spatial distribution of uncertainty in Figures~\ref{fig:a1sd} and \ref{fig:a2sd}. It is also apparent from the uncertainty maps that EVI observation sample size is not the only contributor to $\alpha_1$ and $\alpha_2$ uncertainty. As in any regression, parameter uncertainty is a function of sample size and the inherent variability of the data. Measures of posterior distribution dispersion, e.g., standard deviation, mapped to the pixels accounts for both sample size and data variability. Because the EVI sample size and variability is often pixel specific, it is important to provide pixel-level uncertainty assessment for LSP parameters. 

\begin{figure}[!ht]
\centering
\subfigure[$\alpha_4$ median]{\includegraphics[width = .45\textwidth]{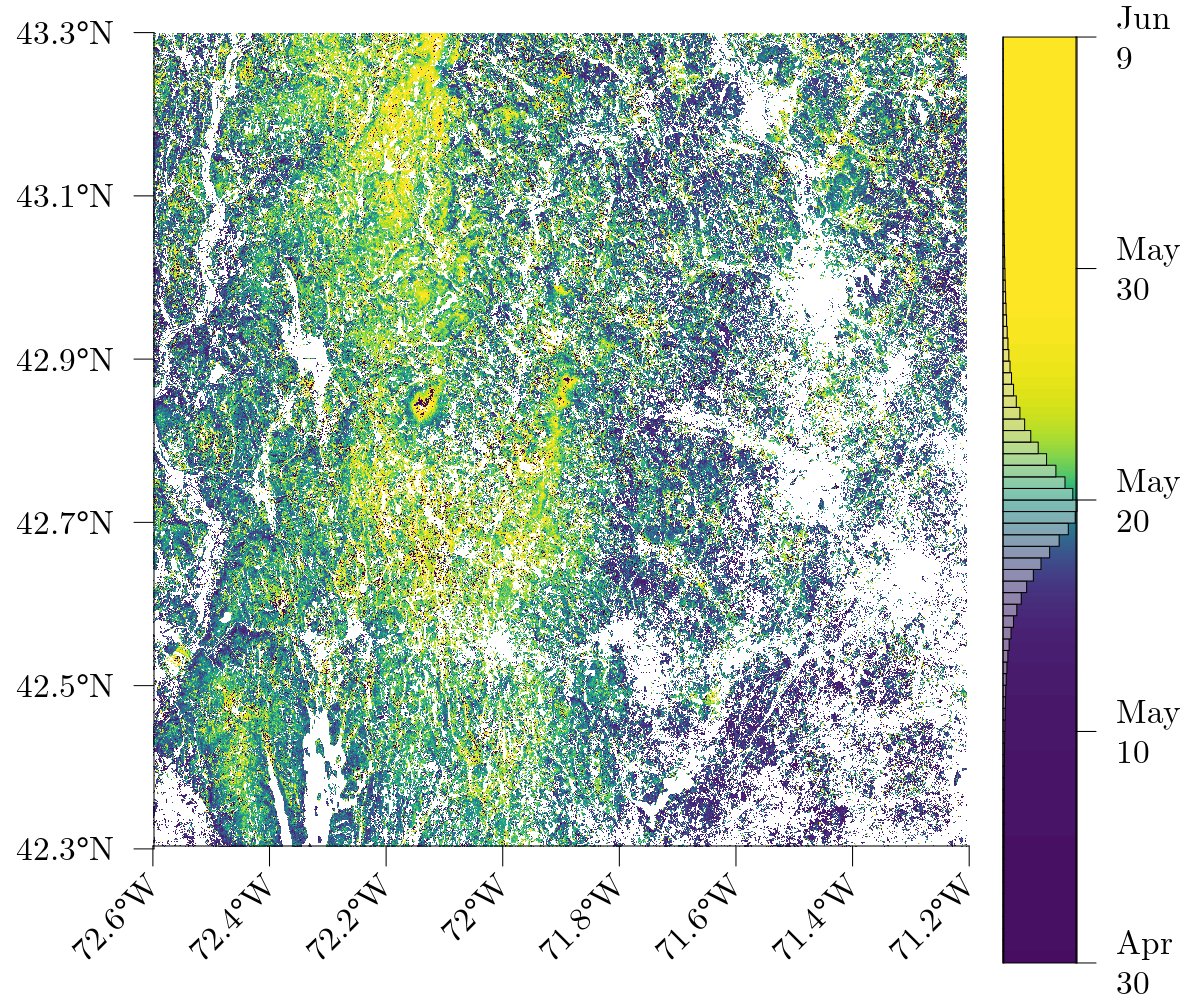}\label{fig:a4median}}
\subfigure[$\alpha_4$ standard deviation]{\includegraphics[width = .45\textwidth]{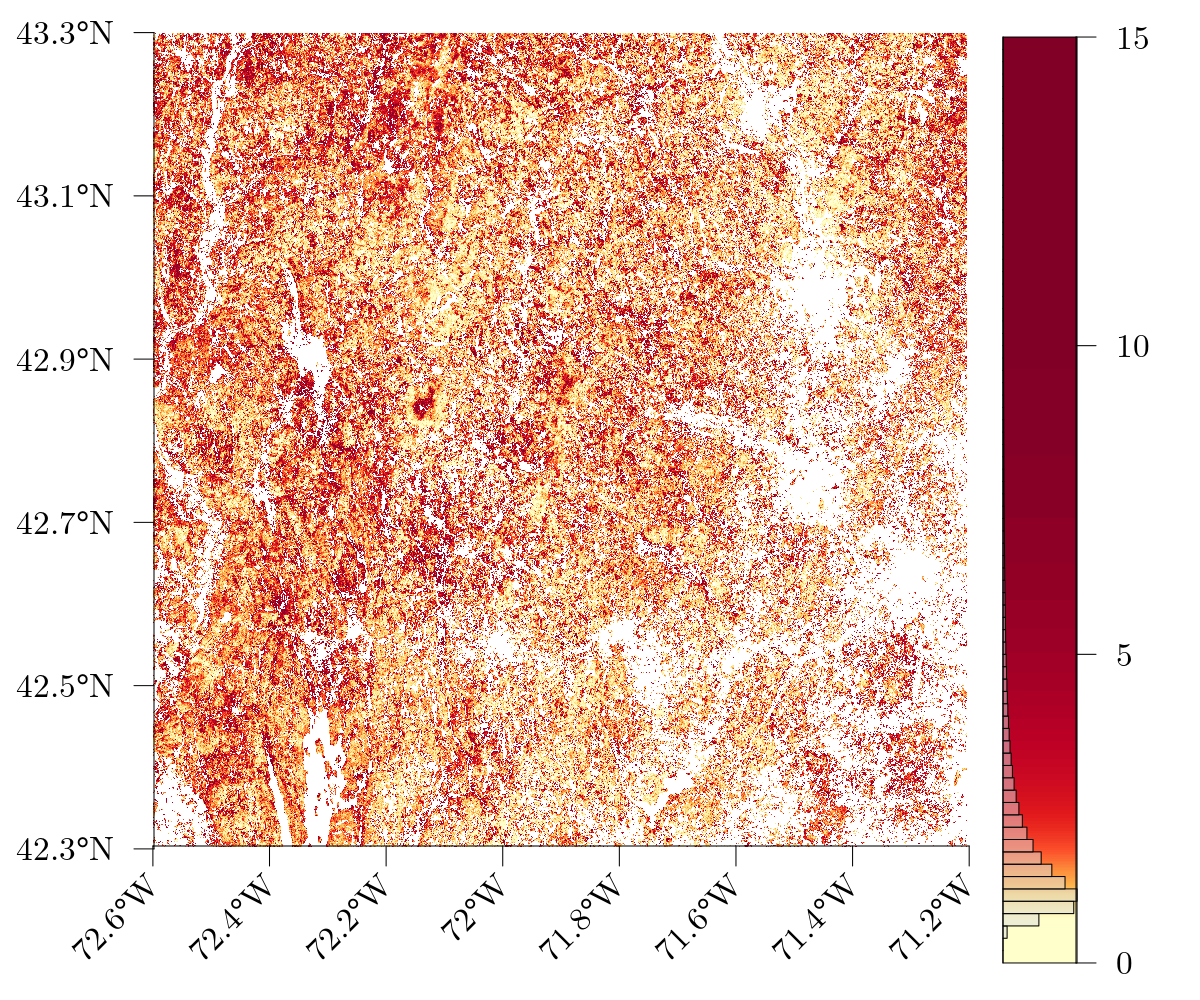}\label{fig:a4sd}}\\
\subfigure[$\alpha_7$ median]{\includegraphics[width = .45\textwidth]{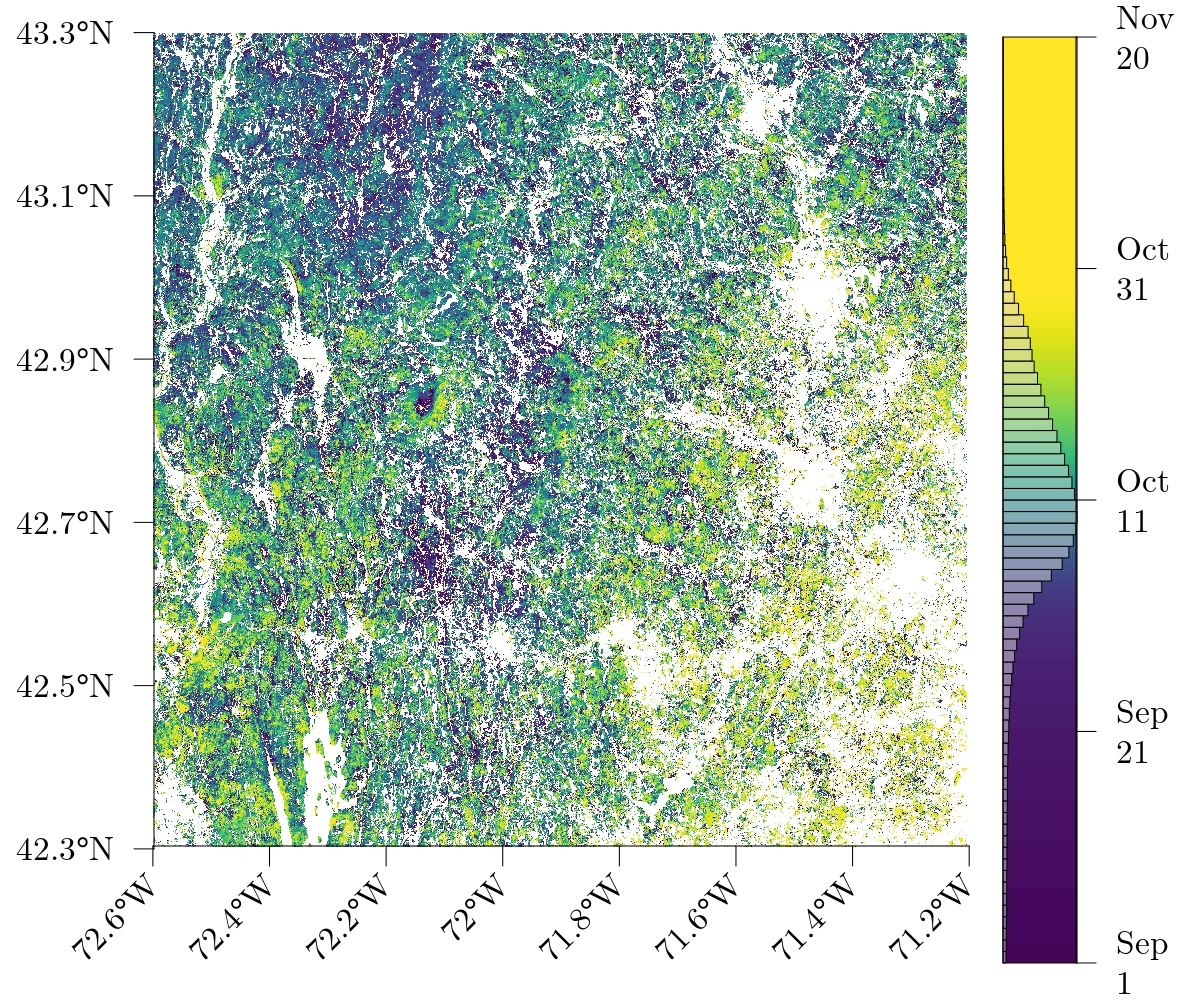}\label{fig:a7median}}
\subfigure[$\alpha_7$ standard deviation]{\includegraphics[width = .45\textwidth]{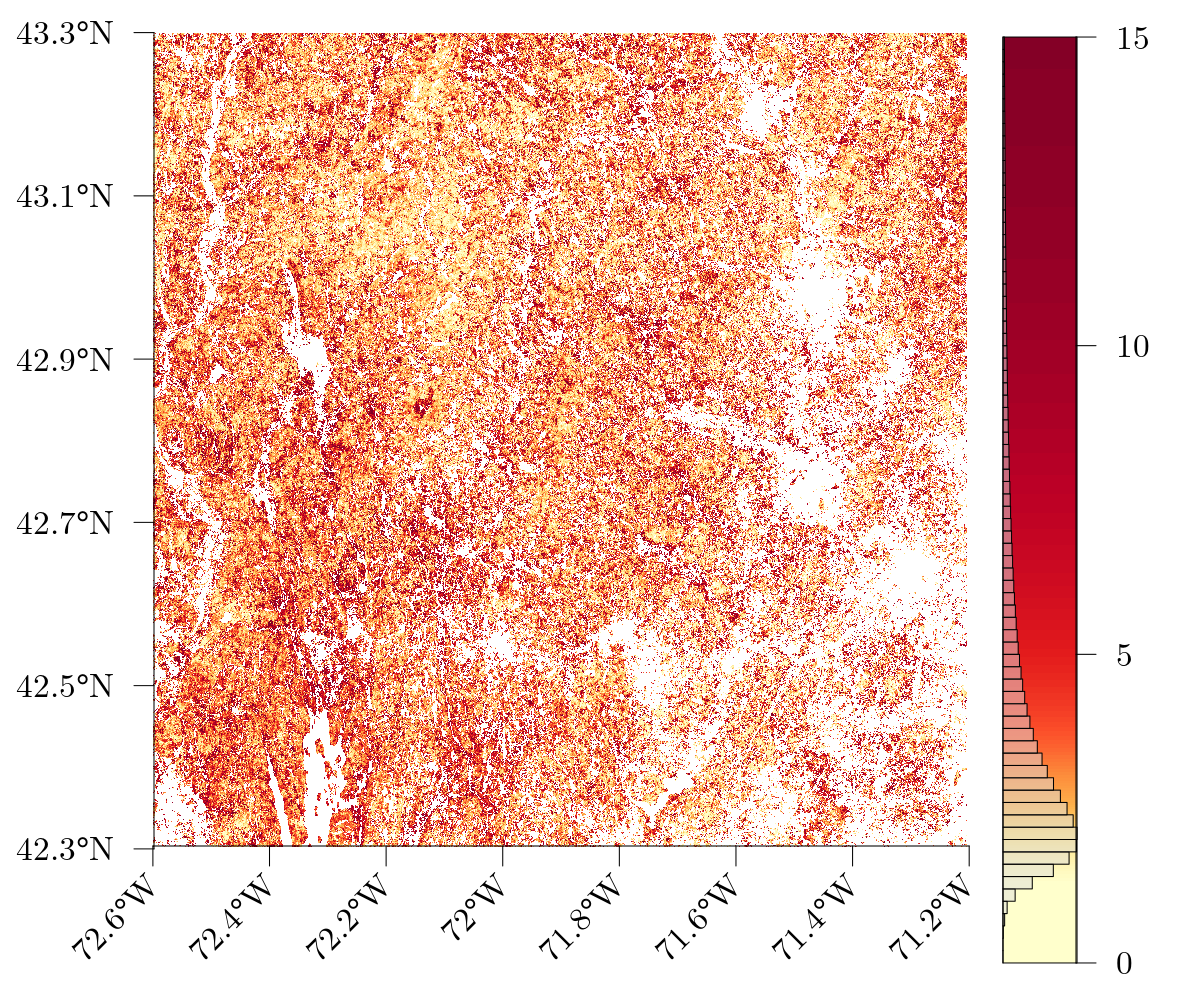}\label{fig:a7sd}}\\
\caption{Maps showing spatial distribution of long-term average (2013-2019) $\alpha_4$ and $\alpha_7$ inflection points and associated uncertainties estimated using the Bayesian LSP model.}\label{fig:alpha4-7-map}
\end{figure}

In Figure~\ref{fig:a4median} we see the start of growing season DOY ($\alpha_4$) generally shifts from later in the year to earlier as we move from west to east. We also notice in Figure~\ref{fig:a7median} the end of growing season DOY shifts ($\alpha_7$) from earlier in the year to later from west to east. Comparing to Figure~\ref{fig:dem}, it becomes clear that these trends are related to elevation. What these trends seem to show is the length of the growing season ($\alpha_7 - \alpha_4$) is inversely related to elevation. It is well understood that vegetation growing seasons tend to start later and end earlier in the year at higher altitudes. This general phenomenon seems to be reflected in the estimated LSP parameter maps.

\subsection{Small area annual estimates and change detection}\label{sec:quabbin-analysis}

\begin{figure}[!ht]
\centering
\subfigure[NAIP July 21, 2014]{\includegraphics[trim=1.95cm 0.25cm 2cm 0cm, clip, width=7cm]{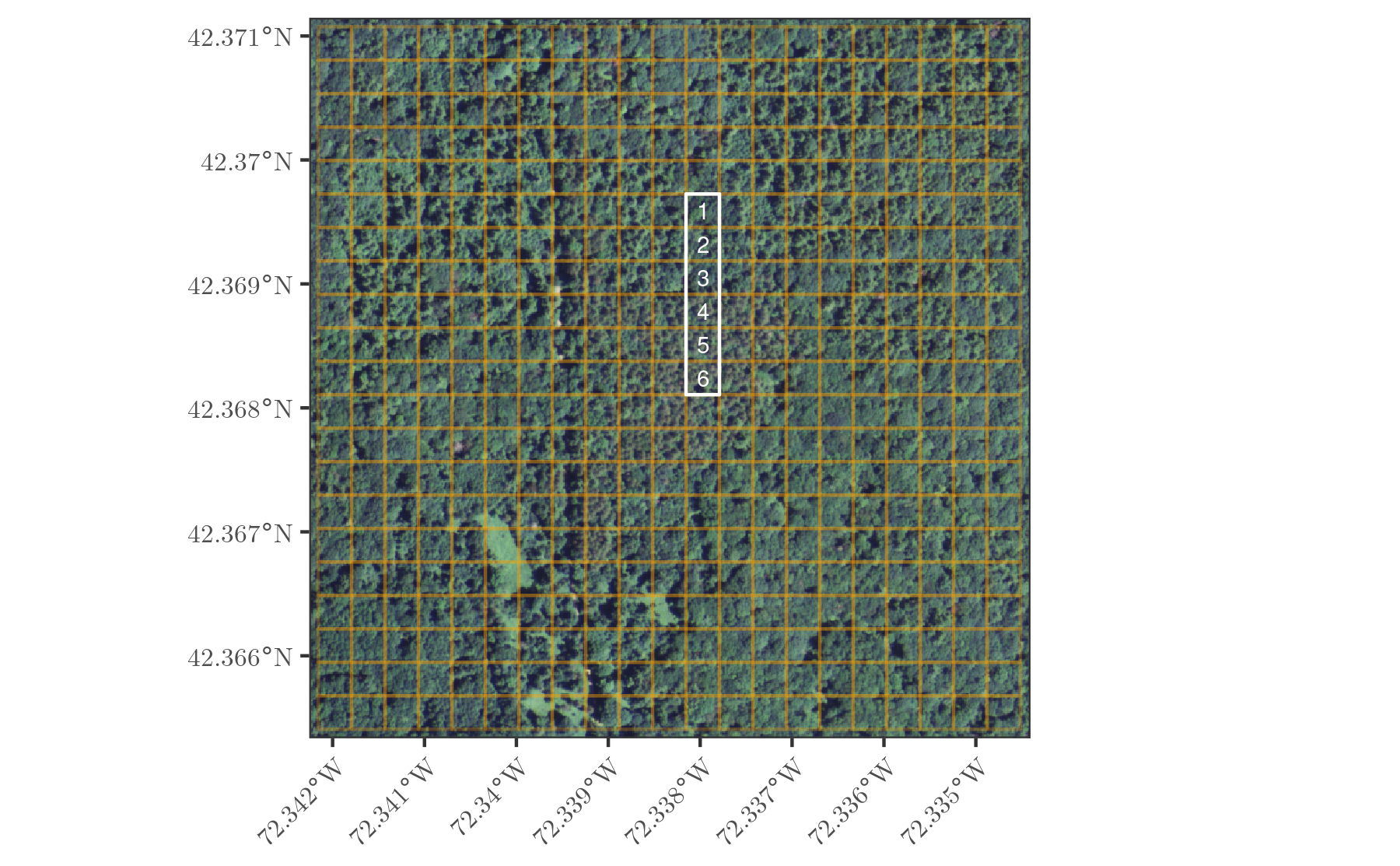}\label{fig:quabbin_naip_2014}}
\subfigure[Area under phenology curve, 2014]{\includegraphics[trim=1.95cm 0.25cm 2cm 0cm, clip, width=7cm]{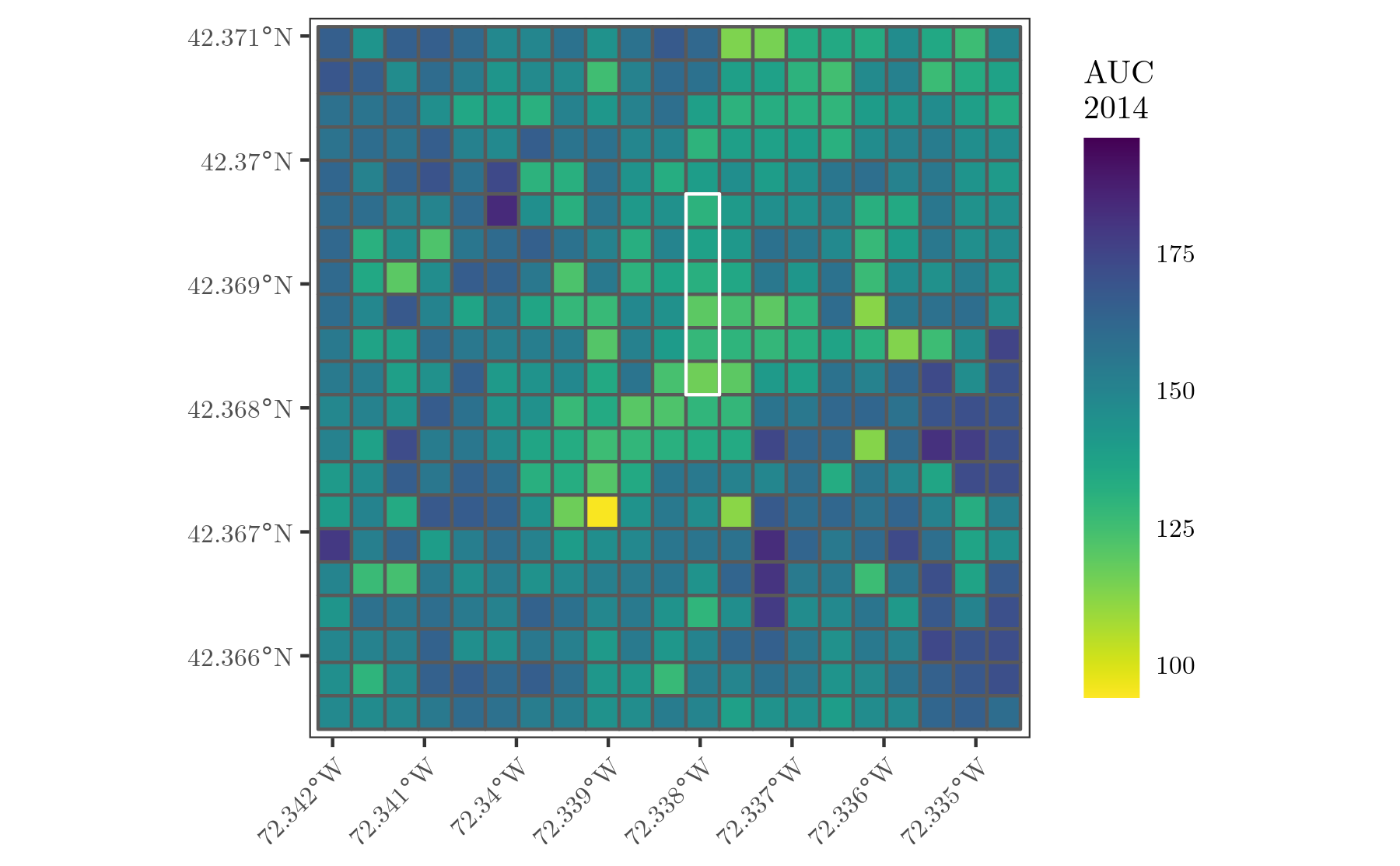}}\\
\subfigure[NAIP August 7, 2016]{\includegraphics[trim=1.95cm 0.25cm 2cm 0cm, clip, width=7cm]{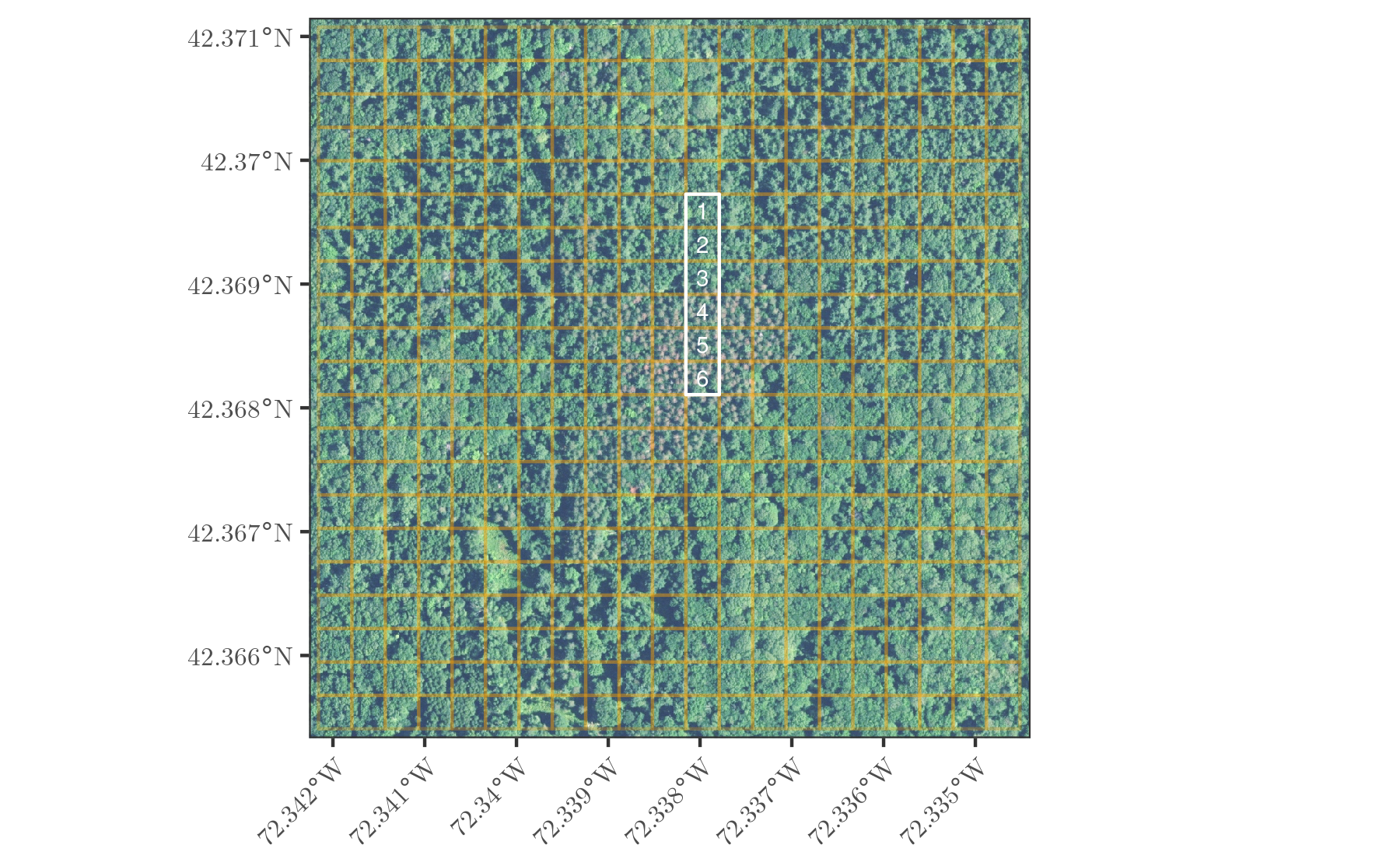}\label{fig:quabbin_naip_2016}}
\subfigure[Area under phenology curve, 2016]{\includegraphics[trim=1.95cm 0.25cm 2cm 0cm, clip, width=7cm]{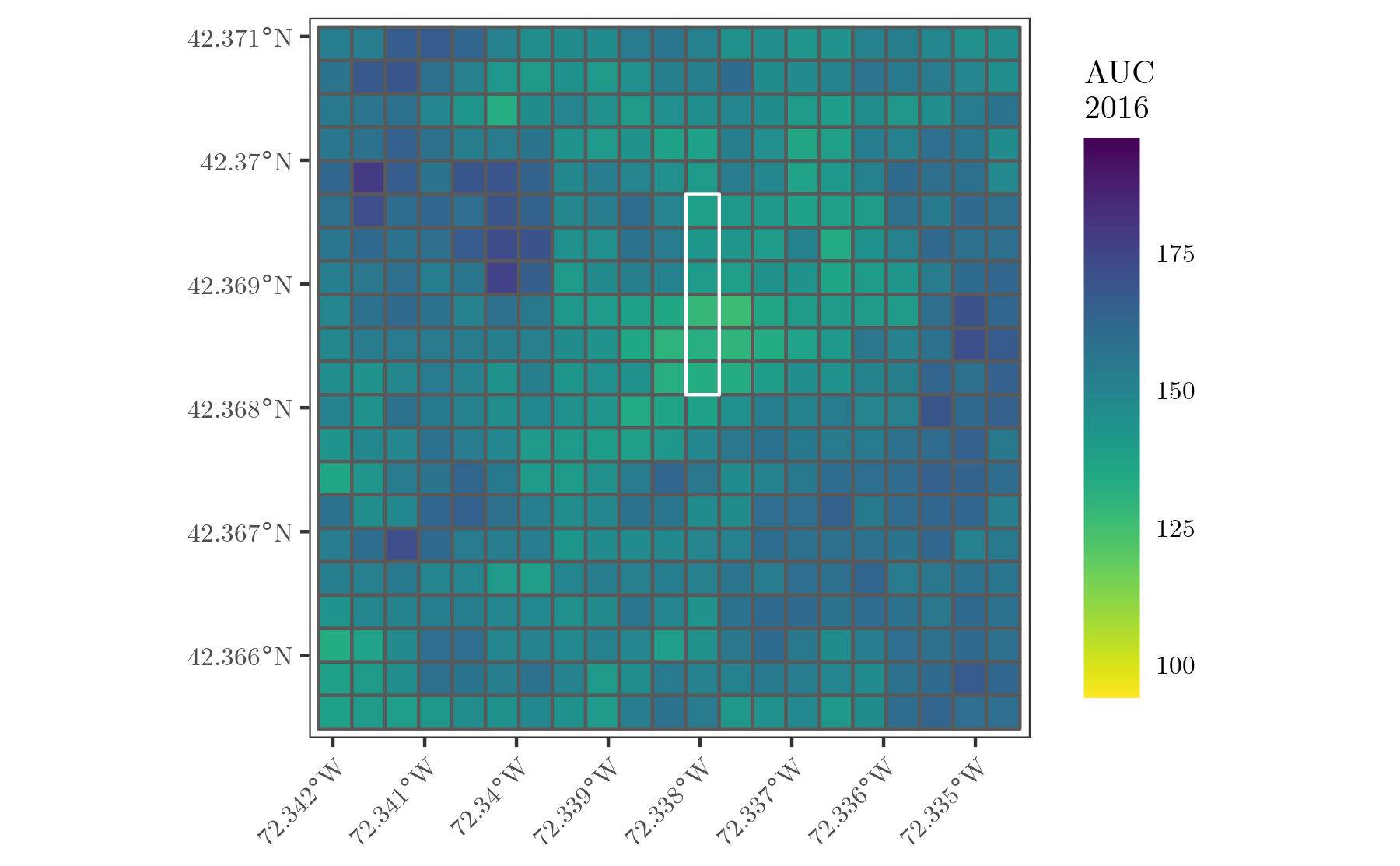}\label{fig:quabbin_auc_2016}}\\
\subfigure[NAIP October 19, 2018]{\includegraphics[trim=1.95cm 0.25cm 2cm 0cm, clip, width=7cm]{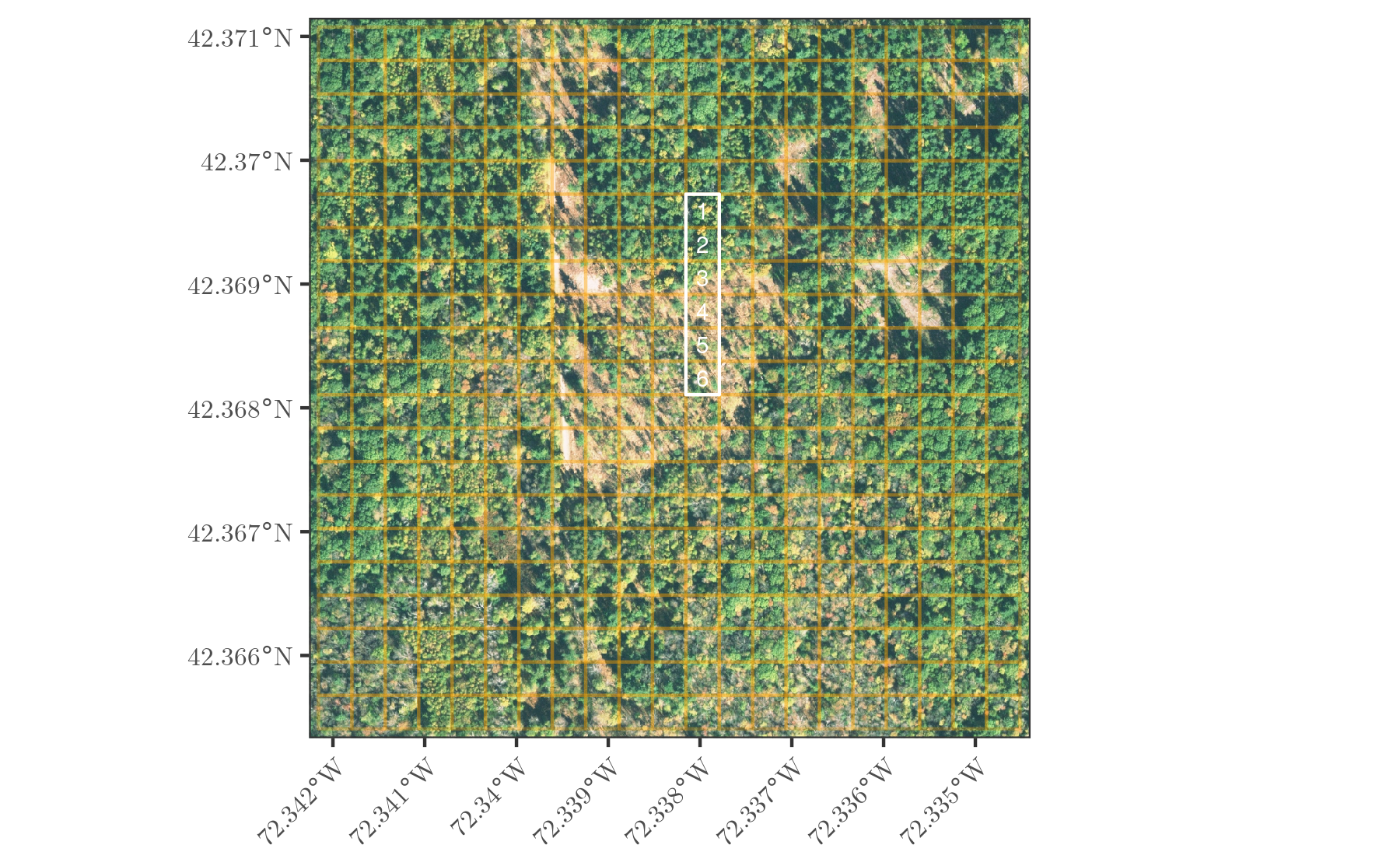}\label{fig:quabbin_naip_2018}}
\subfigure[Area under phenology curve, 2019]{\includegraphics[trim=1.95cm 0.25cm 2cm 0cm, clip, width=7cm]{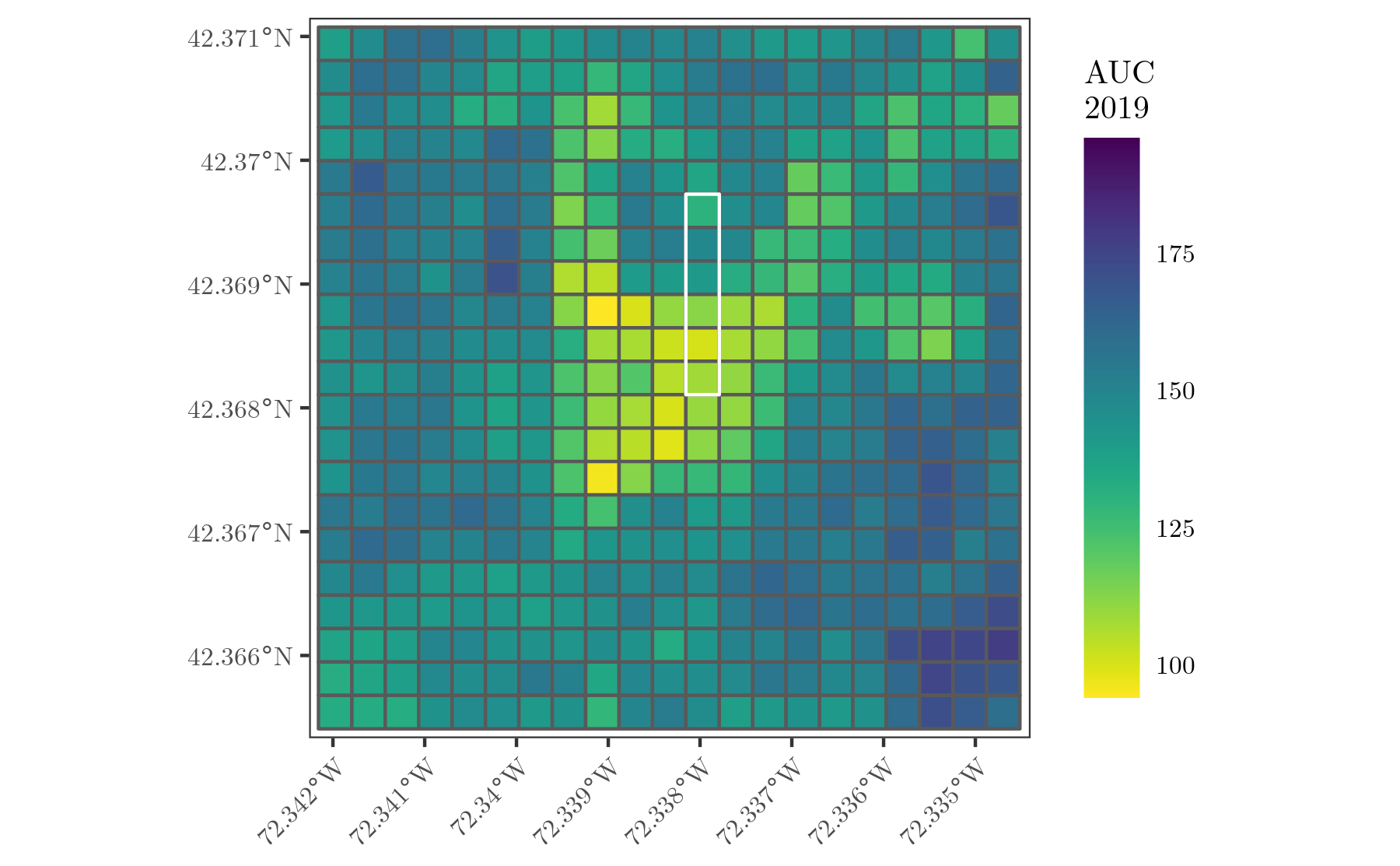}\label{fig:quabbin_auc_2019}}
 \caption{A 144 (21$\times$21) 30 m HLS pixel area of interest on the Quabbin Reservoir Watershed System in central Massachusetts with broader geographic location identified with a green point in Figure~\ref{fig:spatial_coverage}. In the left column, HLS pixels are identified using an orange grid over NAIP imagery. The right column, pixel-level posterior distribution median estimate for the area under the phenology curve from the Beta likelihood model. The white rectangle delineates six pixels for which EVI observations and model results are illustrated in Figure~\ref{fig:quabbin_mod_pix_yr}.  }\label{fig:quabbin_pixels}
\end{figure}

We again consider the phenology function (\ref{eqn:G}) and estimation of its parameters using the three likelihoods within the Bayesian model (\ref{eqn:post}), but unlike Section~\ref{sec:tile-analysis} we estimate models separately for each year. Using 2014, 2016, and 2018 National Agriculture Imagery Program \cite[NAIP][]{naip} 1 m resolution imagery we identified forest areas that underwent substantial disturbance during the study period. Here, we highlight one such area of interest (AOI) identified by a green point in Figure~\ref{fig:spatial_coverage}. The 144 (21$\times$21) 30 m HLS pixel AOI is on the Quabbin Reservoir Watershed System in central Massachusetts which is actively managed by the State's Department of Conservation and Recreation. Figure~\ref{fig:quabbin_naip_2014} shows 2014 NAIP imagery of the area which is dominated by an over-mature red pine (\emph{Pinus resinosa}) plantation planted by the Civilian Conservation Corps in the 1940s. Management records provided by \cite{Beard2020} indicate the area was thinned in 1991 and 1996. Since the 1996 harvest, black birch (\emph{Betula lenta}) took over the understory and occupied about 20 to 30 square feet of basal area per acre by 2018. The residual mature red pine was killed by red pine scale (\emph{Matsucoccus resinosae}) between 2014 and 2016, with mortality clearly visible in the center portion of the 2016 NAIP image in Figure~\ref{fig:quabbin_naip_2016}. In early October 2018 a timber harvest was conducted to salvage the dead and dying red pine. The NAIP imagery shown in Figure~\ref{fig:quabbin_naip_2018} was taken within weeks of the harvest.

The AOI analysis has two parts. First, we consider EVI estimates for a short transect of six pixels delineated by the white box in Figure~\ref{fig:quabbin_naip_2014}. Second, following the methods for sampling from $\Phi$'s posterior distribution in Section~\ref{sec:implementation}, we estimate AUC as $\int_{1}^{365} G(t; \balpha)dt$ for each pixel in the AOI (see Supplementary Material for code implementation).

\begin{figure}[!ht]
    \begin{center}
    \includegraphics[width=16cm]{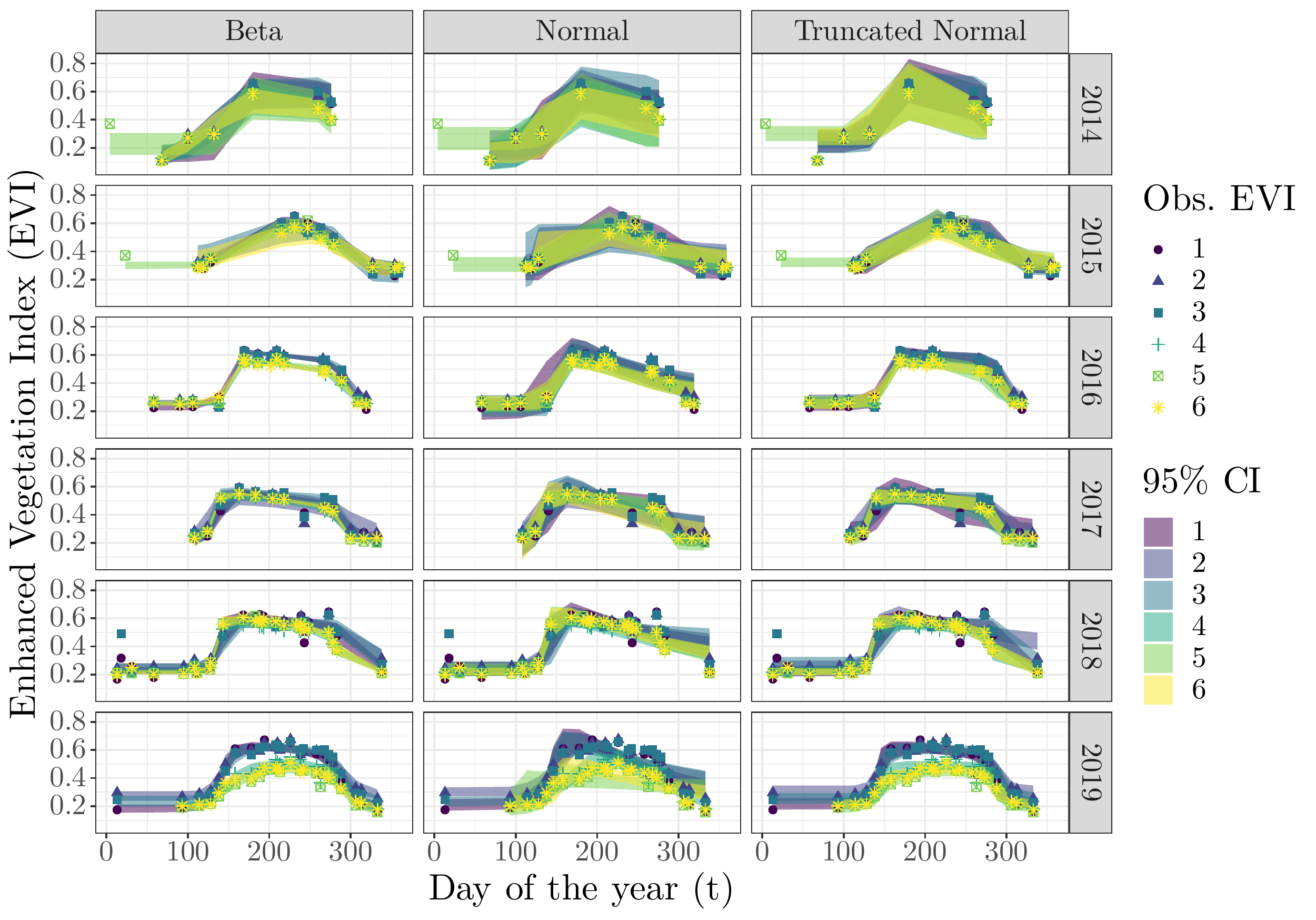}
    \caption{Model~(\ref{eqn:post}) fitted 95\% credible interval (CI) bands for the Beta, Normal, and Truncated Normal likelihoods fit to annual EVI observations from the six pixels identified by the white box and corresponding number labels in Figures~\ref{fig:quabbin_naip_2014} \subref{fig:quabbin_naip_2014}.}\label{fig:quabbin_mod_pix_yr}
    \end{center}
\end{figure}

The six pixel transect was selected to illustrate LSP change from healthy forest in 2014 (Figure~\ref{fig:quabbin_naip_2014}) to partial forest (pixels 1-2) and partial harvest (pixels 3-6) in 2018 (Figure~\ref{fig:quabbin_naip_2018}). Figure~\ref{fig:quabbin_mod_pix_yr} shows the posterior predictive median and 95\% credible interval band for each likelihood fit to each transect pixels' annual EVI observations. As shown in Figure~\ref{fig:time_coverage}, the availability of HLS data increases over the study period. For years 2014-2019, EVI sample sizes for transect pixels range from 6-7, 11-12, 14-15, 12-13, 23-25, and 22-25, respectively. Increases in sample size results in narrowing of credible interval bands from 2014 to 2019 in Figure~\ref{fig:quabbin_mod_pix_yr}, with the most pronounced narrowing occurring between 2014 and 2016. 

Looking at NAIP images in Figure~\ref{fig:quabbin_pixels}, we can see transect pixels 3-6 are harvested in October 2018. In a non-leap year, October is days 274-304 which is well into the fall leaf senescence portion of the LSP curve for 2018 in Figure~\ref{fig:quabbin_mod_pix_yr} and therefore we would not expect to observe a harvest effect on EVI estimates. However, 2019 mid-summer EVI credible interval bands estimated using the Beta and Truncated Normal likelihoods show clear separation between the non-harvested and harvested pixels. In comparison, the credible intervals estimated using the Normal likelihood are a bit wider and hence does not differentiate non-harvest and harvest pixels as clearly in 2019. In general, the Beta distribution provides marginally narrower credible intervals than the Normal and Truncated Normal likelihoods, while still maintaining a nominal credible interval coverage rate of 95\% for observed EVI.

We turn now to the AOI AUC analysis. In general, there should be a positive association between greenness and LSP curve AUC. Following this idea, we would expect pre-harvest AUC to be greater than post-harvest AUC. Indeed, the harvested pixels in the 2018 NAIP image, Figure~\ref{fig:quabbin_naip_2018}, have have lower AUC estimates in Figure~\ref{fig:quabbin_auc_2019}. While the AUC estimates identify harvest area, they do not seem to pick up the red pine mortality seen in the NAIP imagery prior to the harvest (i.e., comparing Figures~\ref{fig:quabbin_naip_2016} and \ref{fig:quabbin_auc_2016}). For completeness, the AUC estimates for all years in the study period are given in Supplementary Figure~\ref{fig:all_quabbin_pixels}.

With regard to pixel-level AUC uncertainty estimates, beyond approximately 12 EVI observations within a given year and pixel there is little change in the posterior distribution's standard deviation. This can be seen when looking at Supplementary Figures~\ref{fig:all_quabbin_sd_pixels} and \ref{fig:all_quabbin_n_pixels}, which map the AUC posterior standard deviation and sample size, respectively. The same general conclusion can be drawn looking at the credible interval band widths in Figure~\ref{fig:quabbin_mod_pix_yr}.

\section{Discussion and conclusions}\label{sec:discussion-summary}

We propose a new approach to estimate LSP parameters using a Bayesian model that builds on work by \citet{melaas2013} and \citet{senf2017}. We examined the Bayesian LSP model's performance using EVI time series' generated from HLS reflectance data. Composite long-term average LSP parameters were estimated for all NLCD forest class pixels in one HLS tile using three different likelihoods. We found there was negligible difference in parameter estimates from the different likelihoods indicating that, when VI sample size is sufficiently large and observed values are far from the VI's bounds, a Normal likelihood can be used without compromised inference even though its support is not bounded. From a computational standpoint, evaluating the Normal likelihood is much faster relative to evaluating the Truncated Normal or Beta. This decreased computing time becomes especially important in a MCMC setting where the likelihood needs to be evaluated for each iteration in the sampler. This decrease in run time to deliver posterior samples from model parameters can lead to substantial compute time savings when larger datasets are considered, e.g., continental-scale analyses. As a point of reference, to deliver 15,000 post convergence MCMC samples for the $\sim$7 million tile pixels run on a 64 core computer, the Beta likelihood took $\sim$3 days whereas the Normal likelihood took $\sim$2.5 days. However, in the AOI case study, we saw that when VI time series sample sizes are small, choosing the Beta or Truncated Normal likelihood delivered improved inference via narrower credible interval bands.

Examining LSP parameter estimates revealed relationships between NLCD forest type classes and LSP curve minimums, amplitudes, and maximums. The tendency for deciduous forests to have lower dormant season greenness, more significant spring green-ups, and higher maximum growing season greenness compared coniferous forests was clearly identified in the LSP parameter estimates. This result suggests that 30 m resolution LSP estimates from the Bayesian LSP model might be useful in ecological studies where vegetation phenology is thought to play a role.   

We showed that disparity in sample size and inherent variability in LSP observations can impact LSP parameter estimates. Current and future large remote sensing time series datasets useful for LSP modeling, such as HLS, will have highly varying sample sizes and levels of data noise from pixel to pixel due to sensor path overlap and cloud cover among other factors. If we are going to use data like HLS to make LSP estimates, it will be important to obtain pixel-level uncertainties to identify when and where LSP estimates are made with sufficient certainty. It is common for LSP estimates to be used in downstream analyses. The MCMC sampling mechanism implemented here allows for easy propagation of LSP parameter estimate uncertainty in subsequent analyses. The uncertainty propagation process was demonstrated in our AUC calculations for the annual LSP curve estimation case study.

The \citet{melaas2013} and \citet{senf2017} studies provide a methodology to account for interannual variability when using VI observations pooled across years. We did not attempt to account for interannual variability in our first case study where pooled VI time series' were used. This choice was largely influenced by computing time. While we have made significant attempts to speed up fitting times for the Bayesian LSP model, as with many MCMC based Bayesian model-fitting approaches, this model still requires substantial run time for even moderate sized datasets. 

While not pursued here, our posited Bayesian model can accommodate random effects for LSP parameters to account for interannual variability with pooled data. We note that, considering the HLS dataset specifically, partial pooling models to account for interannual variability (as was used in \citet{senf2017}) may not be necessary with post 2017 data (when OLI and both MSI sensors are fully operational). As seen in the second case study, LSP curves were estimated with sufficient certainty to track variation in LSP parameters from year-to-year for the selected individual pixels. Examining pixel-level uncertainties suggested that there is likely a sufficient number of observations each year so that pooling across years may not be needed to increase sample sizes. 

The \R package companion to the article, \pkg{rsBayes} (version 0.1.1), provides first-cut functionality for users to apply our Bayesian LSP approach to their own datasets. Considerable effort was made to develop efficient fitting algorithms written in \texttt{C} with an eye toward making this Bayesian LSP estimation approach viable for continental or global scale applications. As computing technology continues to improve at an exponential rate, we believe the modeling approach described here will be applicable at these scales in the near future. 

\section*{Acknowledgments}
The work of the first and second author was supported, in part, by the USDA Forest Service Forest Inventory and Analysis (FIA) and Forest Health Monitoring (FHM) programs, and National Aeronautics and Space Administration's Carbon Monitoring System project. The first author's work was also supported through the Minnesota Agricultural Research, Education and Extension Tech Transfer program (AGREETT). The second author received additional support from the National Science Foundation (NSF) NSF/EF 1253225 and NSF/DMS 1916395. The third author's work was supported by the NSF Graduate Research Fellowship and University of Minnesota Doctoral Dissertation Fellowship.

\clearpage

\section*{Supplementary material}\label{sec:appendix}
\beginsupplement

\subsection{Supplementary figures}
\begin{figure}[!h]
\centering
\subfigure[$\alpha_3$ median]{\includegraphics[width = .45\textwidth]{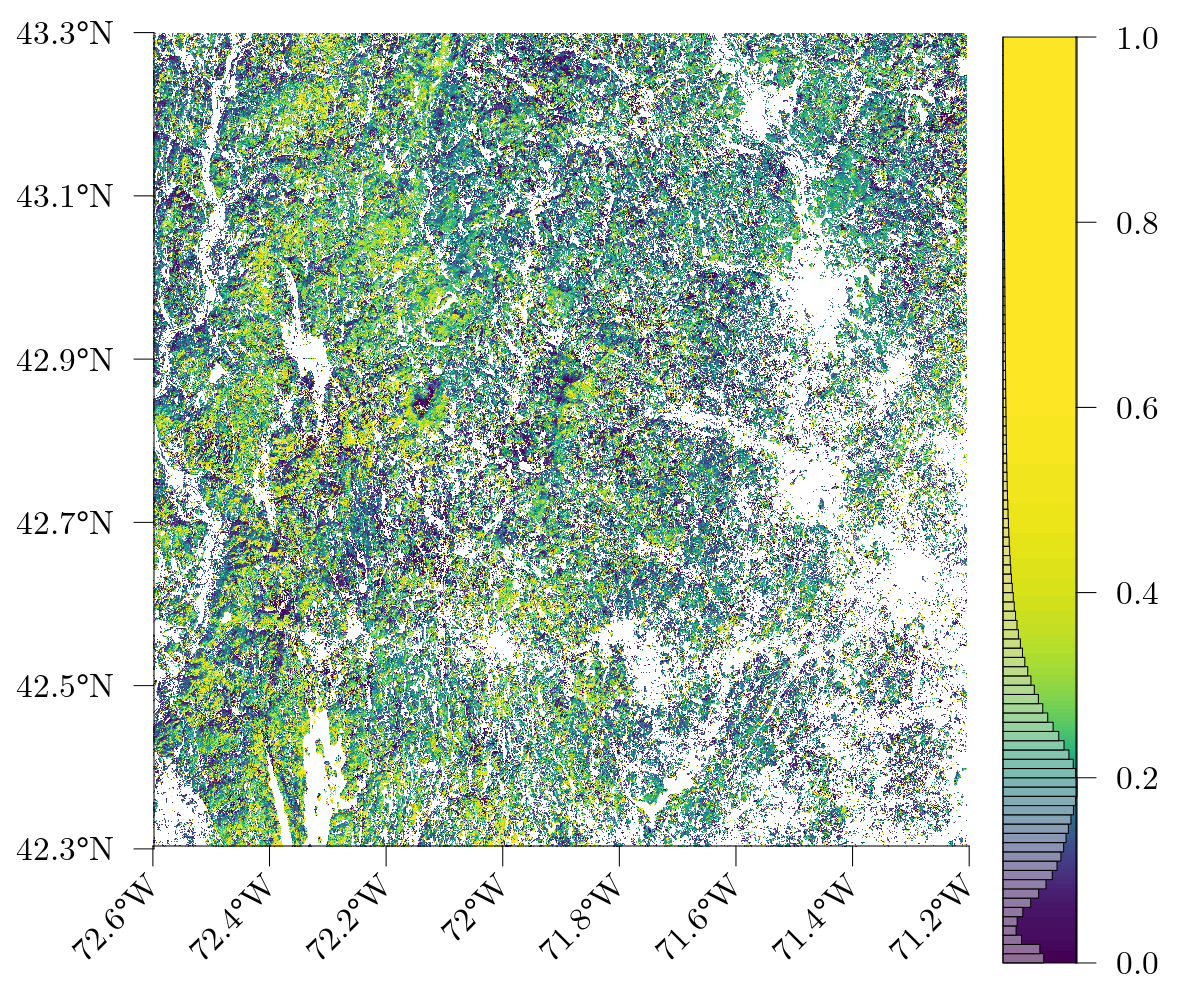}}
\subfigure[$\alpha_3$ standard deviation]{\includegraphics[width = .45\textwidth]{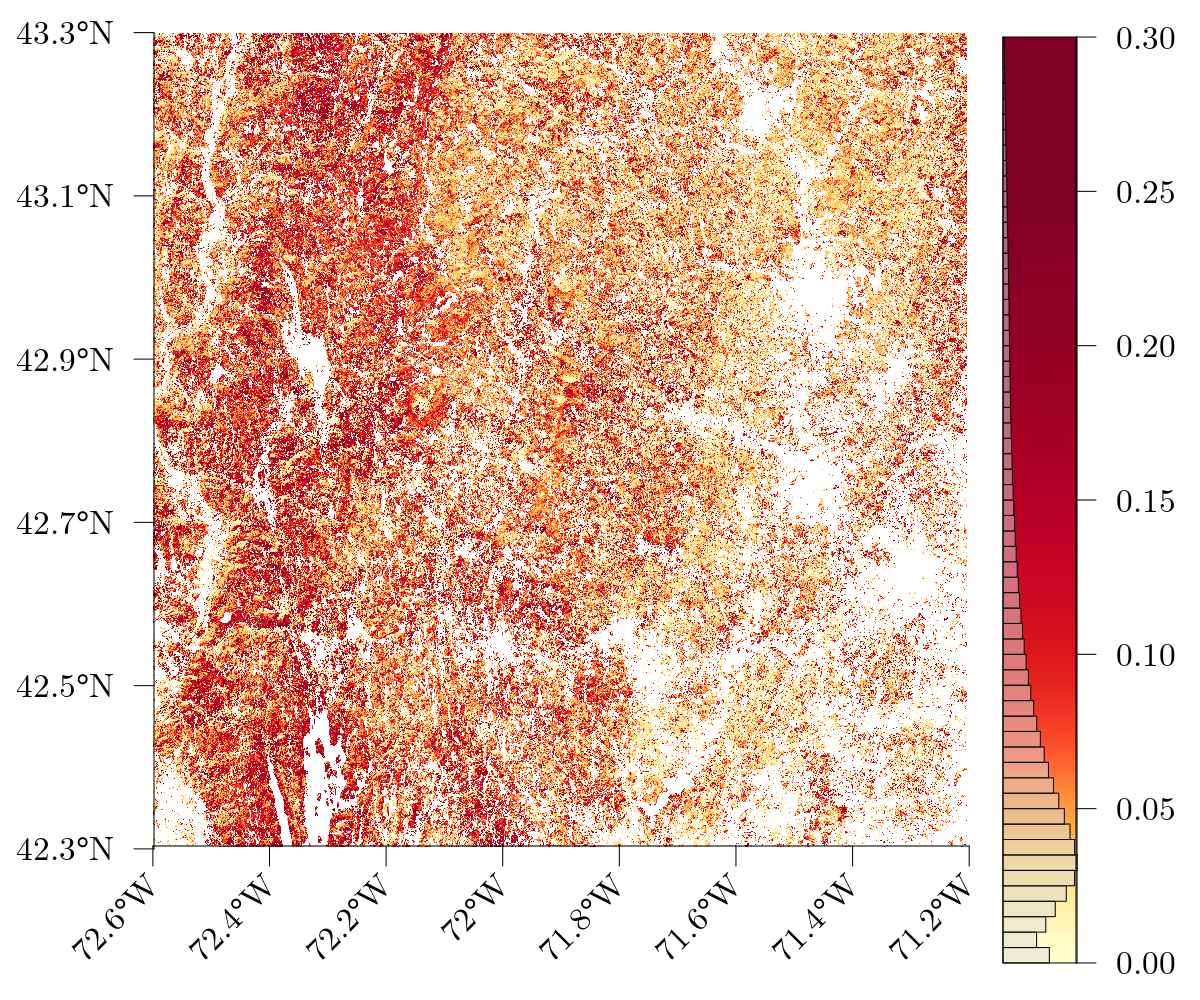}}\\
\subfigure[$\alpha_6$ median]{\includegraphics[width = .45\textwidth]{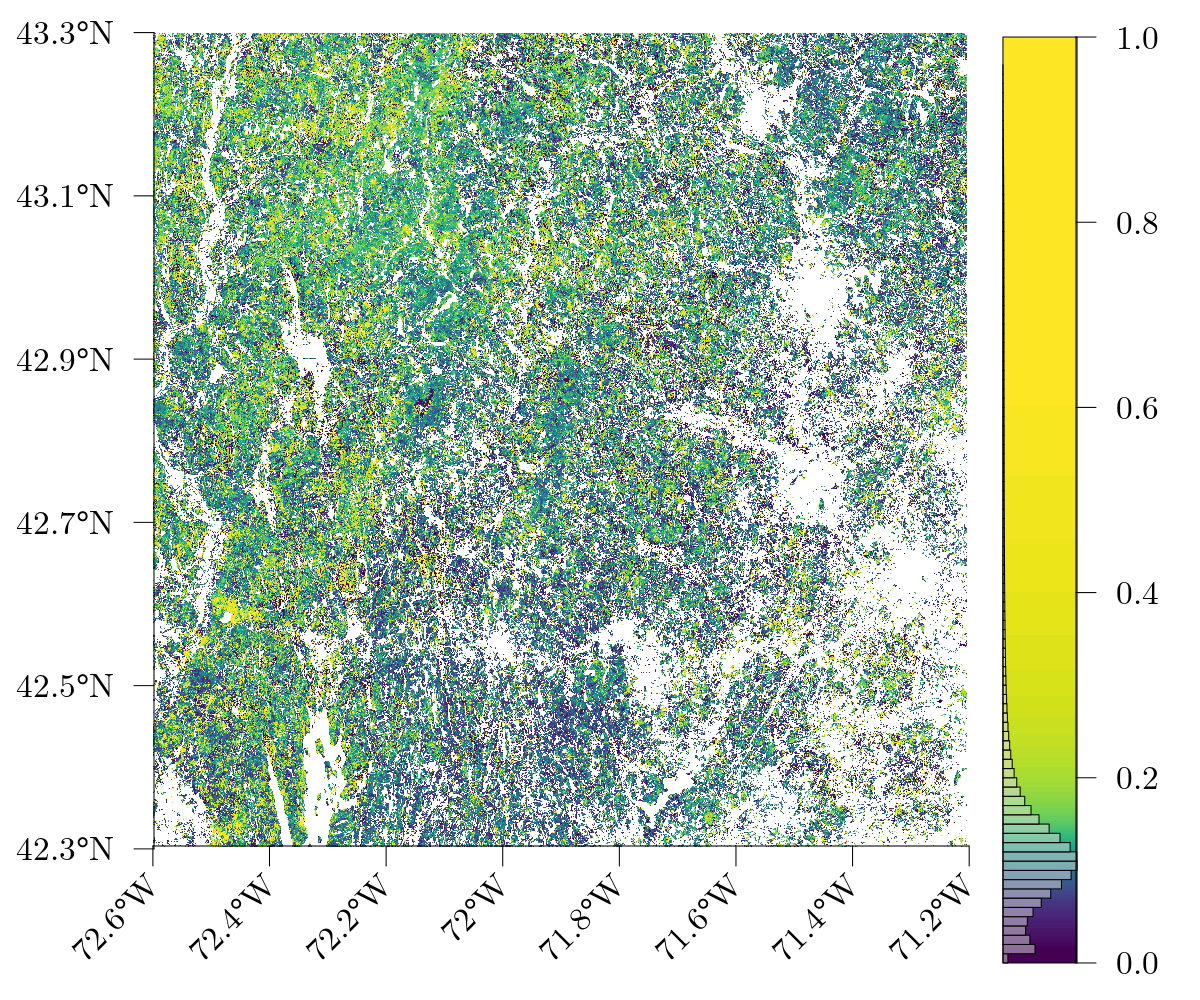}}
\subfigure[$\alpha_6$ standard deviation]{\includegraphics[width = .45\textwidth]{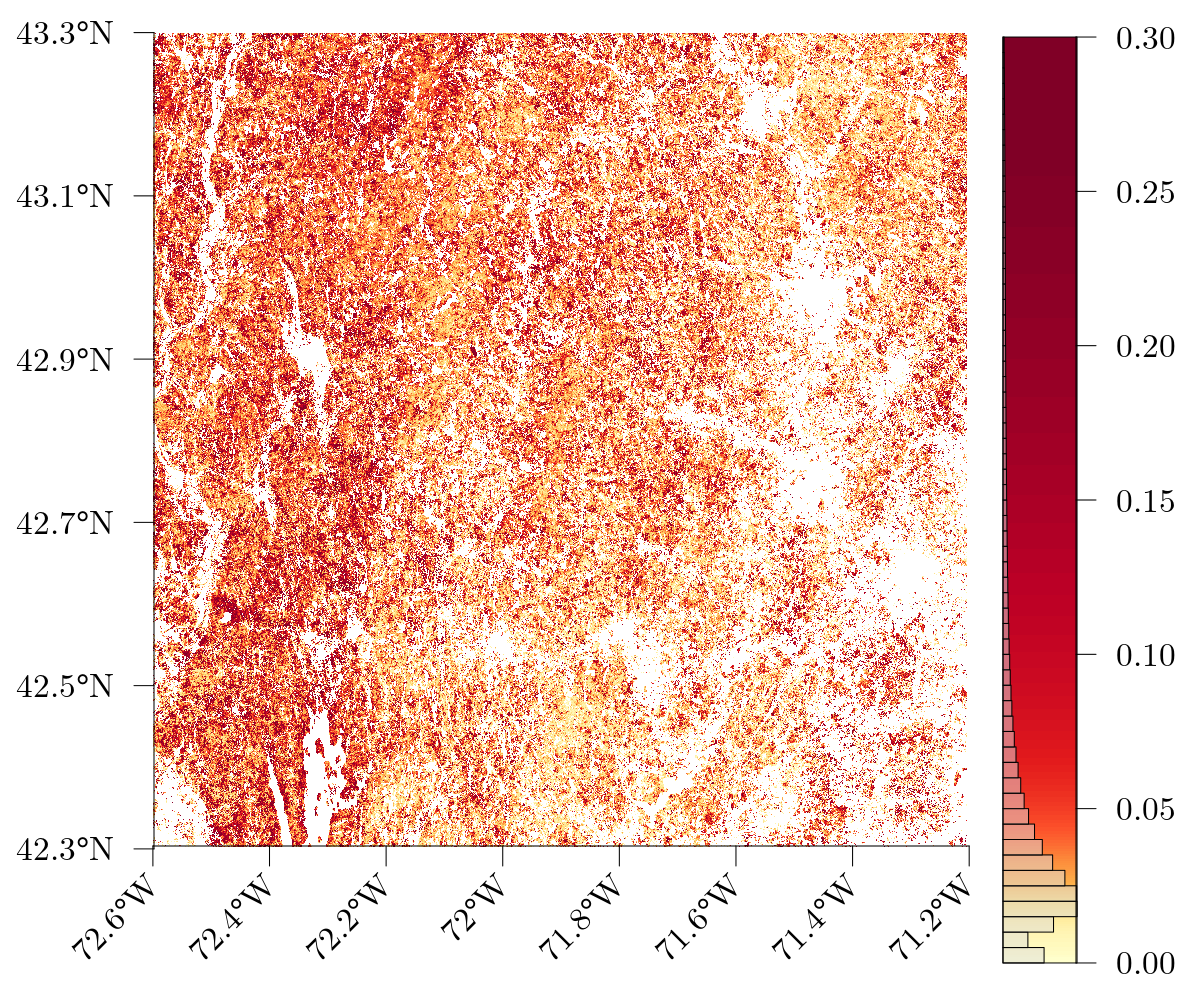}}\\
\caption{Maps showing spatial distribution of long-term average (2013-2019) $\alpha_3$ and $\alpha_6$ point and uncertainty estimates using the Bayesian LSP model.}\label{fig:alpha3-6-map}
\end{figure}

\begin{figure}[!h]
\centering
\subfigure[$\alpha_5$ median]{\includegraphics[width = .45\textwidth]{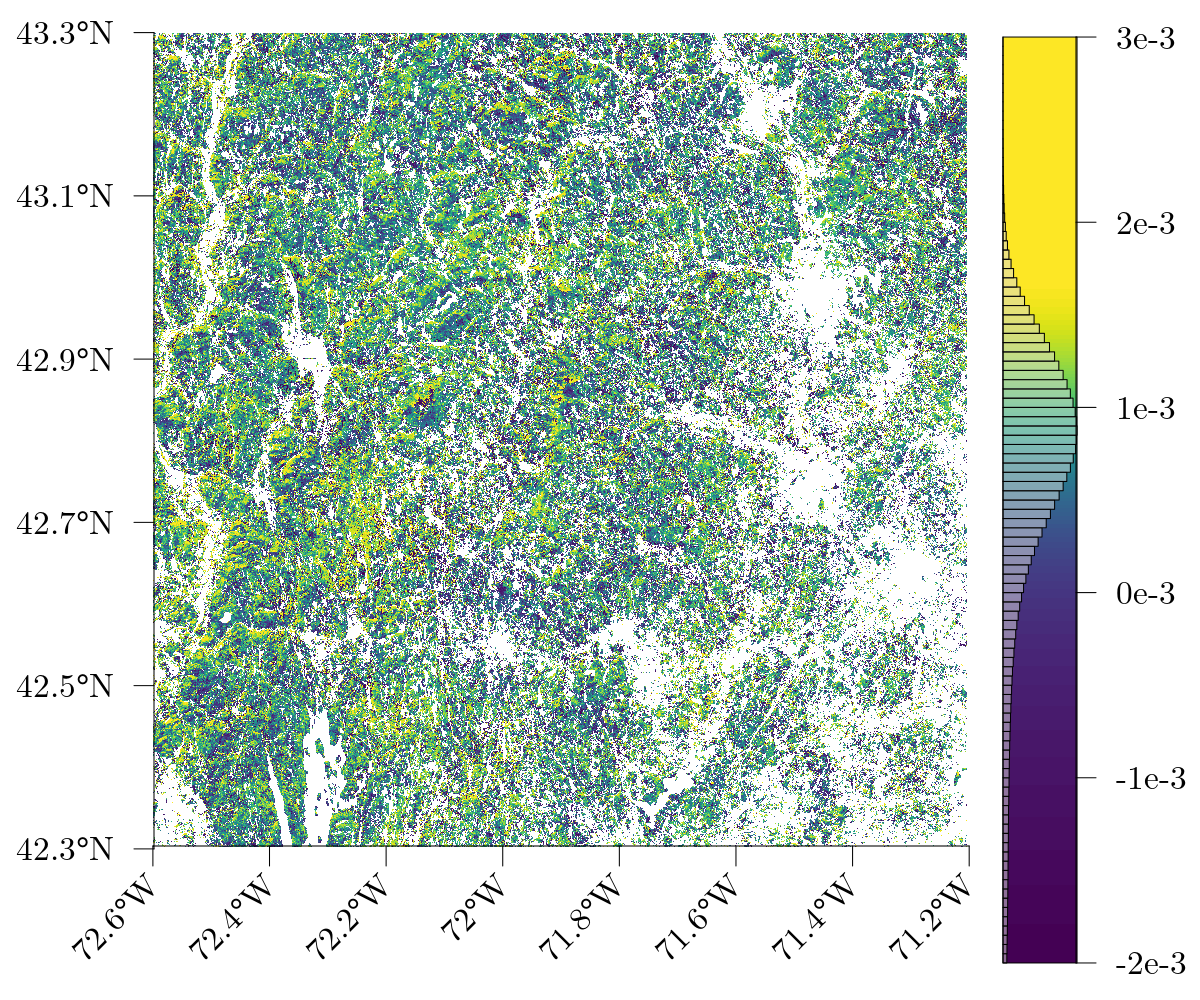}}
\subfigure[$\alpha_5$ standard deviation]{\includegraphics[width = .45\textwidth]{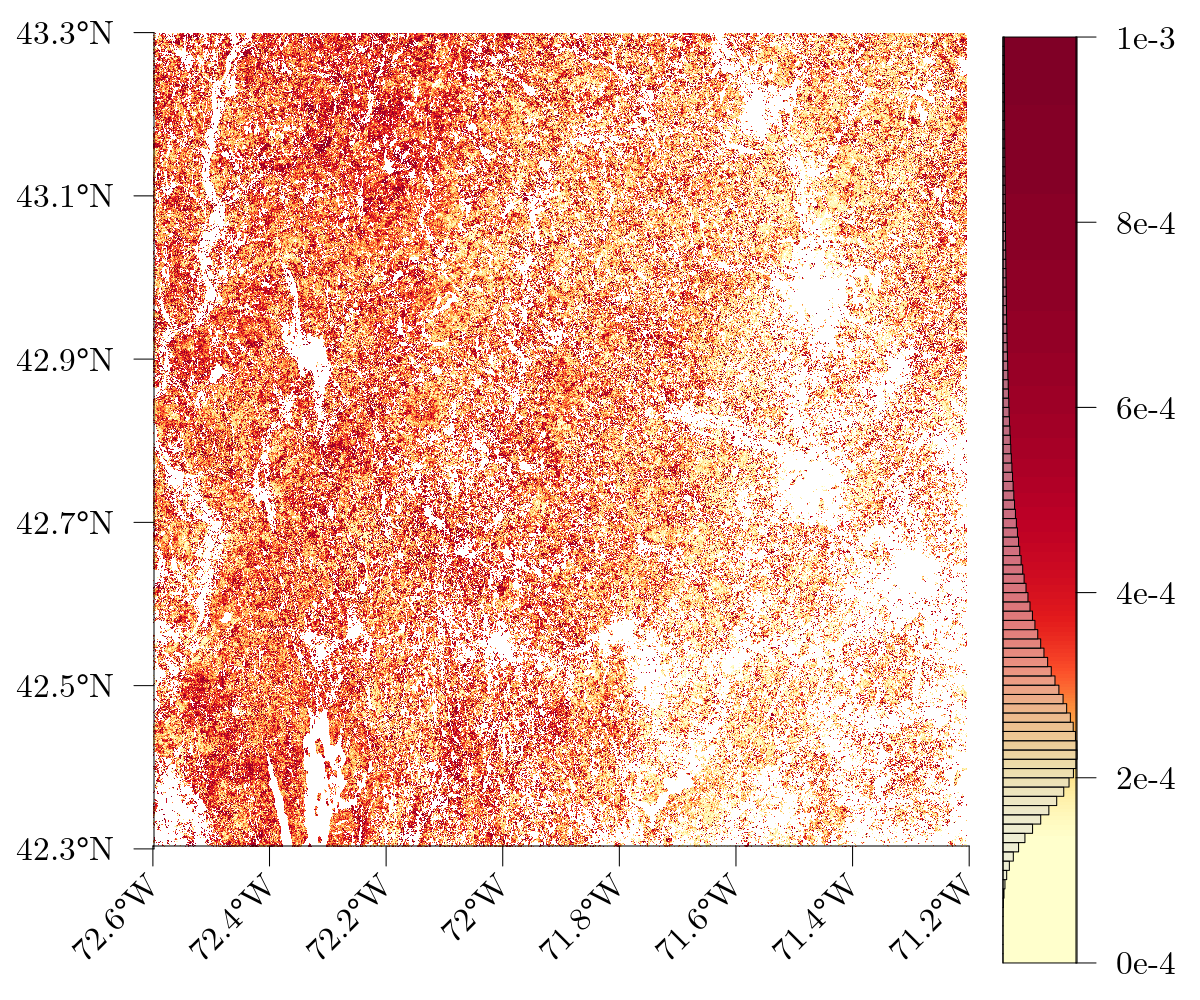}}\\
\subfigure[$\sigma^2$ median]{\includegraphics[width = .45\textwidth]{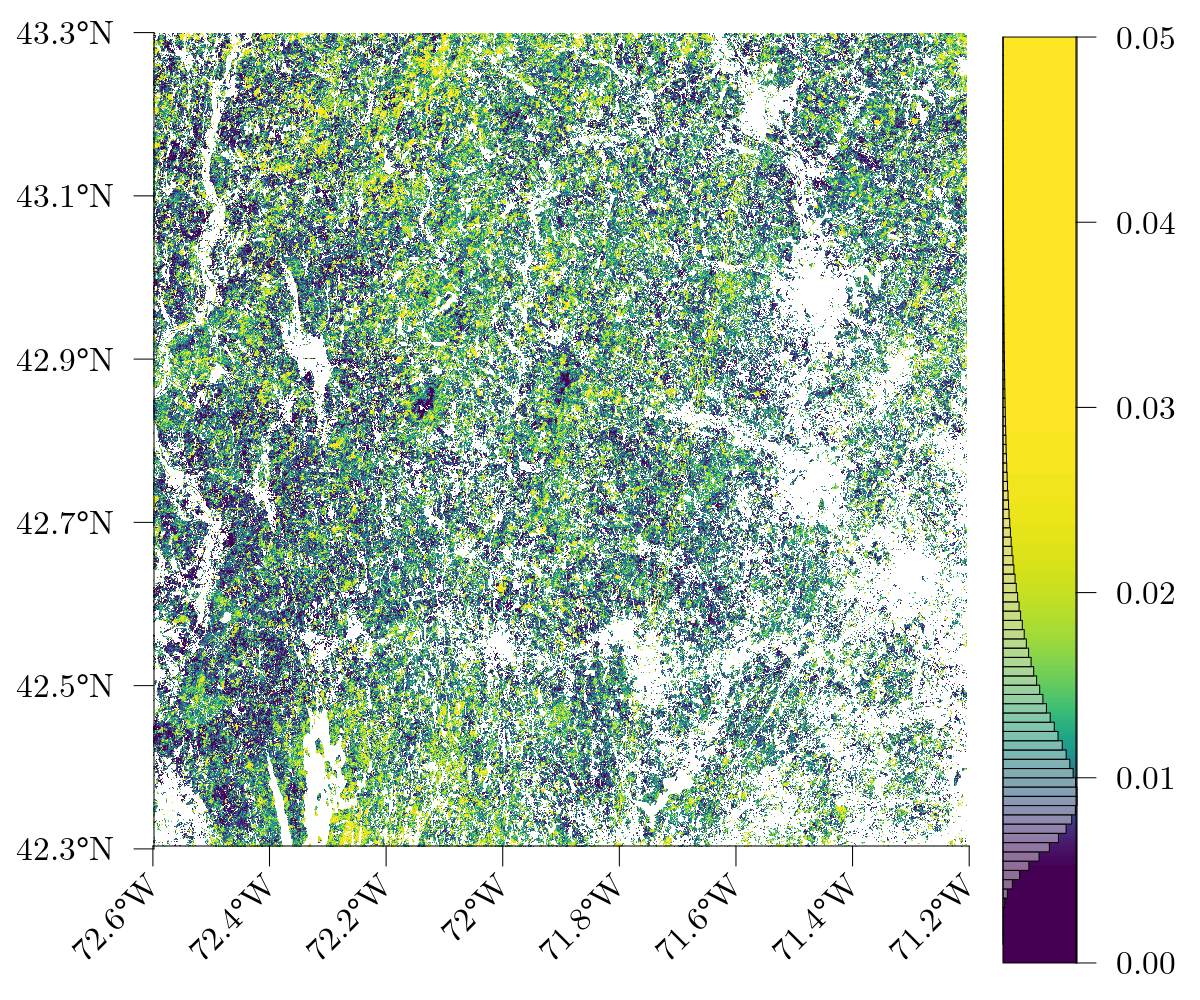}}
\subfigure[$\sigma^2$ standard deviation]{\includegraphics[width = .45\textwidth]{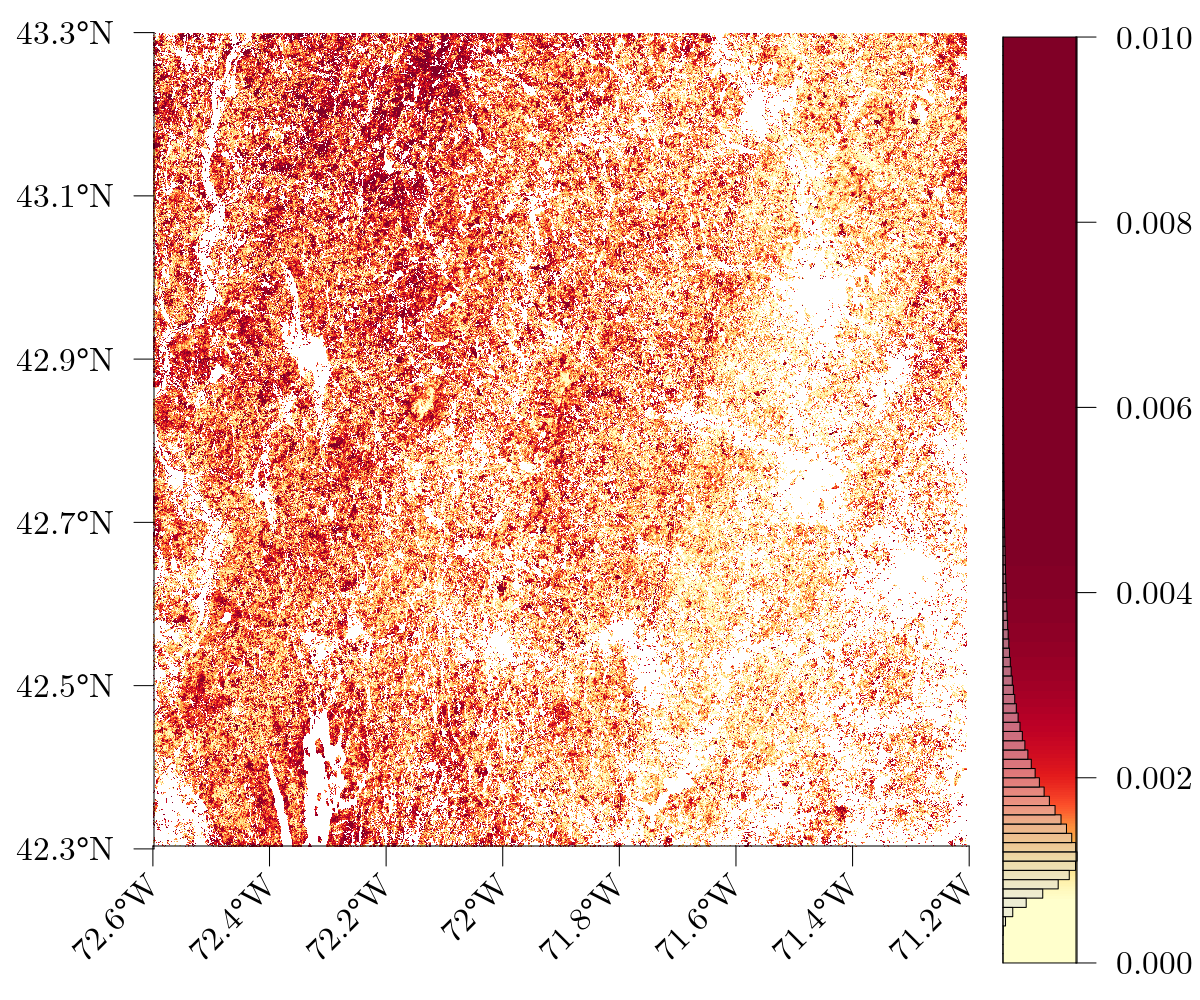}}\\
\caption{Maps showing spatial distribution of long-term average (2013-2019) $\alpha_2$ and $\alpha_5$ point and uncertainty estimates using the Bayesian LSP model.}\label{fig:alpha5-s2-map}
\end{figure}

\begin{figure}[!ht]
\centering
\includegraphics[trim=1.95cm 0.25cm 2cm 0cm, clip, width=4.5cm]{figures/auc_14_pixel_grid.png}
\includegraphics[trim=1.95cm 0.25cm 2cm 0cm, clip, width=4.5cm]{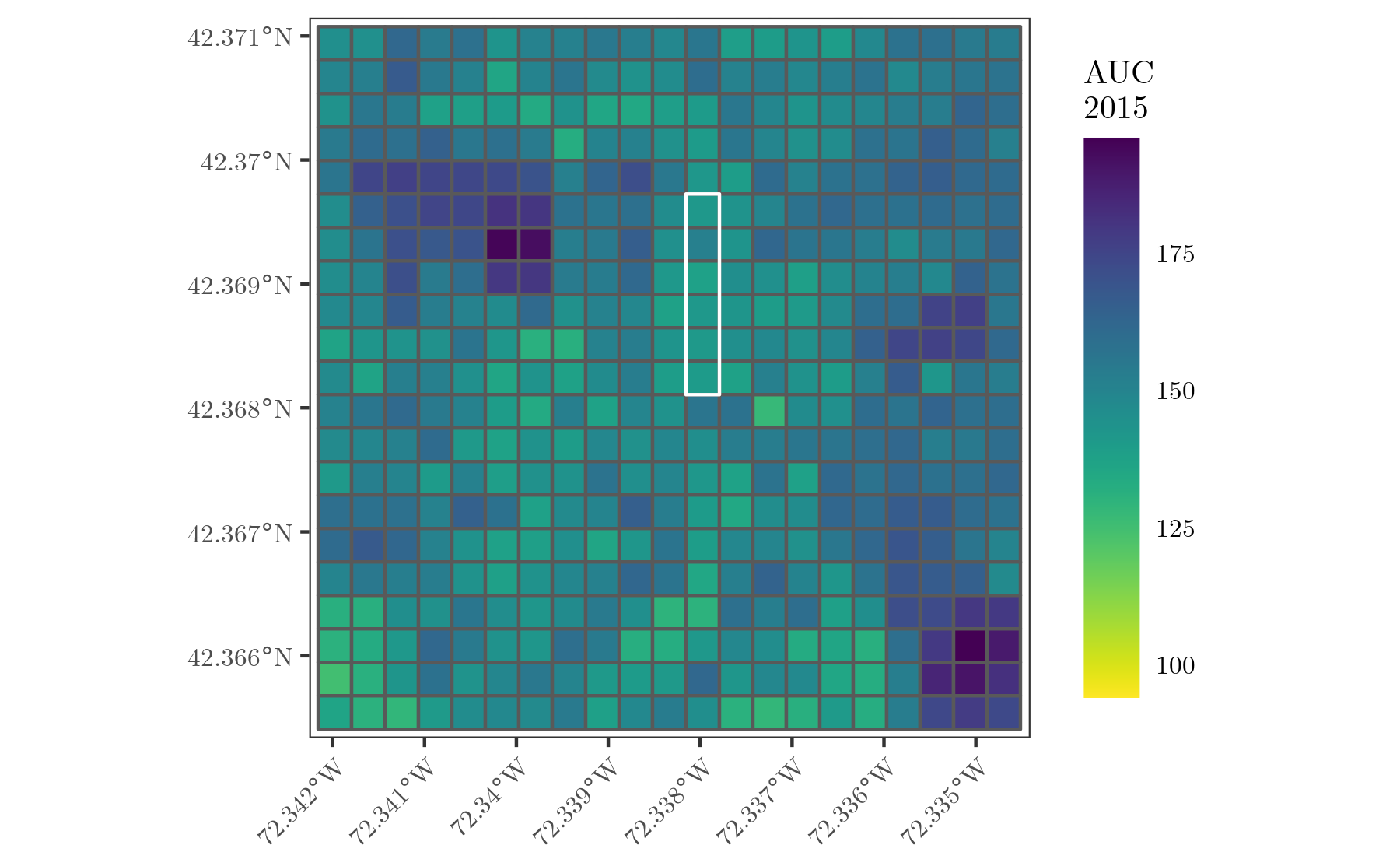}
\includegraphics[trim=1.95cm 0.25cm 2cm 0cm, clip, width=4.5cm]{figures/auc_16_pixel_grid.png}
\includegraphics[trim=1.95cm 0.25cm 2cm 0cm, clip, width=4.5cm]{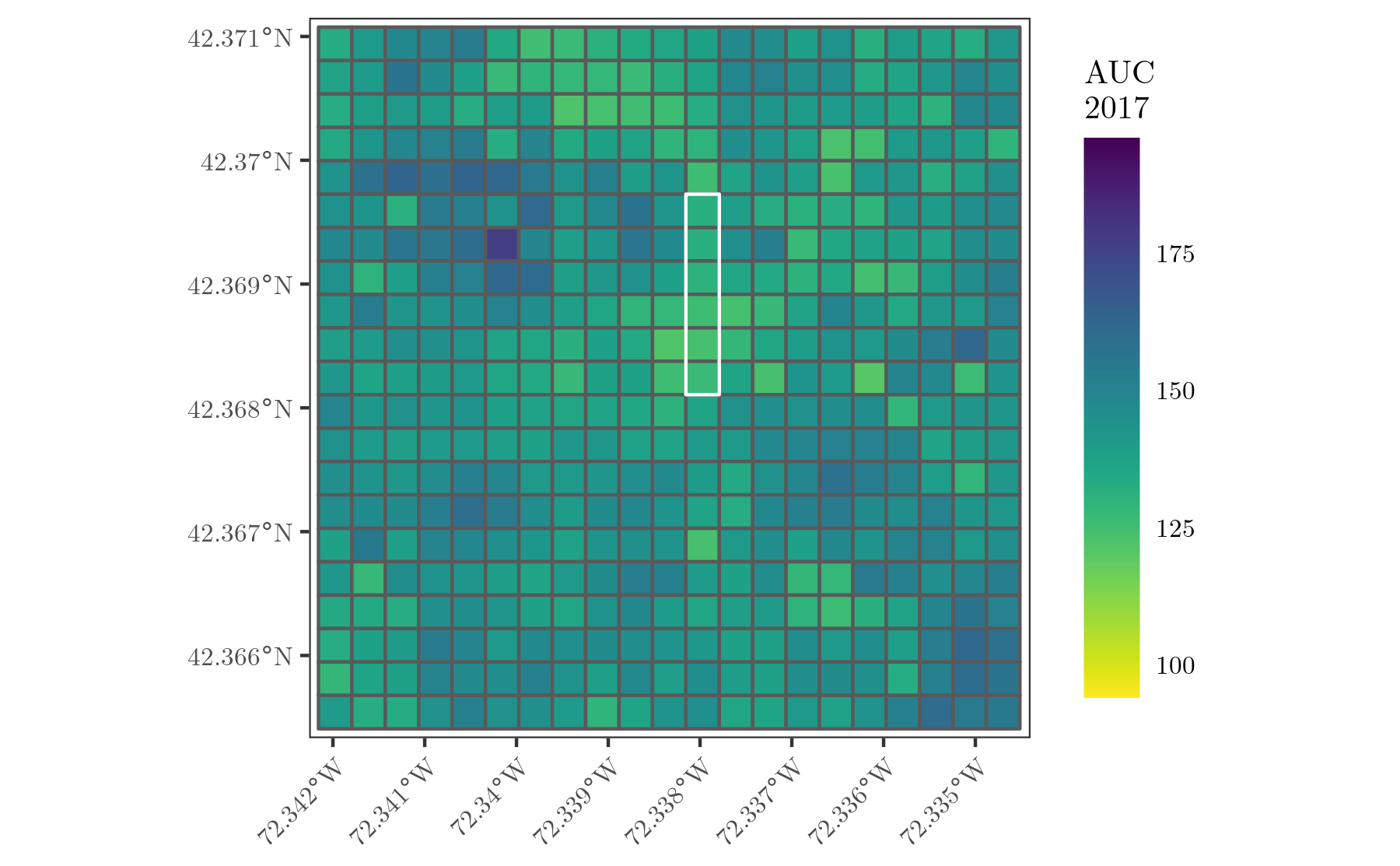}
\includegraphics[trim=1.95cm 0.25cm 2cm 0cm, clip, width=4.5cm]{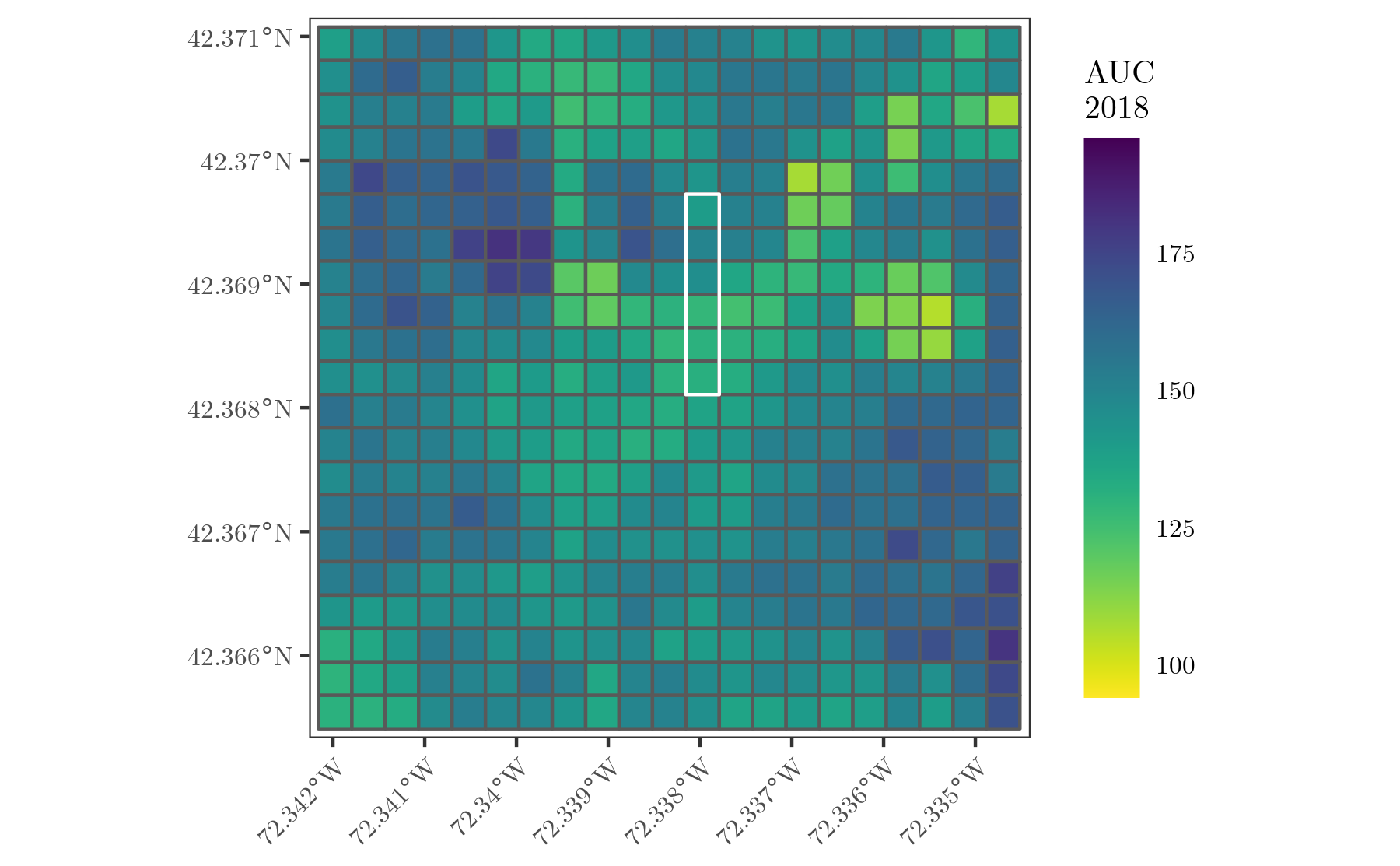}
\includegraphics[trim=1.95cm 0.25cm 2cm 0cm, clip, width=4.5cm]{figures/auc_19_pixel_grid.png}
\caption{Supplement to Figure~\ref{fig:quabbin_pixels} that provides pixel-level area under the curve (AUC) posterior distribution median estimates for all years in the study period.}\label{fig:all_quabbin_pixels}
\end{figure}

\begin{figure}[!ht]
\centering
\includegraphics[trim=1.95cm 0.25cm 2cm 0cm, clip, width=4.5cm]{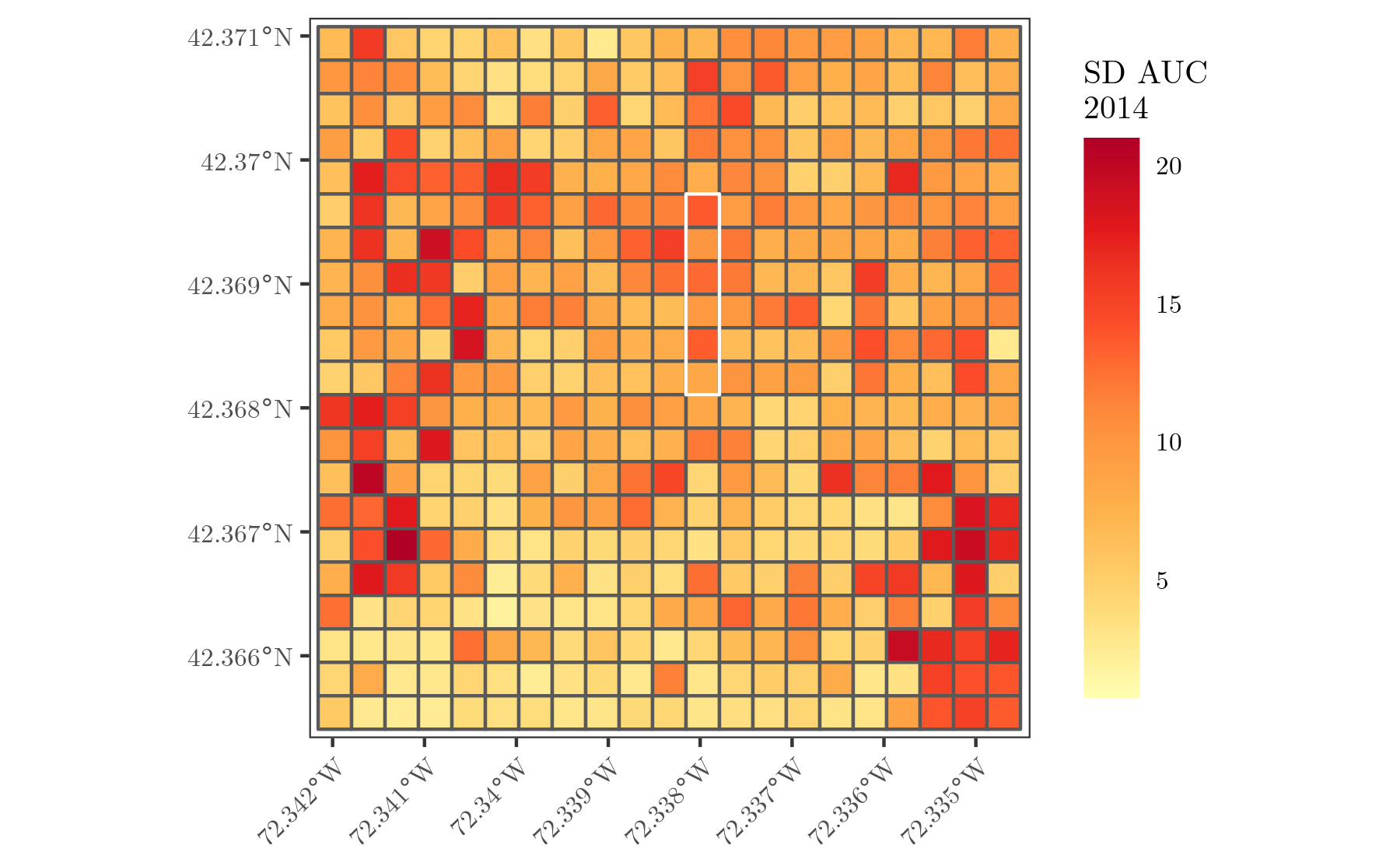}
\includegraphics[trim=1.95cm 0.25cm 2cm 0cm, clip, width=4.5cm]{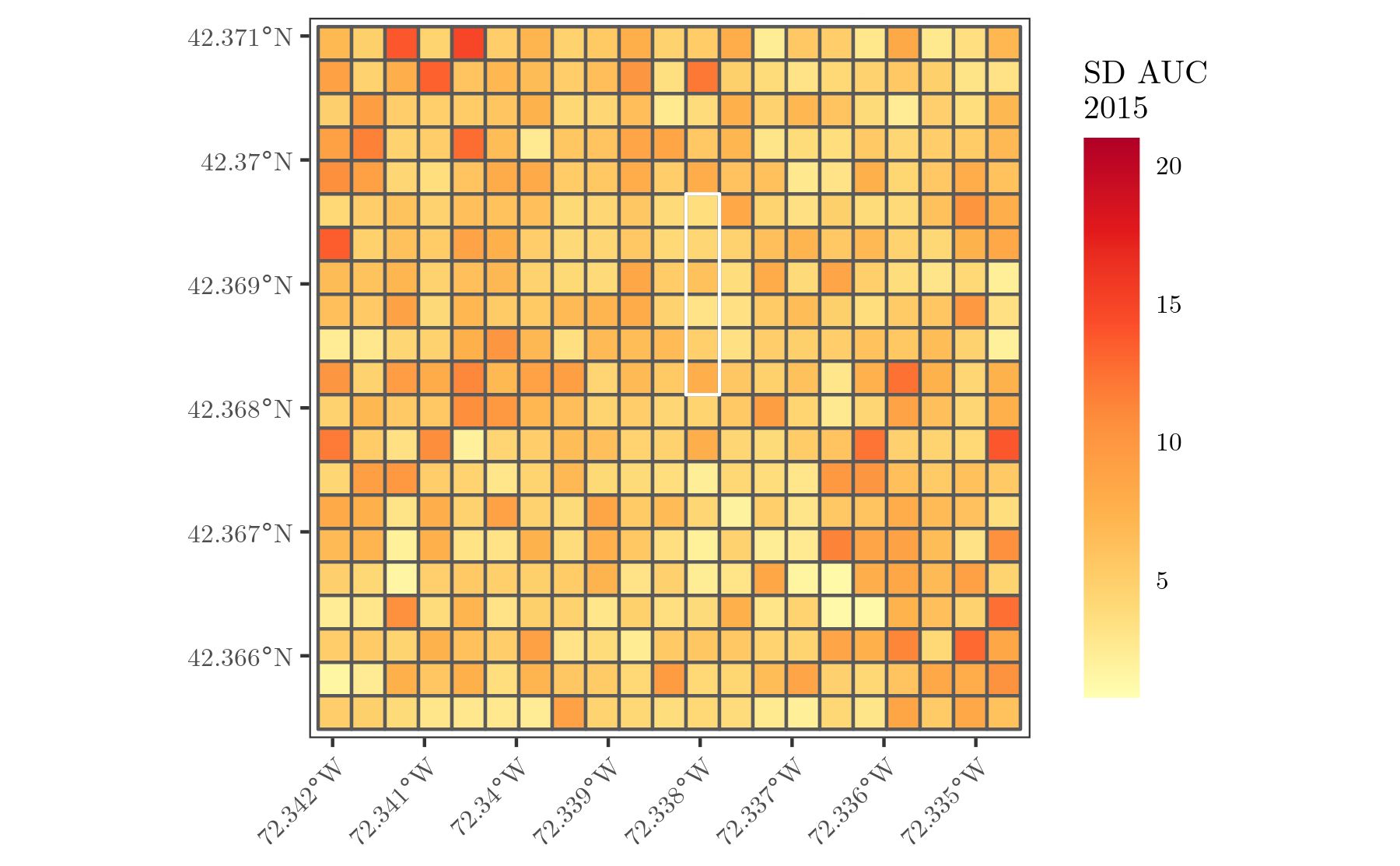}
\includegraphics[trim=1.95cm 0.25cm 2cm 0cm, clip, width=4.5cm]{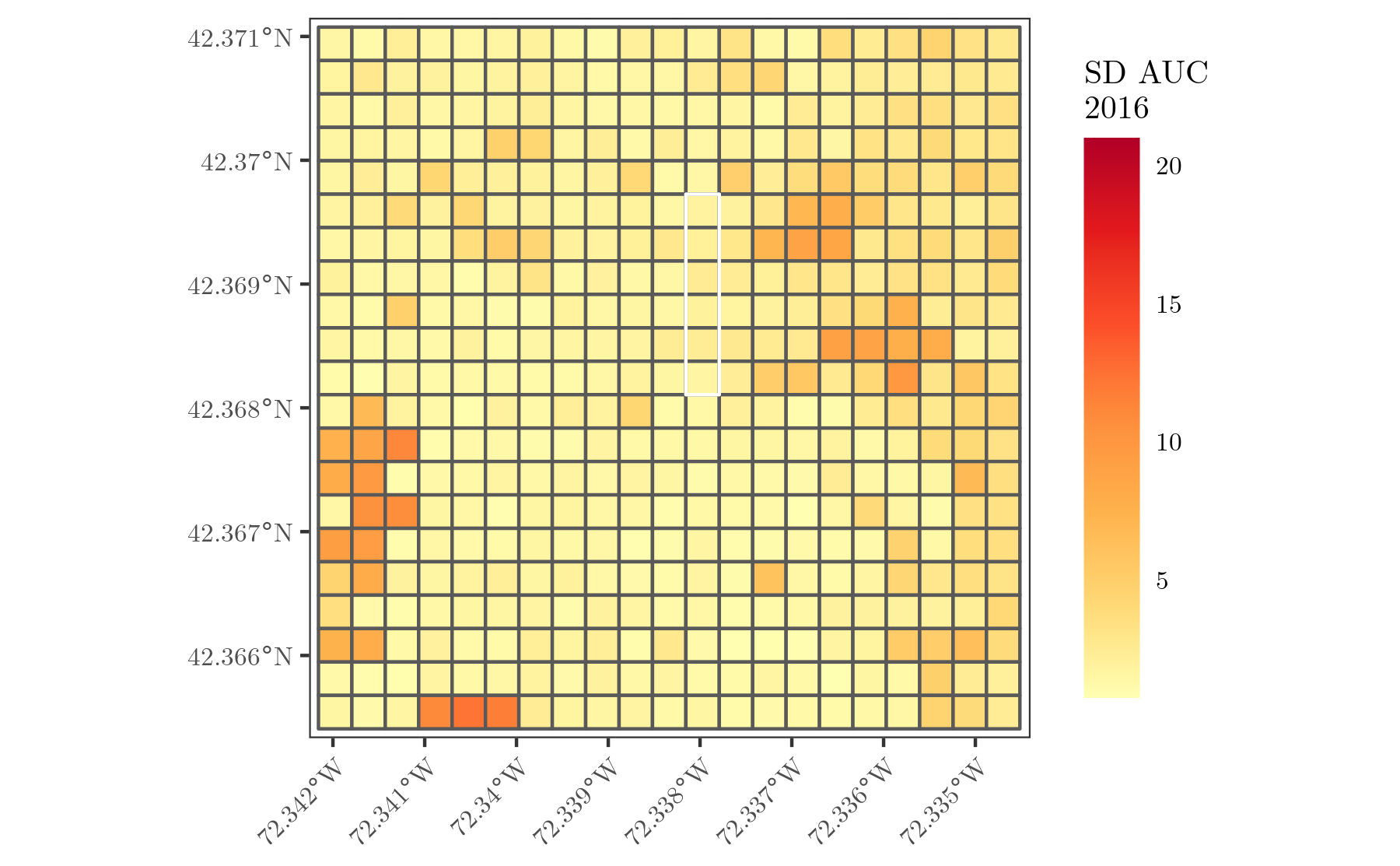}
\includegraphics[trim=1.95cm 0.25cm 2cm 0cm, clip, width=4.5cm]{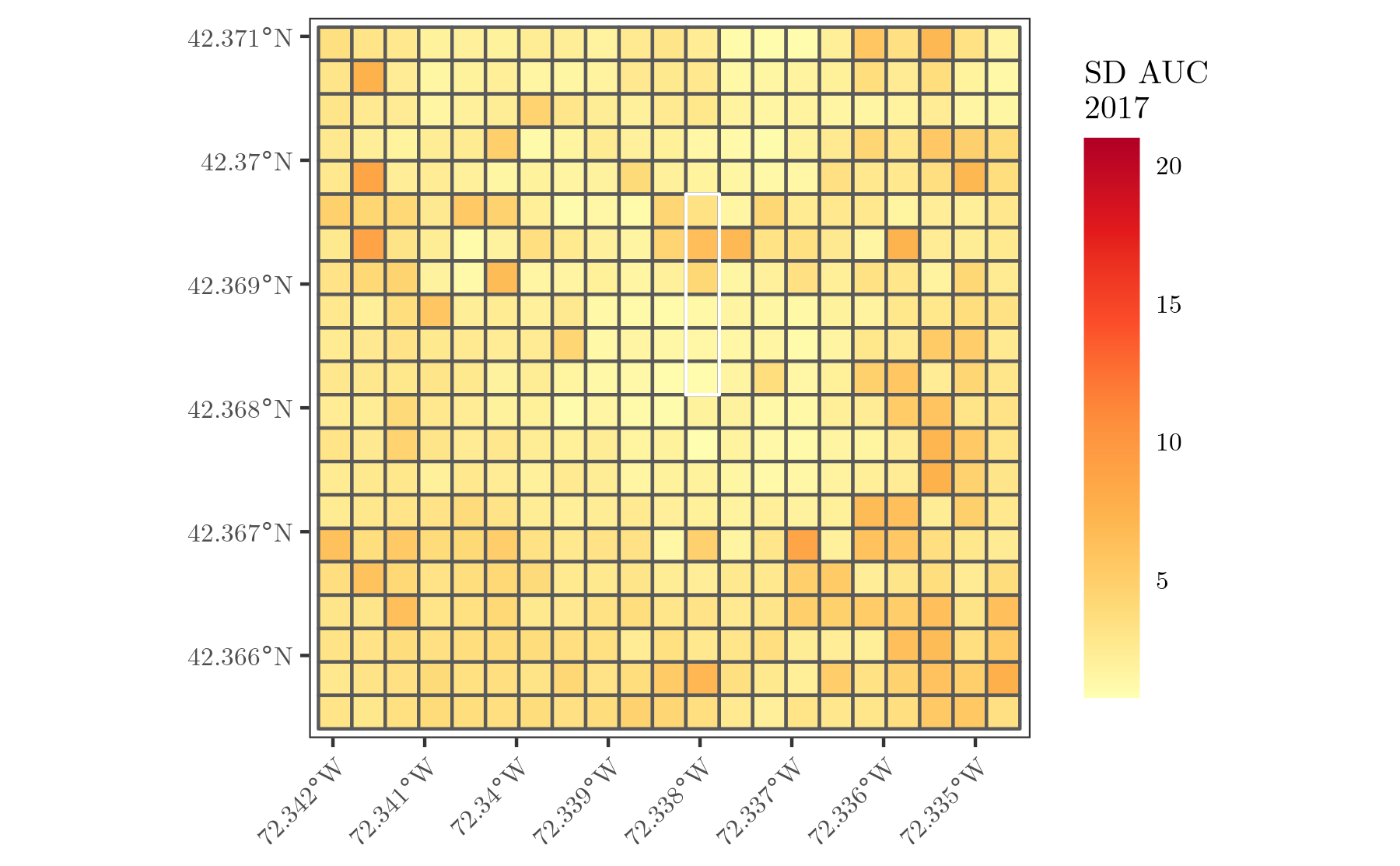}
\includegraphics[trim=1.95cm 0.25cm 2cm 0cm, clip, width=4.5cm]{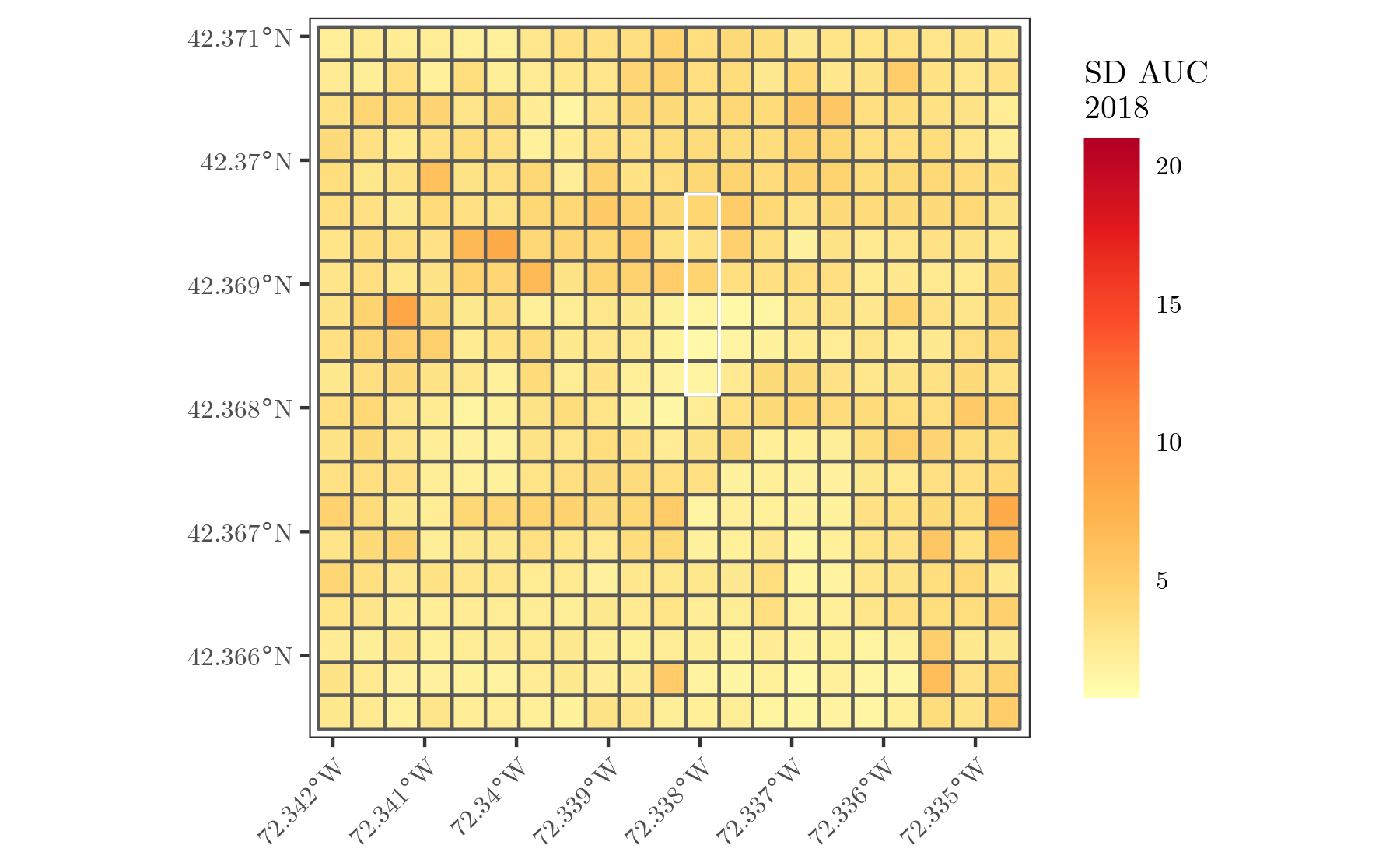}
\includegraphics[trim=1.95cm 0.25cm 2cm 0cm, clip, width=4.5cm]{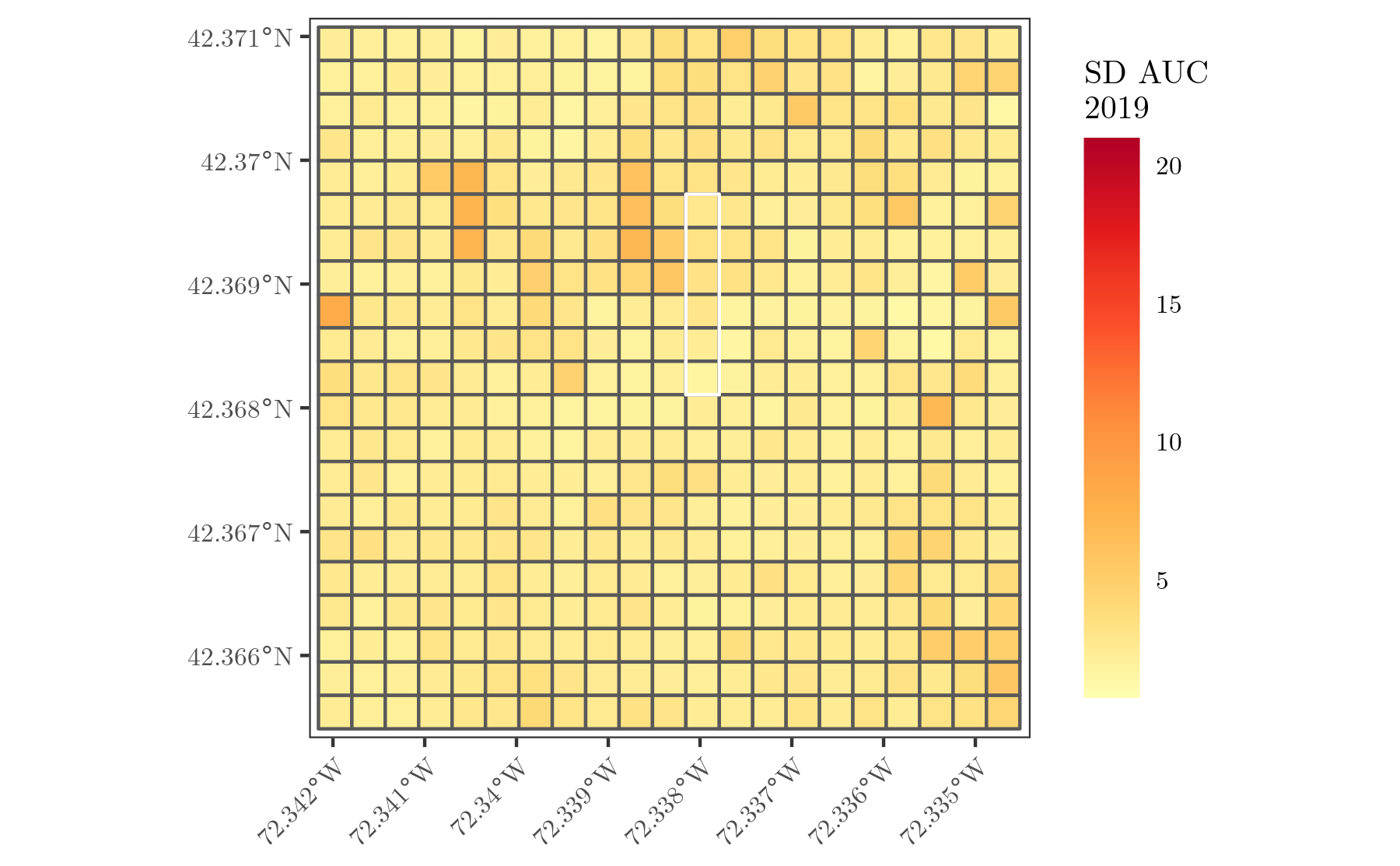}
\caption{Supplement to Figures~\ref{fig:quabbin_pixels} and \ref{fig:all_quabbin_pixels} that provides pixel-level area under the curve (AUC) posterior distribution standard deviation (SD) estimates for all years in the study period.}\label{fig:all_quabbin_sd_pixels}
\end{figure}

\begin{figure}[!ht]
\centering
\includegraphics[trim=1.95cm 0.25cm 2cm 0cm, clip, width=4.5cm]{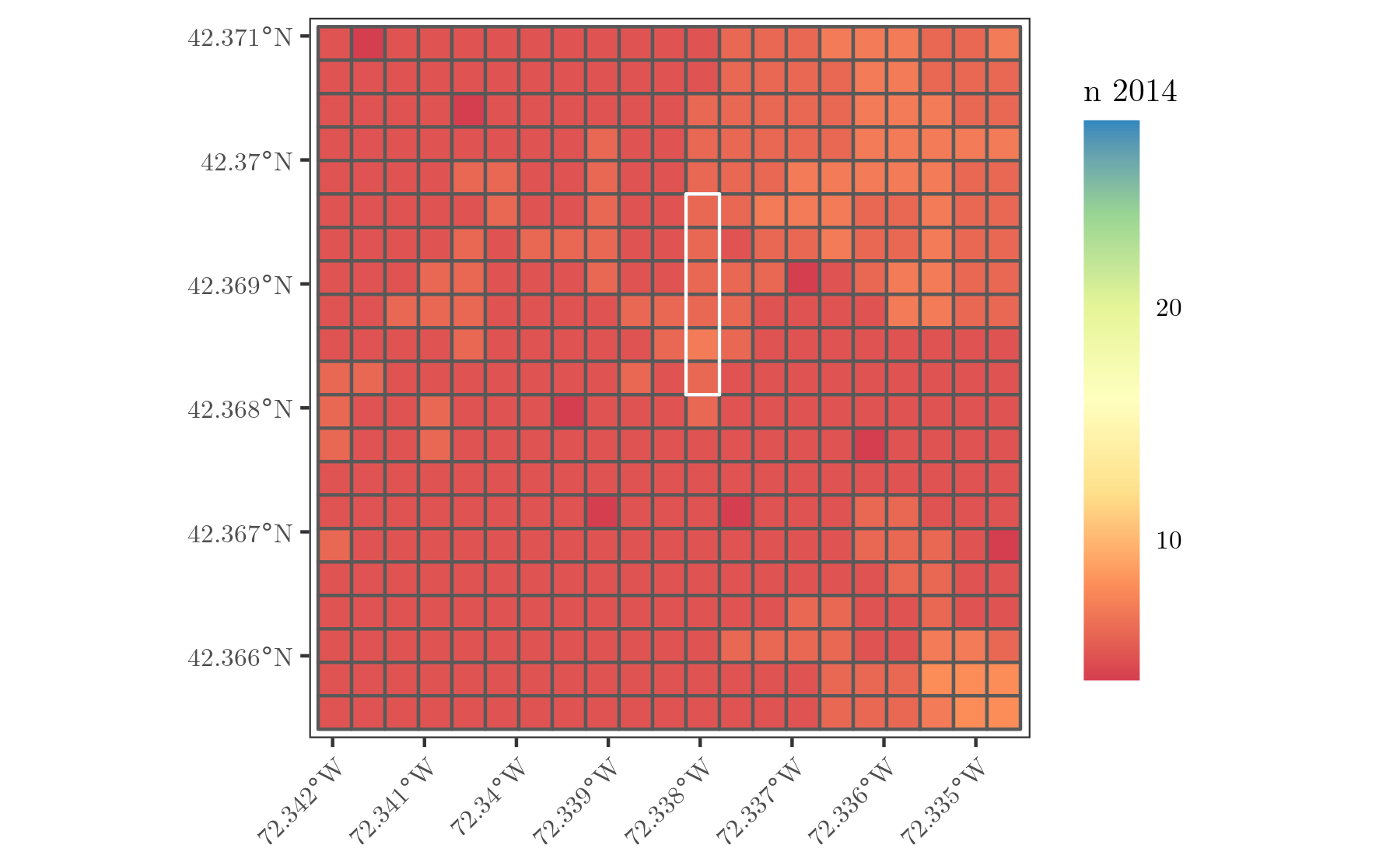}
\includegraphics[trim=1.95cm 0.25cm 2cm 0cm, clip, width=4.5cm]{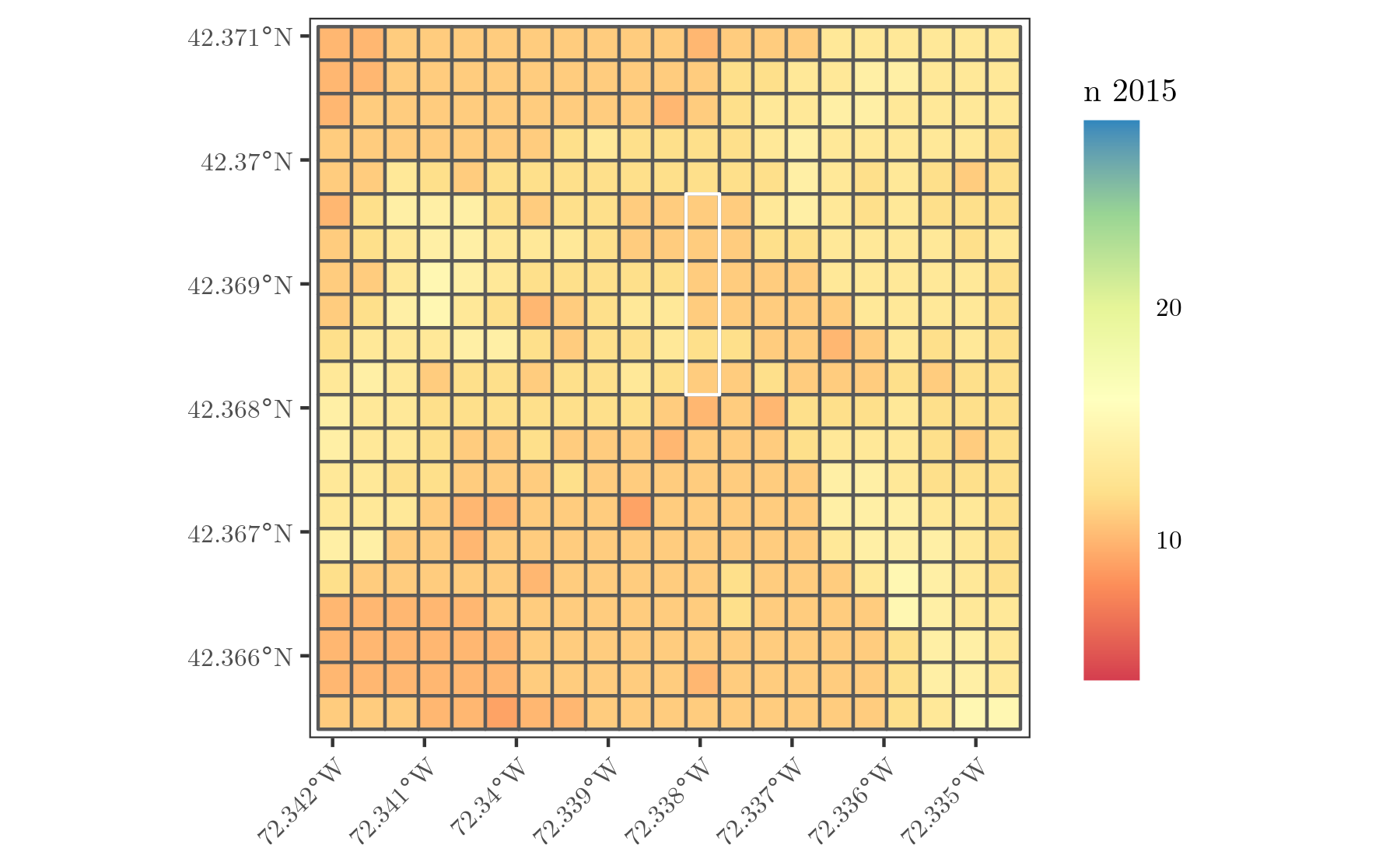}
\includegraphics[trim=1.95cm 0.25cm 2cm 0cm, clip, width=4.5cm]{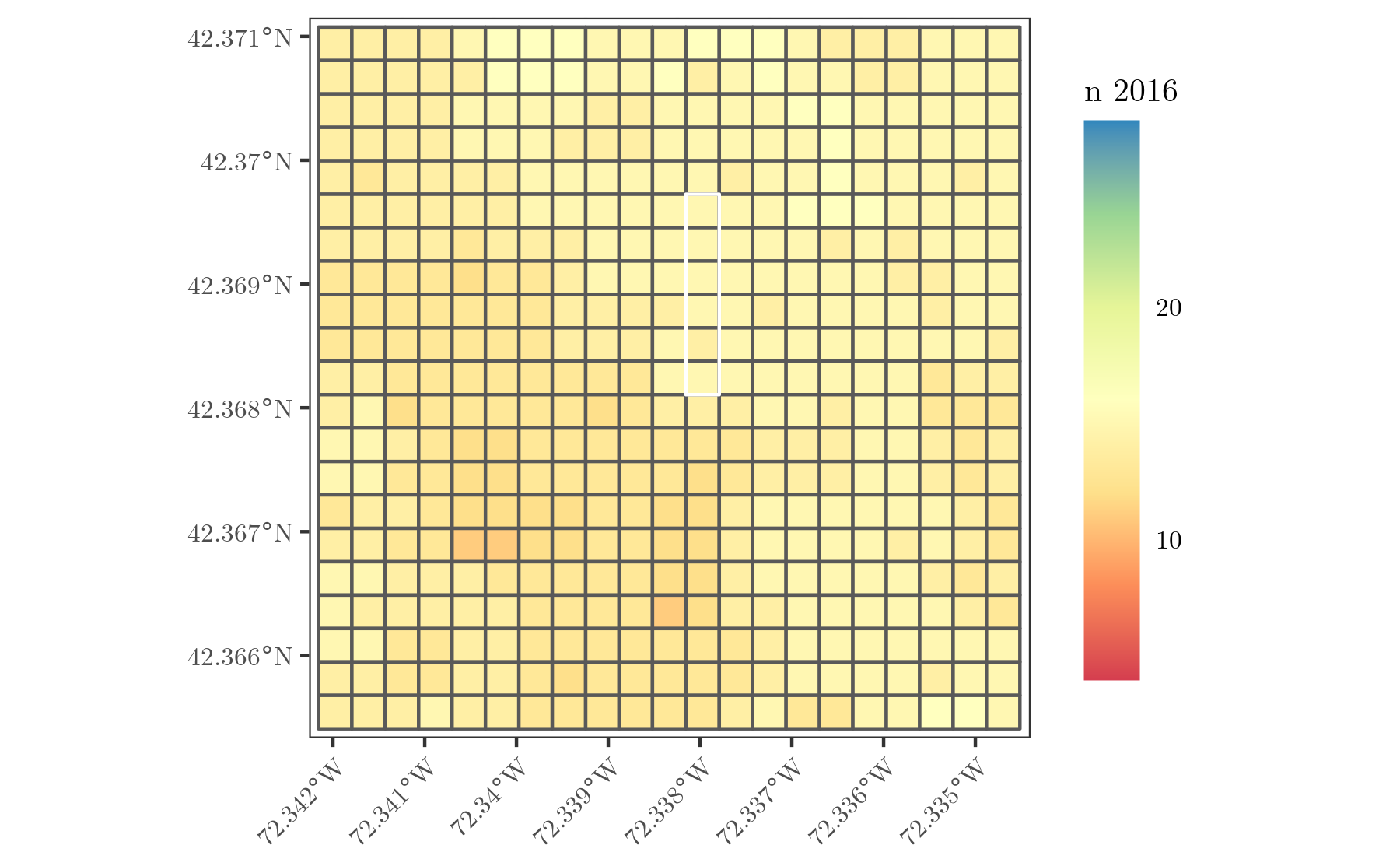}
\includegraphics[trim=1.95cm 0.25cm 2cm 0cm, clip, width=4.5cm]{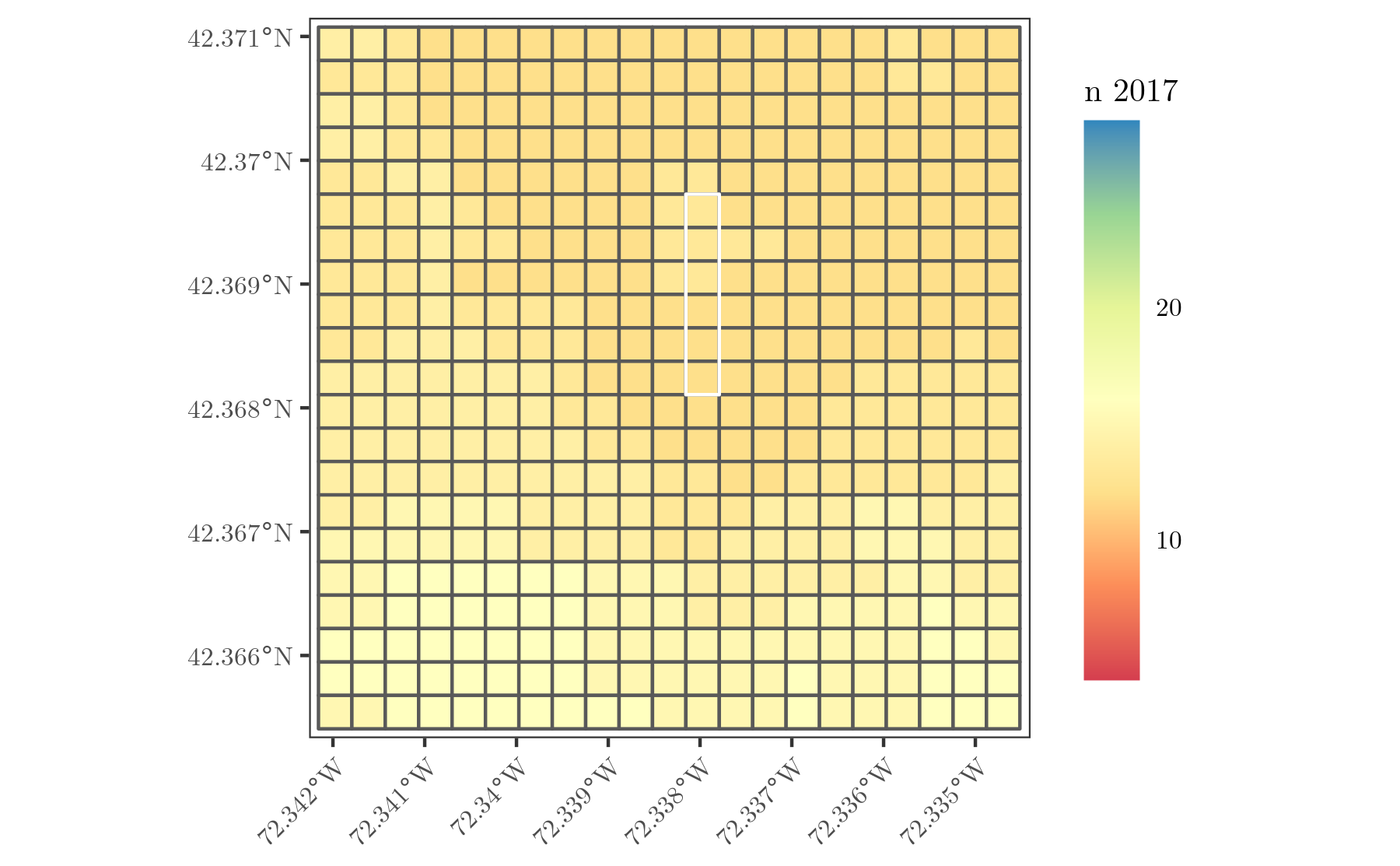}
\includegraphics[trim=1.95cm 0.25cm 2cm 0cm, clip, width=4.5cm]{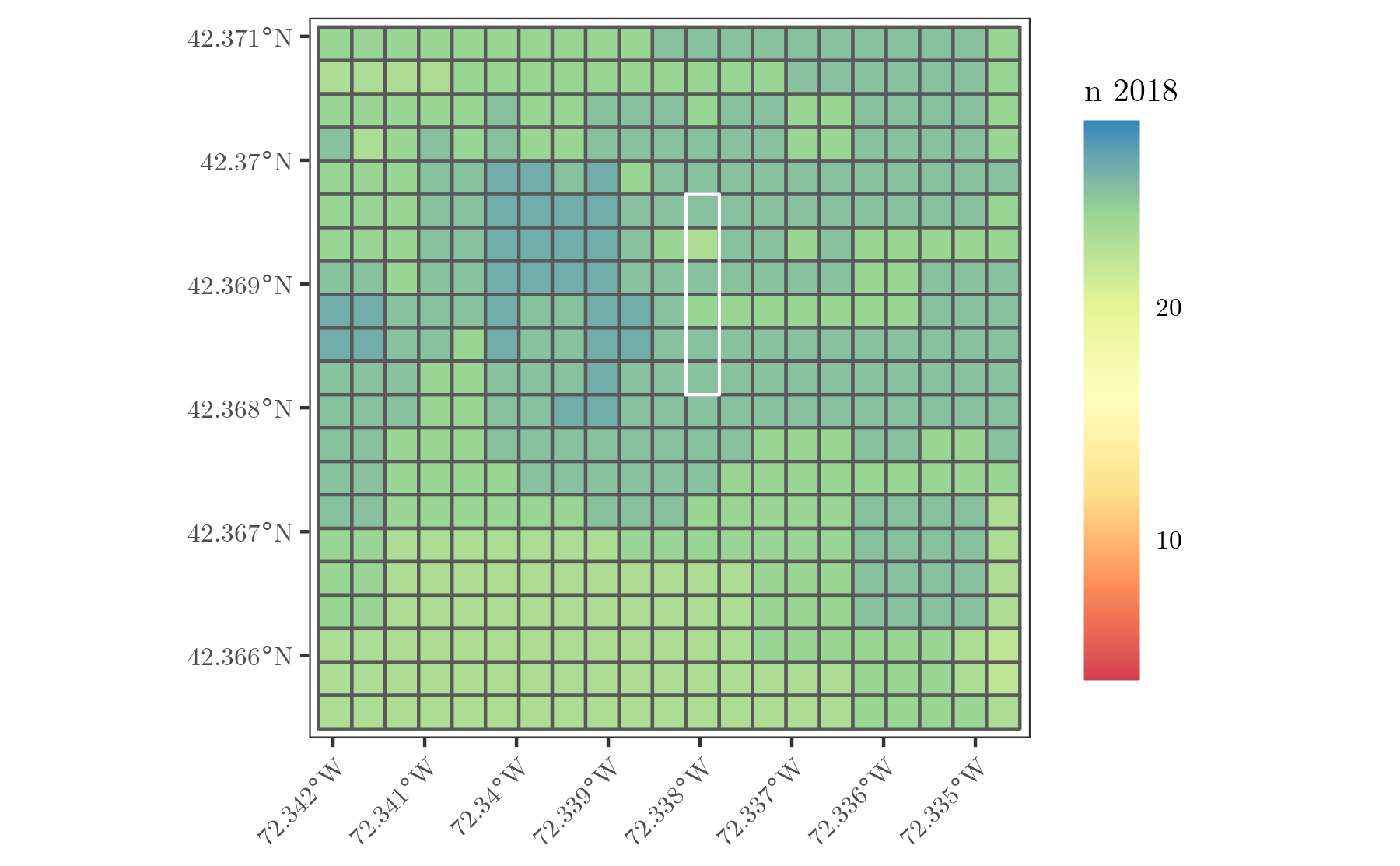}
\includegraphics[trim=1.95cm 0.25cm 2cm 0cm, clip, width=4.5cm]{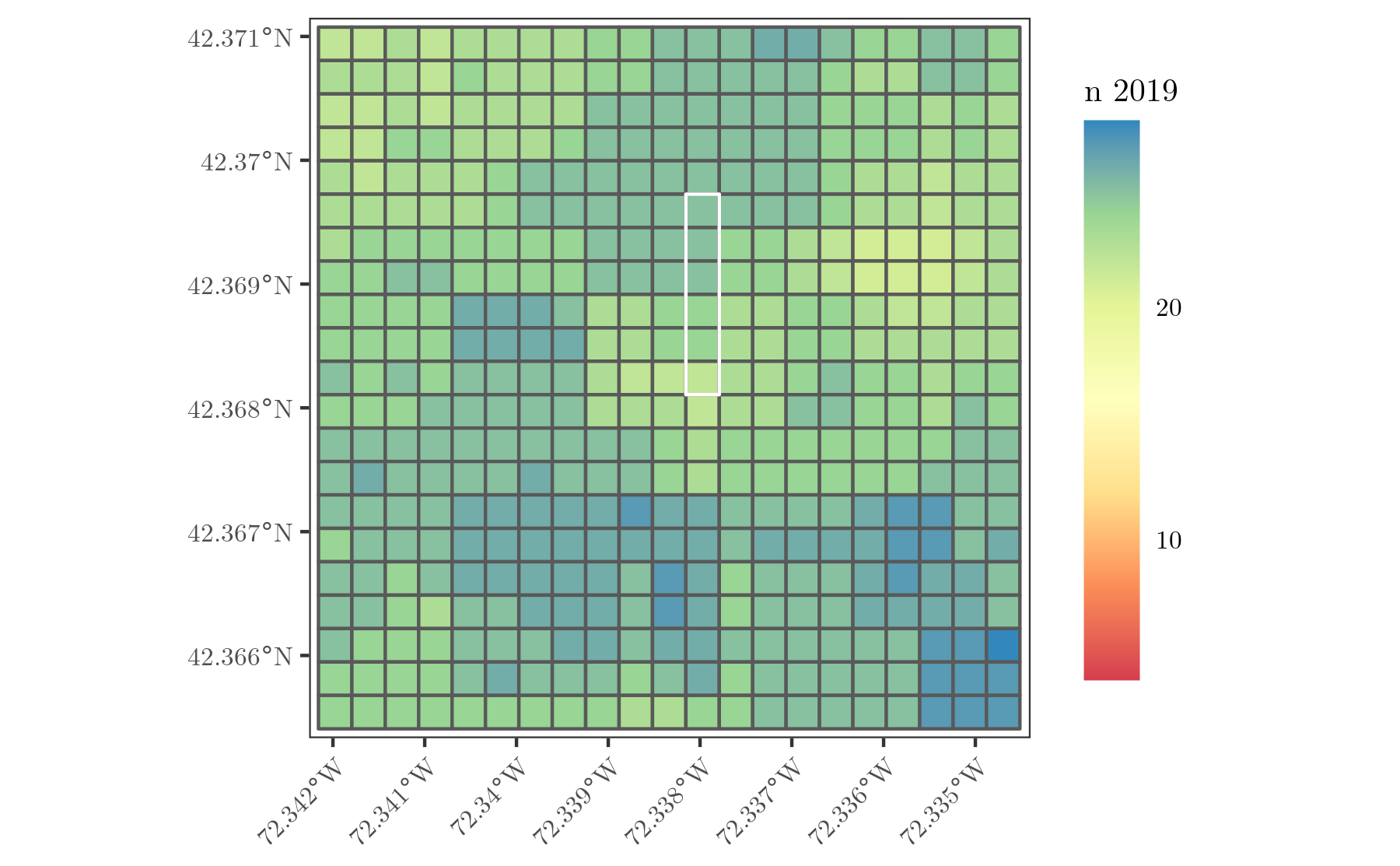}
\caption{Supplement to Figures~\ref{fig:quabbin_pixels}, \ref{fig:all_quabbin_pixels}, and \ref{fig:all_quabbin_pixels} that provides pixel-level sample size ($n$) used to estimate area under the curve (AUC) posterior distribution all years in the study period.}\label{fig:all_quabbin_n_pixels}
\end{figure}

\clearpage

\subsection{Reproduce portions of manuscript Section 3.3 using \code{pheno}}

The \R~code below reproduces key portions of the analysis in manuscript Section 3.3. We begin by loading the necessary packages and scripts. The current version of \pkg{rsBayes} 0.1.1 is required and can be obtained from its git repository using the code below.

\begin{knitrout}
\definecolor{shadecolor}{rgb}{0.969, 0.969, 0.969}\color{fgcolor}\begin{kframe}
\begin{alltt}
\hlstd{R> }\hlkwd{library}\hlstd{(devtools)}
\hlstd{R> }\hlstd{devtools}\hlopt{::}\hlkwd{install_github}\hlstd{(}\hlstr{"finleya/rsBayes"}\hlstd{)}
\end{alltt}
\end{kframe}
\end{knitrout}

\begin{knitrout}
\definecolor{shadecolor}{rgb}{0.969, 0.969, 0.969}\color{fgcolor}\begin{kframe}
\begin{alltt}
\hlstd{R> }\hlkwd{library}\hlstd{(rsBayes)}
\hlstd{R> }
\hlstd{R> }\hlcom{##for plotting}
\hlstd{R> }\hlkwd{library}\hlstd{(sp)}
\hlstd{R> }\hlkwd{library}\hlstd{(sf)}
\hlstd{R> }\hlkwd{library}\hlstd{(dplyr)}
\hlstd{R> }\hlkwd{library}\hlstd{(ggplot2)}
\hlstd{R> }\hlkwd{library}\hlstd{(cowplot)}
\hlstd{R> }\hlkwd{library}\hlstd{(viridis)}
\end{alltt}
\end{kframe}
\end{knitrout}

The HLS EVI data used in manuscript Section 3.3 is available in the \textsf{R} \textbf{rsBayes} package's \texttt{aoi\_hls} dataset. 
\begin{knitrout}
\definecolor{shadecolor}{rgb}{0.969, 0.969, 0.969}\color{fgcolor}\begin{kframe}
\begin{alltt}
\hlstd{R> }\hlkwd{head}\hlstd{(aoi_hls)}
\end{alltt}
\begin{verbatim}
## # A tibble: 6 x 7
##   pixel      x       y sat    year   doy   evi
##   <dbl>  <dbl>   <dbl> <fct> <int> <int> <dbl>
## 1     1 718875 4694385 L30    2014     4    NA
## 2     1 718875 4694385 L30    2014    13    NA
## 3     1 718875 4694385 L30    2014    20    NA
## 4     1 718875 4694385 L30    2014    29    NA
## 5     1 718875 4694385 L30    2014    36    NA
## 6     1 718875 4694385 L30    2014    45    NA
\end{verbatim}
\end{kframe}
\end{knitrout}

The \pkg{rsBayes} package \code{pheno} function implements all methods in manuscript Section 2. The code below generates MCMC samples for model parameters for each pixel of 2019 EVI data.
\begin{knitrout}
\definecolor{shadecolor}{rgb}{0.969, 0.969, 0.969}\color{fgcolor}\begin{kframe}
\begin{alltt}
\hlstd{R> }\hlstd{pixels.2019} \hlkwb{<-} \hlstd{aoi_hls} \hlopt{%>%} \hlkwd{filter}\hlstd{(year} \hlopt{==} \hlnum{2019}\hlstd{)}
\hlstd{R> }
\hlstd{R> }\hlstd{n.pixels} \hlkwb{<-} \hlkwd{length}\hlstd{(}\hlkwd{unique}\hlstd{(pixels.2019}\hlopt{$}\hlstd{pixel))}
\hlstd{R> }
\hlstd{R> }\hlstd{pixel.list} \hlkwb{<-} \hlkwd{list}\hlstd{()} \hlcom{##save each pixel's pheno model output}
\hlstd{R> }
\hlstd{R> }\hlkwd{set.seed}\hlstd{(}\hlnum{1}\hlstd{)}
\hlstd{R> }
\hlstd{R> }\hlkwa{for}\hlstd{(i} \hlkwa{in} \hlnum{1}\hlopt{:}\hlstd{n.pixels)\{}

\hlstd{+ }  \hlstd{pixel.dat} \hlkwb{<-} \hlkwd{subset}\hlstd{(pixels.2019, pixel} \hlopt{==} \hlstd{i)} \hlcom{##get data for the i-th pixel}

\hlstd{+ }  \hlstd{pixel.list[[i]]} \hlkwb{<-} \hlkwd{pheno}\hlstd{(evi}\hlopt{~}\hlstd{doy,} \hlkwc{data}\hlstd{=pixel.dat,} \hlkwc{family}\hlstd{=}\hlstr{"beta"}\hlstd{,}
\hlstd{+ }                           \hlkwc{starting}\hlstd{=}\hlkwd{list}\hlstd{(}\hlkwc{alpha.1}\hlstd{=}\hlnum{0.2}\hlstd{,} \hlkwc{alpha.2}\hlstd{=}\hlnum{0.5}\hlstd{,} \hlkwc{alpha.3}\hlstd{=}\hlnum{0.25}\hlstd{,}
\hlstd{+ }                                         \hlkwc{alpha.4}\hlstd{=}\hlnum{100}\hlstd{,} \hlkwc{alpha.5}\hlstd{=}\hlnum{0.0001}\hlstd{,} \hlkwc{alpha.6}\hlstd{=}\hlnum{0.25}\hlstd{,}
\hlstd{+ }                                         \hlkwc{alpha.7}\hlstd{=}\hlnum{200}\hlstd{,} \hlkwc{sigma.sq}\hlstd{=}\hlnum{0.001}\hlstd{),}
\hlstd{+ }                           \hlkwc{tuning}\hlstd{=}\hlkwd{list}\hlstd{(}\hlkwc{alpha.1}\hlstd{=}\hlnum{0.001}\hlstd{,} \hlkwc{alpha.2}\hlstd{=}\hlnum{0.01}\hlstd{,} \hlkwc{alpha.3}\hlstd{=}\hlnum{0.01}\hlstd{,}
\hlstd{+ }                                       \hlkwc{alpha.4}\hlstd{=}\hlnum{0.5}\hlstd{,} \hlkwc{alpha.5}\hlstd{=}\hlnum{0.0001}\hlstd{,} \hlkwc{alpha.6}\hlstd{=}\hlnum{0.01}\hlstd{,}
\hlstd{+ }                                       \hlkwc{alpha.7}\hlstd{=}\hlnum{1}\hlstd{,} \hlkwc{sigma.sq}\hlstd{=}\hlnum{0.1}\hlstd{),}
\hlstd{+ }                           \hlkwc{priors}\hlstd{=}\hlkwd{list}\hlstd{(}\hlkwc{alpha}\hlstd{=}\hlkwd{list}\hlstd{(}\hlkwc{alpha.5}\hlstd{=}\hlkwd{c}\hlstd{(}\hlopt{-}\hlnum{0.001}\hlstd{,} \hlnum{0.001}\hlstd{)),}
\hlstd{+ }                                       \hlkwc{sigma.sq.IG}\hlstd{=}\hlkwd{c}\hlstd{(}\hlnum{2}\hlstd{,} \hlnum{0.001}\hlstd{)),}
\hlstd{+ }                           \hlkwc{n.samples}\hlstd{=}\hlnum{50000}\hlstd{,}
\hlstd{+ }                           \hlkwc{sub.sample}\hlstd{=}\hlkwd{list}\hlstd{(}\hlstr{"start"}\hlstd{=}\hlnum{25000}\hlstd{,} \hlkwc{thin}\hlstd{=}\hlnum{25}\hlstd{),} \hlkwc{verbose} \hlstd{=} \hlnum{FALSE}\hlstd{)}

\hlstd{+ }\hlstd{\}}
\end{alltt}
\end{kframe}
\end{knitrout}

Note, each object in the \code{pixel.list} is a \code{pheno} model class that holds the respective pixel model information and MCMC samples. The code below checks the class then prints parameter estimates for the first pixel.
\begin{knitrout}
\definecolor{shadecolor}{rgb}{0.969, 0.969, 0.969}\color{fgcolor}\begin{kframe}
\begin{alltt}
\hlstd{R> }\hlkwd{class}\hlstd{(pixel.list[[}\hlnum{1}\hlstd{]])}
\end{alltt}
\begin{verbatim}
## [1] "pheno"
\end{verbatim}
\begin{alltt}
\hlstd{R> }\hlkwd{summary}\hlstd{(pixel.list[[}\hlnum{1}\hlstd{]])}
\end{alltt}
\begin{verbatim}
## Chain sub.sample:
## start = 1
## end = 1001
## thin = 1
## samples size = 1001
##          2.5%     25%      50%      75%      97.5%   
## alpha.1  0.1788   0.1969   0.2081   0.2160   0.2262  
## alpha.2  0.4914   0.5656   0.5948   0.6155   0.6490  
## alpha.3  0.0715   0.0965   0.1131   0.1365   0.1771  
## alpha.4  133.8739 136.5704 137.6818 138.6741 140.8830
## alpha.5  0.0004   0.0007   0.0009   0.0009   0.0010  
## alpha.6  0.0712   0.0940   0.1060   0.1228   0.1699  
## alpha.7  286.5546 290.2408 292.1101 294.1779 298.4399
## sigma.sq 0.0020   0.0028   0.0034   0.0041   0.0067
\end{verbatim}
\end{kframe}
\end{knitrout}

Now using the post convergence MCMC samples compute posterior samples for the area under the curve (AUC) then generate several summary statistics for the AUC posterior.
\begin{knitrout}
\definecolor{shadecolor}{rgb}{0.969, 0.969, 0.969}\color{fgcolor}\begin{kframe}
\begin{alltt}
\hlstd{R> }\hlstd{G} \hlkwb{<-} \hlkwa{function}\hlstd{(}\hlkwc{t}\hlstd{,} \hlkwc{alpha}\hlstd{)\{}
\hlstd{+ }    \hlkwd{ifelse}\hlstd{(t} \hlopt{<=} \hlstd{(alpha[}\hlnum{3}\hlstd{]}\hlopt{*}\hlstd{alpha[}\hlnum{4}\hlstd{]}\hlopt{+}\hlstd{alpha[}\hlnum{6}\hlstd{]}\hlopt{*}\hlstd{alpha[}\hlnum{7}\hlstd{])}\hlopt{/}\hlstd{(alpha[}\hlnum{3}\hlstd{]}\hlopt{+}\hlstd{alpha[}\hlnum{6}\hlstd{]),}
\hlstd{+ }           \hlstd{alpha[}\hlnum{1}\hlstd{]} \hlopt{+} \hlstd{(alpha[}\hlnum{2}\hlstd{]} \hlopt{-} \hlstd{alpha[}\hlnum{5}\hlstd{]}\hlopt{*}\hlstd{t)}\hlopt{/}\hlstd{(}\hlnum{1.0} \hlopt{+} \hlkwd{exp}\hlstd{(}\hlopt{-}\hlstd{alpha[}\hlnum{3}\hlstd{]}\hlopt{*}\hlstd{(t}\hlopt{-}\hlstd{alpha[}\hlnum{4}\hlstd{]))),}
\hlstd{+ }           \hlstd{alpha[}\hlnum{1}\hlstd{]} \hlopt{+} \hlstd{(alpha[}\hlnum{2}\hlstd{]} \hlopt{-} \hlstd{alpha[}\hlnum{5}\hlstd{]}\hlopt{*}\hlstd{t)}\hlopt{/}\hlstd{(}\hlnum{1.0} \hlopt{+} \hlkwd{exp}\hlstd{(}\hlopt{-}\hlstd{alpha[}\hlnum{6}\hlstd{]}\hlopt{*}\hlstd{(alpha[}\hlnum{7}\hlstd{]}\hlopt{-}\hlstd{t)))}
\hlstd{+ }           \hlstd{)}
\hlstd{+ }\hlstd{\}}
\hlstd{R> }
\hlstd{R> }\hlstd{my.summary} \hlkwb{<-} \hlkwa{function}\hlstd{(}\hlkwc{x}\hlstd{)\{}
\hlstd{+ }  \hlstd{q} \hlkwb{<-} \hlkwd{quantile}\hlstd{(x,} \hlkwc{probs}\hlstd{=}\hlkwd{c}\hlstd{(}\hlnum{0.5}\hlstd{,} \hlnum{0.05}\hlstd{,} \hlnum{0.975}\hlstd{))}
\hlstd{+ }  \hlkwd{c}\hlstd{(}\hlkwd{mean}\hlstd{(x),} \hlkwd{sd}\hlstd{(x), q[}\hlnum{1}\hlstd{], q[}\hlnum{3}\hlstd{]}\hlopt{-}\hlstd{q[}\hlnum{2}\hlstd{])}
\hlstd{+ }\hlstd{\}}
\hlstd{R> }
\hlstd{R> }\hlstd{post.auc} \hlkwb{<-} \hlkwd{t}\hlstd{(}\hlkwd{as.data.frame}\hlstd{(}\hlkwd{lapply}\hlstd{(pixel.list,} \hlkwa{function}\hlstd{(}\hlkwc{x}\hlstd{)\{}
\hlstd{+ }  \hlkwd{my.summary}\hlstd{(}\hlkwd{apply}\hlstd{(x}\hlopt{$}\hlstd{p.theta.samples,} \hlnum{1}\hlstd{,} \hlkwa{function}\hlstd{(}\hlkwc{v}\hlstd{)\{}
\hlstd{+ }    \hlkwd{integrate}\hlstd{(G,} \hlkwc{lower} \hlstd{=} \hlnum{1}\hlstd{,} \hlkwc{upper} \hlstd{=} \hlnum{365}\hlstd{, v)[}\hlnum{1}\hlstd{]}\hlopt{$}\hlstd{value\}))\})))}
\hlstd{R> }
\hlstd{R> }\hlkwd{rownames}\hlstd{(post.auc)} \hlkwb{<-} \hlkwa{NULL}
\hlstd{R> }\hlkwd{colnames}\hlstd{(post.auc)} \hlkwb{<-} \hlkwd{c}\hlstd{(}\hlstr{"auc.mean"}\hlstd{,} \hlstr{"auc.sd"}\hlstd{,} \hlstr{"auc.median"}\hlstd{,} \hlstr{"auc.q95.width"}\hlstd{)}
\end{alltt}
\end{kframe}
\end{knitrout}

The next bit of code joins the pixel coordinates with their respective posterior summaries, converts them to spatial objects, then plots a few of the summaries. Note the plot labeled ``Posterior median'' is the same as manuscript Figure 9(f). The other figures are provided to illustrate how other summary statistics are generated and mapped for the given posterior.
\begin{knitrout}
\definecolor{shadecolor}{rgb}{0.969, 0.969, 0.969}\color{fgcolor}\begin{kframe}
\begin{alltt}
\hlstd{R> }\hlcom{##get the coordinates for each pixel}
\hlstd{R> }\hlstd{df} \hlkwb{<-} \hlstd{pixels.2019} \hlopt{%>%} \hlkwd{group_by}\hlstd{(pixel)} \hlopt{%>%} \hlkwd{slice}\hlstd{(}\hlnum{1}\hlstd{)} \hlopt{%>%} \hlkwd{select}\hlstd{(pixel, x, y)}
\hlstd{R> }
\hlstd{R> }\hlcom{#join with posterior summaries and create spatial object}
\hlstd{R> }\hlstd{df} \hlkwb{<-} \hlkwd{cbind}\hlstd{(}\hlkwd{as.data.frame}\hlstd{(df), post.auc)}
\hlstd{R> }
\hlstd{R> }\hlkwd{coordinates}\hlstd{(df)} \hlkwb{<-} \hlkwd{c}\hlstd{(}\hlstr{"x"}\hlstd{,}\hlstr{"y"}\hlstd{)}
\hlstd{R> }\hlkwd{proj4string}\hlstd{(df)} \hlkwb{<-} \hlstr{"+proj=utm +zone=18 +ellps=WGS84 +units=m +no_defs"}
\hlstd{R> }\hlstd{sf.df} \hlkwb{<-} \hlkwd{as}\hlstd{(df,} \hlstr{"sf"}\hlstd{)}
\hlstd{R> }
\hlstd{R> }\hlcom{##make plots}
\hlstd{R> }\hlstd{size} \hlkwb{<-} \hlnum{2.5}
\hlstd{R> }\hlstd{auc.mean} \hlkwb{<-} \hlkwd{ggplot}\hlstd{(sf.df)} \hlopt{+}
\hlstd{+ }            \hlkwd{geom_sf}\hlstd{(}\hlkwc{data} \hlstd{= sf.df,} \hlkwd{aes}\hlstd{(}\hlkwc{color} \hlstd{= auc.mean),} \hlkwc{size}\hlstd{=size,} \hlkwc{shape}\hlstd{=}\hlnum{15}\hlstd{)} \hlopt{+}
\hlstd{+ }            \hlkwd{scale_color_viridis}\hlstd{(}\hlstr{"AUC\textbackslash{}n2019"}\hlstd{,} \hlkwc{direction} \hlstd{=} \hlopt{-}\hlnum{1}\hlstd{)} \hlopt{+}
\hlstd{+ }            \hlkwd{ggtitle}\hlstd{(}\hlstr{"Posteior mean"}\hlstd{)} \hlopt{+}
\hlstd{+ }            \hlkwd{theme_bw}\hlstd{()} \hlopt{+} \hlkwd{theme}\hlstd{(}\hlkwc{axis.text.x} \hlstd{=} \hlkwd{element_text}\hlstd{(}\hlkwc{angle} \hlstd{=} \hlnum{45}\hlstd{,} \hlkwc{hjust} \hlstd{=} \hlnum{1}\hlstd{))}
\hlstd{R> }
\hlstd{R> }\hlstd{auc.sd} \hlkwb{<-} \hlkwd{ggplot}\hlstd{(sf.df)} \hlopt{+}
\hlstd{+ }          \hlkwd{geom_sf}\hlstd{(}\hlkwc{data} \hlstd{= sf.df,} \hlkwd{aes}\hlstd{(}\hlkwc{color} \hlstd{= auc.sd),} \hlkwc{size}\hlstd{=size,} \hlkwc{shape}\hlstd{=}\hlnum{15}\hlstd{)} \hlopt{+}
\hlstd{+ }          \hlkwd{scale_color_viridis}\hlstd{(}\hlstr{"AUC\textbackslash{}n2019"}\hlstd{,} \hlkwc{option}\hlstd{=}\hlstr{"A"}\hlstd{,} \hlkwc{direction} \hlstd{=} \hlopt{-}\hlnum{1}\hlstd{)} \hlopt{+}
\hlstd{+ }          \hlkwd{ggtitle}\hlstd{(}\hlstr{"Posterior standard deviation"}\hlstd{)} \hlopt{+}
\hlstd{+ }          \hlkwd{theme_bw}\hlstd{()} \hlopt{+} \hlkwd{theme}\hlstd{(}\hlkwc{axis.text.x} \hlstd{=} \hlkwd{element_text}\hlstd{(}\hlkwc{angle} \hlstd{=} \hlnum{45}\hlstd{,} \hlkwc{hjust} \hlstd{=} \hlnum{1}\hlstd{))}
\hlstd{R> }
\hlstd{R> }\hlstd{auc.median} \hlkwb{<-} \hlkwd{ggplot}\hlstd{(sf.df)} \hlopt{+}
\hlstd{+ }              \hlkwd{geom_sf}\hlstd{(}\hlkwc{data} \hlstd{= sf.df,} \hlkwd{aes}\hlstd{(}\hlkwc{color} \hlstd{= auc.median),} \hlkwc{size}\hlstd{=size,} \hlkwc{shape}\hlstd{=}\hlnum{15}\hlstd{)} \hlopt{+}
\hlstd{+ }              \hlkwd{scale_color_viridis}\hlstd{(}\hlstr{"AUC\textbackslash{}n2019"}\hlstd{,} \hlkwc{direction} \hlstd{=} \hlopt{-}\hlnum{1}\hlstd{)} \hlopt{+}
\hlstd{+ }              \hlkwd{ggtitle}\hlstd{(}\hlstr{"Posterior median"}\hlstd{)} \hlopt{+}
\hlstd{+ }              \hlkwd{theme_bw}\hlstd{()} \hlopt{+} \hlkwd{theme}\hlstd{(}\hlkwc{axis.text.x} \hlstd{=} \hlkwd{element_text}\hlstd{(}\hlkwc{angle} \hlstd{=} \hlnum{45}\hlstd{,} \hlkwc{hjust} \hlstd{=} \hlnum{1}\hlstd{))}
\hlstd{R> }
\hlstd{R> }\hlstd{auc.q95.width} \hlkwb{<-} \hlkwd{ggplot}\hlstd{(sf.df)} \hlopt{+}
\hlstd{+ }                \hlkwd{geom_sf}\hlstd{(}\hlkwc{data} \hlstd{= sf.df,} \hlkwd{aes}\hlstd{(}\hlkwc{color} \hlstd{= auc.q95.width),} \hlkwc{size}\hlstd{=size,} \hlkwc{shape}\hlstd{=}\hlnum{15}\hlstd{)} \hlopt{+}
\hlstd{+ }                \hlkwd{scale_color_viridis}\hlstd{(}\hlstr{"AUC\textbackslash{}n2019"}\hlstd{,} \hlkwc{option}\hlstd{=}\hlstr{"A"}\hlstd{,} \hlkwc{direction} \hlstd{=} \hlopt{-}\hlnum{1}\hlstd{)} \hlopt{+}
\hlstd{+ }                \hlkwd{ggtitle}\hlstd{(}\hlstr{"Posterior 95% CI width"}\hlstd{)} \hlopt{+}
\hlstd{+ }                \hlkwd{theme_bw}\hlstd{()} \hlopt{+} \hlkwd{theme}\hlstd{(}\hlkwc{axis.text.x} \hlstd{=} \hlkwd{element_text}\hlstd{(}\hlkwc{angle} \hlstd{=} \hlnum{45}\hlstd{,} \hlkwc{hjust} \hlstd{=} \hlnum{1}\hlstd{))}
\hlstd{R> }
\hlstd{R> }\hlkwd{plot_grid}\hlstd{(auc.mean, auc.sd, auc.median, auc.q95.width,} \hlkwc{align}\hlstd{=}\hlstr{"hv"}\hlstd{)}
\end{alltt}
\end{kframe}

{\centering \includegraphics[width=\maxwidth]{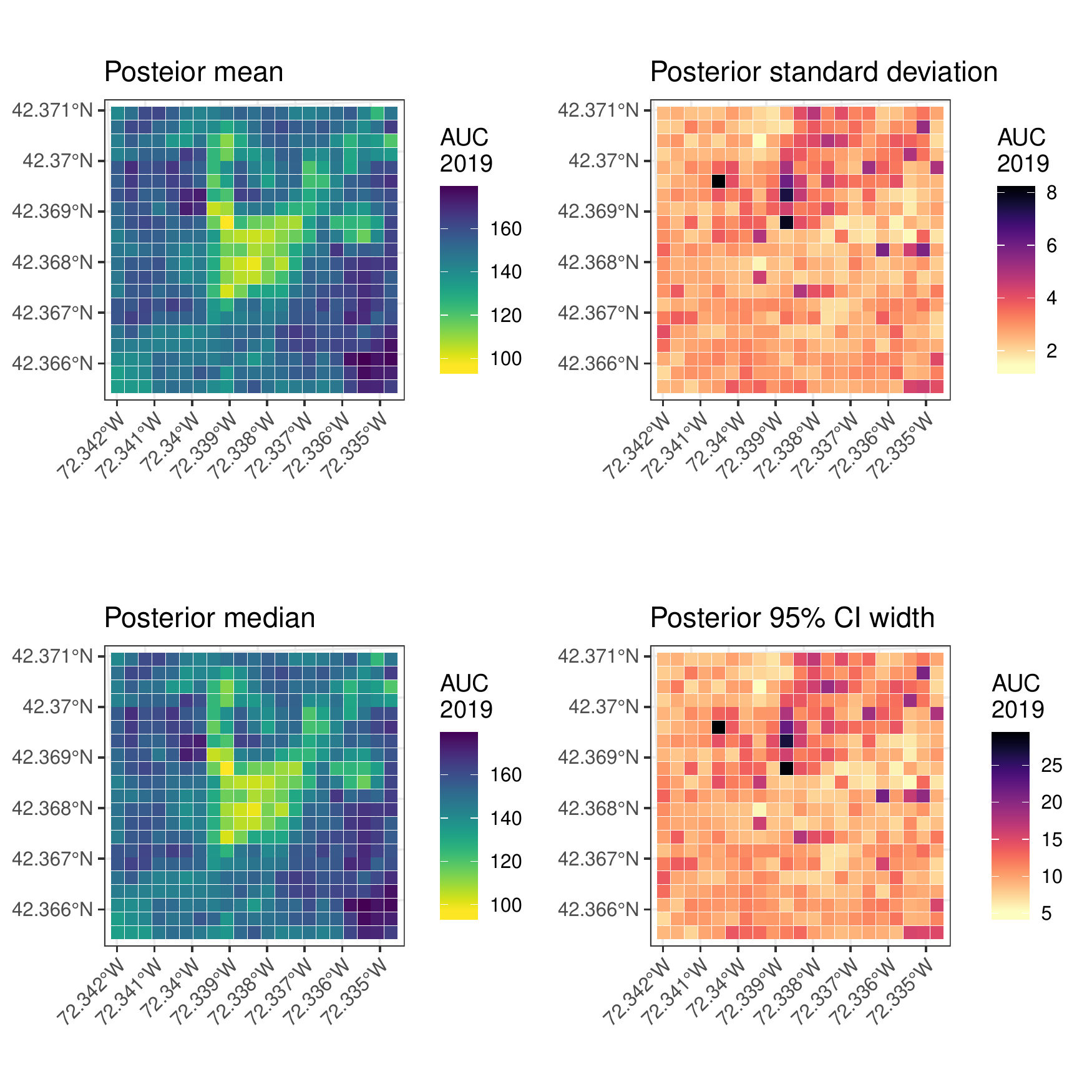} 

}

\end{knitrout}

\pagebreak
\subsection{Run \code{phenoBrick} for \code{RasterBrick} object using multiple CPU cores}

The \R~code below uses the \code{phenoBrick} function to fit the Bayesian LSP model presented in manuscript Section 2. The \R~code that defines the \code{phenoBrick} function is given in Section \ref{phenoBrick} of this supplement. \R-like documentation for \code{phenoBrick} is provided in Section \ref{documentation} of this supplement. We use the AOI for manuscript Section 3.3 as our example dataset---this time formatted as a \code{RasterBrick} object.

The following code chunk loads the necessary packages and the \code{phenoBrick} function. Note that the code defining the \code{phenoBrick} function is suppressed in this portion of the supplement to make the code in this document easier to follow.
\begin{knitrout}
\definecolor{shadecolor}{rgb}{0.969, 0.969, 0.969}\color{fgcolor}\begin{kframe}
\begin{alltt}
\hlstd{R> }\hlkwd{library}\hlstd{(rsBayes)}
\hlstd{R> }
\hlstd{R> }\hlcom{## needed for phenoBrick function}
\hlstd{R> }\hlkwd{library}\hlstd{(raster)}
\hlstd{R> }\hlkwd{library}\hlstd{(doSNOW)}
\hlstd{R> }\hlkwd{library}\hlstd{(progress)}
\hlstd{R> }
\hlstd{R> }\hlcom{## needed for plotting}
\hlstd{R> }\hlkwd{library}\hlstd{(sf)}
\hlstd{R> }\hlkwd{library}\hlstd{(ggplot2)}
\hlstd{R> }\hlkwd{library}\hlstd{(cowplot)}
\hlstd{R> }\hlkwd{library}\hlstd{(viridis)}
\hlstd{R> }
\hlstd{R> }\hlcom{## phenoBrick function code suppressed for Supplementary Material pdf}
\end{alltt}
\end{kframe}
\end{knitrout}

The next piece of code loads the \code{aoi\_hls\_rast} and \code{aoi\_hls\_rast\_info} data objects. The \code{aoi\_hls\_rast} object holds the same VI data as \code{rsBayes::aoi\_hls} only in a \code{RasterBrick} format. The \code{aoi\_hls\_rast\_info} object holds the DOY metadata for each of the 651 bands in \code{aoi\_hls\_rast}. Row one of \code{aoi\_hls\_rast\_info} corresponds to band 1 of \code{aoi\_hls\_rast} and so on.
\begin{knitrout}
\definecolor{shadecolor}{rgb}{0.969, 0.969, 0.969}\color{fgcolor}\begin{kframe}
\begin{alltt}
\hlstd{R> }\hlkwd{class}\hlstd{(aoi_hls_rast)}
\end{alltt}
\begin{verbatim}
## [1] "RasterBrick"
## attr(,"package")
## [1] "raster"
\end{verbatim}
\begin{alltt}
\hlstd{R> }\hlkwd{nlayers}\hlstd{(aoi_hls_rast)}
\end{alltt}
\begin{verbatim}
## [1] 651
\end{verbatim}
\begin{alltt}
\hlstd{R> }\hlkwd{head}\hlstd{(aoi_hls_rast_info)}
\end{alltt}
\begin{verbatim}
##   sat year doy
## 1 L30 2013 106
## 2 L30 2013 113
## 3 L30 2013 122
## 4 L30 2013 138
## 5 L30 2013 145
## 6 L30 2013 154
\end{verbatim}
\begin{alltt}
\hlstd{R> }\hlkwd{dim}\hlstd{(aoi_hls_rast_info)}
\end{alltt}
\begin{verbatim}
## [1] 651   3
\end{verbatim}
\end{kframe}
\end{knitrout}

\pagebreak

The following code sets the necessary starting, tuning, and prior values for \code{phenoBrick} in the same list format used for \code{pheno}. The code below also sets the number of MCMC samples and subsets the resulting MCMC samples for subsequent summarization. The number of CPU cores for \code{phenoBrick} to use is also set here. 
\begin{knitrout}
\definecolor{shadecolor}{rgb}{0.969, 0.969, 0.969}\color{fgcolor}\begin{kframe}
\begin{alltt}
\hlstd{R> }\hlcom{## set starting values}
\hlstd{R> }\hlstd{starting} \hlkwb{<-} \hlkwd{list}\hlstd{(}\hlkwc{alpha.1}\hlstd{=}\hlnum{0.2}\hlstd{,} \hlkwc{alpha.2}\hlstd{=}\hlnum{0.5}\hlstd{,} \hlkwc{alpha.3}\hlstd{=}\hlnum{0.25}\hlstd{,} \hlkwc{alpha.4}\hlstd{=}\hlnum{100}\hlstd{,} \hlkwc{alpha.5}\hlstd{=}\hlnum{0.0001}\hlstd{,}
\hlstd{+ }                 \hlkwc{alpha.6}\hlstd{=}\hlnum{0.25}\hlstd{,} \hlkwc{alpha.7}\hlstd{=}\hlnum{200}\hlstd{,} \hlkwc{sigma.sq}\hlstd{=}\hlnum{0.001}\hlstd{)}
\hlstd{R> }\hlcom{## set tuning values}
\hlstd{R> }\hlstd{tuning} \hlkwb{<-} \hlkwd{list}\hlstd{(}\hlkwc{alpha.1}\hlstd{=}\hlnum{0.001}\hlstd{,} \hlkwc{alpha.2}\hlstd{=}\hlnum{0.001}\hlstd{,} \hlkwc{alpha.3}\hlstd{=}\hlnum{0.005}\hlstd{,}\hlkwc{alpha.4}\hlstd{=}\hlnum{0.5}\hlstd{,}\hlkwc{alpha.5}\hlstd{=}\hlnum{0.0001}\hlstd{,}
\hlstd{+ }               \hlkwc{alpha.6}\hlstd{=}\hlnum{0.005}\hlstd{,} \hlkwc{alpha.7}\hlstd{=}\hlnum{0.5}\hlstd{,} \hlkwc{sigma.sq}\hlstd{=}\hlnum{0.03}\hlstd{)}
\hlstd{R> }\hlcom{## set priors}
\hlstd{R> }\hlstd{priors} \hlkwb{<-} \hlkwd{list}\hlstd{(}\hlkwc{alpha}\hlstd{=}\hlkwd{list}\hlstd{(}\hlkwc{alpha.5}\hlstd{=}\hlkwd{c}\hlstd{(}\hlopt{-}\hlnum{0.001}\hlstd{,} \hlnum{0.001}\hlstd{)),} \hlkwc{sigma.sq.IG}\hlstd{=}\hlkwd{c}\hlstd{(}\hlnum{2}\hlstd{,} \hlnum{0.001}\hlstd{))}
\hlstd{R> }\hlcom{## set number of MCMC samples}
\hlstd{R> }\hlstd{n.samples}  \hlkwb{<-} \hlnum{50000}
\hlstd{R> }\hlstd{sub.sample} \hlkwb{<-} \hlkwd{list}\hlstd{(}\hlstr{"start"} \hlstd{=} \hlnum{25001}\hlstd{,} \hlstr{"thin"} \hlstd{=} \hlnum{25}\hlstd{)}
\hlstd{R> }\hlcom{## set the number of CPU cores phenoBrick should use}
\hlstd{R> }\hlstd{n.cores}  \hlkwb{<-} \hlnum{2}
\hlstd{R> }\hlkwd{set.seed}\hlstd{(}\hlnum{1}\hlstd{)}
\hlstd{R> }\hlstd{modBrick} \hlkwb{<-} \hlkwd{phenoBrick}\hlstd{(}\hlkwc{VI.rast} \hlstd{= aoi_hls_rast,} \hlkwc{doy} \hlstd{= aoi_hls_rast_info}\hlopt{$}\hlstd{doy,} \hlkwc{family} \hlstd{=} \hlstr{"beta"}\hlstd{,}
\hlstd{+ }                       \hlkwc{starting} \hlstd{= starting,} \hlkwc{tuning} \hlstd{= tuning,} \hlkwc{priors} \hlstd{= priors,}
\hlstd{+ }                       \hlkwc{n.samples} \hlstd{= n.samples,}\hlkwc{sub.sample} \hlstd{= sub.sample,}\hlkwc{n.cores} \hlstd{= n.cores)}
\end{alltt}
\end{kframe}
\end{knitrout}

The \code{modBrick} object is a list of 9 \code{RasterBrick} objects. The first 8 hold the subsetted MCMC samples for the $\alpha$ and $\sigma^2$ parameters---1 for each parameter. The Metropolis acceptance rates for each pixel are held in \code{modBrick\$acc.rate}. The following figure maps the acceptance rate (left) and shows a histogram of the pixel acceptance rates (right). It is evident that only a few pixels fall outside the 20-50 percent range.
\begin{knitrout}
\definecolor{shadecolor}{rgb}{0.969, 0.969, 0.969}\color{fgcolor}\begin{kframe}
\begin{alltt}
\hlstd{R> }\hlkwd{par}\hlstd{(}\hlkwc{mfrow} \hlstd{=} \hlkwd{c}\hlstd{(}\hlnum{1}\hlstd{,}\hlnum{2}\hlstd{))}
\hlstd{R> }\hlkwd{plot}\hlstd{(modBrick}\hlopt{$}\hlstd{acc.rate[[}\hlnum{1}\hlstd{]],} \hlkwc{main} \hlstd{=} \hlstr{"acceptance rate (%)"}\hlstd{)}
\hlstd{R> }\hlkwd{hist}\hlstd{(modBrick}\hlopt{$}\hlstd{acc.rate[[}\hlnum{1}\hlstd{]],} \hlkwc{main} \hlstd{=} \hlstr{"acceptance rate (%)"}\hlstd{,} \hlkwc{xlab} \hlstd{=} \hlstr{""}\hlstd{)}
\end{alltt}
\end{kframe}

{\centering \includegraphics[width=\maxwidth]{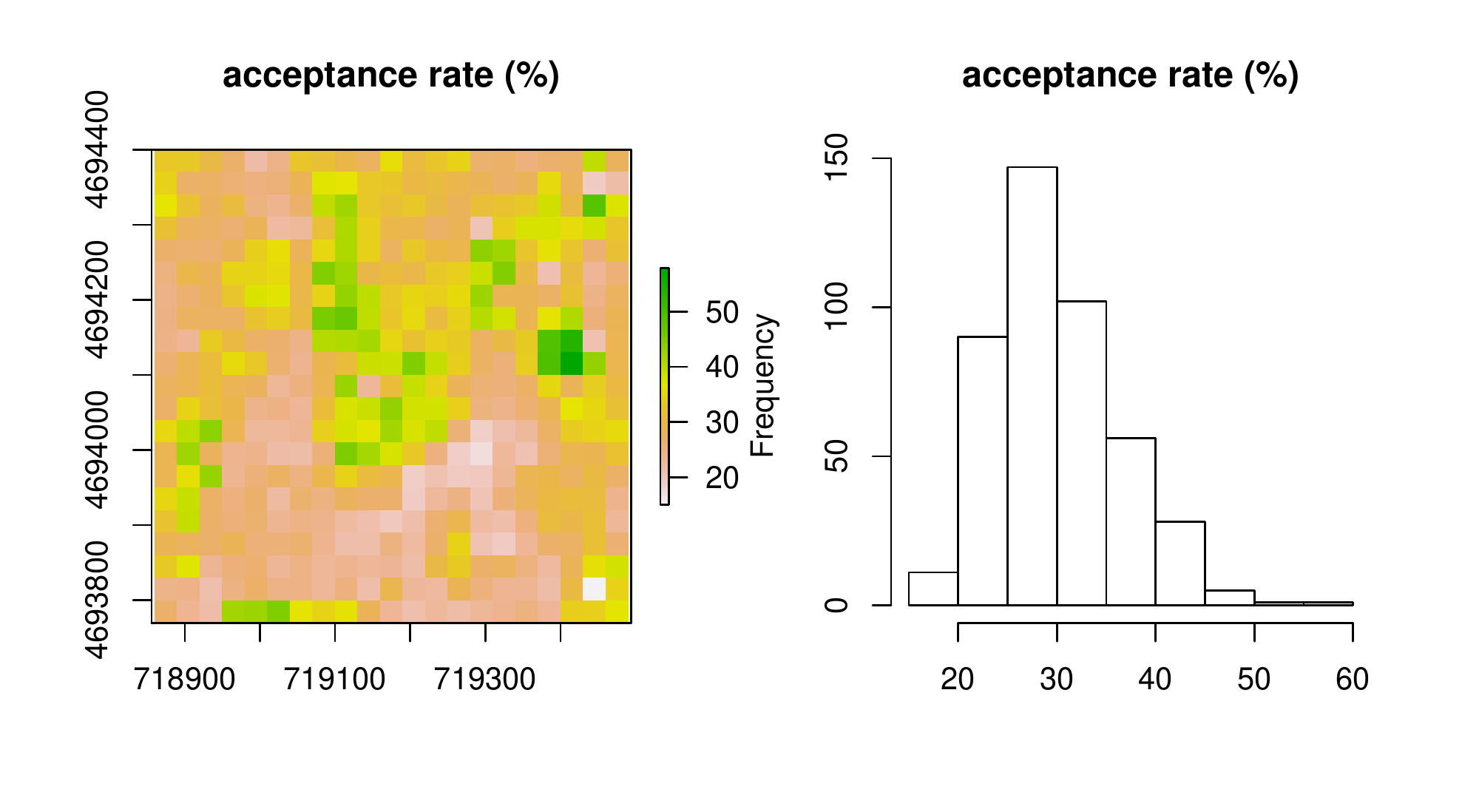}

}

\end{knitrout}

The next piece of code calculates posterior medians and standard deviations for $\alpha_1$ and for season length ($\alpha_7 - \alpha_4$).
\begin{knitrout}
\definecolor{shadecolor}{rgb}{0.969, 0.969, 0.969}\color{fgcolor}\begin{kframe}
\begin{alltt}
\hlstd{R> }\hlcom{## calculate posterior medians for alpha.1}
\hlstd{R> }\hlstd{alpha1.median} \hlkwb{<-} \hlkwd{stackApply}\hlstd{(modBrick}\hlopt{$}\hlstd{alpha1.samps,}\hlnum{1}\hlstd{,median)}
\hlstd{R> }
\hlstd{R> }\hlcom{## calculate posterior standard deviations for alpha.1}
\hlstd{R> }\hlstd{alpha1.sd} \hlkwb{<-} \hlkwd{stackApply}\hlstd{(modBrick}\hlopt{$}\hlstd{alpha1.samps,}\hlnum{1}\hlstd{,sd)}
\hlstd{R> }
\hlstd{R> }\hlcom{## calculate posterior medians for season length (alpha.7 - alpha.4)}
\hlstd{R> }\hlstd{season.length.median} \hlkwb{<-}\hlkwd{stackApply}\hlstd{(modBrick}\hlopt{$}\hlstd{alpha7.samps}\hlopt{-}\hlstd{modBrick}\hlopt{$}\hlstd{alpha4.samps,}\hlnum{1}\hlstd{,median)}
\hlstd{R> }
\hlstd{R> }\hlcom{## calculate posterior standard deviations for season length (alpha.7 - alpha.4)}
\hlstd{R> }\hlstd{season.length.sd} \hlkwb{<-} \hlkwd{stackApply}\hlstd{(modBrick}\hlopt{$}\hlstd{alpha7.samps}\hlopt{-}\hlstd{modBrick}\hlopt{$}\hlstd{alpha4.samps,}\hlnum{1}\hlstd{,sd)}
\end{alltt}
\end{kframe}
\end{knitrout}

The following converts the summary \code{RasterLayer} objects to \code{sf} objects for plotting purposes.
\begin{knitrout}
\definecolor{shadecolor}{rgb}{0.969, 0.969, 0.969}\color{fgcolor}\begin{kframe}
\begin{alltt}
\hlstd{R> }\hlcom{## convert to sf for ggplot}
\hlstd{R> }\hlstd{alpha1.median} \hlkwb{<-} \hlkwd{as}\hlstd{(}\hlkwd{as}\hlstd{(alpha1.median,} \hlstr{"SpatialPixelsDataFrame"}\hlstd{),}\hlstr{"sf"}\hlstd{)}
\hlstd{R> }\hlkwd{names}\hlstd{(alpha1.median)[}\hlnum{1}\hlstd{]} \hlkwb{<-} \hlstr{"alpha1.median"}
\hlstd{R> }
\hlstd{R> }\hlstd{alpha1.sd} \hlkwb{<-} \hlkwd{as}\hlstd{(}\hlkwd{as}\hlstd{(alpha1.sd,} \hlstr{"SpatialPixelsDataFrame"}\hlstd{),}\hlstr{"sf"}\hlstd{)}
\hlstd{R> }\hlkwd{names}\hlstd{(alpha1.sd)[}\hlnum{1}\hlstd{]} \hlkwb{<-} \hlstr{"alpha1.sd"}
\hlstd{R> }
\hlstd{R> }\hlstd{season.length.median} \hlkwb{<-} \hlkwd{as}\hlstd{(}\hlkwd{as}\hlstd{(season.length.median,} \hlstr{"SpatialPixelsDataFrame"}\hlstd{),}\hlstr{"sf"}\hlstd{)}
\hlstd{R> }\hlkwd{names}\hlstd{(season.length.median)[}\hlnum{1}\hlstd{]} \hlkwb{<-} \hlstr{"season.length.median"}
\hlstd{R> }
\hlstd{R> }\hlstd{season.length.sd} \hlkwb{<-} \hlkwd{as}\hlstd{(}\hlkwd{as}\hlstd{(season.length.sd,} \hlstr{"SpatialPixelsDataFrame"}\hlstd{),}\hlstr{"sf"}\hlstd{)}
\hlstd{R> }\hlkwd{names}\hlstd{(season.length.sd)[}\hlnum{1}\hlstd{]} \hlkwb{<-} \hlstr{"season.length.sd"}
\end{alltt}
\end{kframe}
\end{knitrout}

The next code chunk generates maps of the posterior medians and standard deviations for $\alpha_1$ and season length ($\alpha_7 - \alpha_4$).
\begin{knitrout}
\definecolor{shadecolor}{rgb}{0.969, 0.969, 0.969}\color{fgcolor}\begin{kframe}
\begin{alltt}
\hlstd{R> }\hlcom{## make maps}
\hlstd{R> }\hlstd{size} \hlkwb{<-} \hlnum{2.5}
\hlstd{R> }
\hlstd{R> }\hlstd{alpha1.median.plt} \hlkwb{<-} \hlkwd{ggplot}\hlstd{(alpha1.median)} \hlopt{+}
\hlstd{+ }    \hlkwd{geom_sf}\hlstd{(}\hlkwc{data} \hlstd{= alpha1.median,} \hlkwd{aes}\hlstd{(}\hlkwc{color} \hlstd{= alpha1.median),}
\hlstd{+ }            \hlkwc{size}\hlstd{=size,} \hlkwc{shape}\hlstd{=}\hlnum{15}\hlstd{)} \hlopt{+}
\hlstd{+ }    \hlkwd{scale_color_viridis}\hlstd{(}\hlstr{"alpha 1"}\hlstd{,} \hlkwc{direction} \hlstd{=} \hlopt{-}\hlnum{1}\hlstd{)} \hlopt{+}
\hlstd{+ }    \hlkwd{ggtitle}\hlstd{(}\hlstr{"Posterior median"}\hlstd{)} \hlopt{+}
\hlstd{+ }    \hlkwd{theme_bw}\hlstd{()} \hlopt{+} \hlkwd{theme}\hlstd{(}\hlkwc{axis.text.x} \hlstd{=} \hlkwd{element_text}\hlstd{(}\hlkwc{angle} \hlstd{=} \hlnum{45}\hlstd{,} \hlkwc{hjust} \hlstd{=} \hlnum{1}\hlstd{))}
\hlstd{R> }
\hlstd{R> }\hlstd{alpha1.sd.plt} \hlkwb{<-} \hlkwd{ggplot}\hlstd{(alpha1.sd)} \hlopt{+}
\hlstd{+ }    \hlkwd{geom_sf}\hlstd{(}\hlkwc{data} \hlstd{= alpha1.sd,} \hlkwd{aes}\hlstd{(}\hlkwc{color} \hlstd{= alpha1.sd),}
\hlstd{+ }            \hlkwc{size}\hlstd{=size,} \hlkwc{shape}\hlstd{=}\hlnum{15}\hlstd{)} \hlopt{+}
\hlstd{+ }    \hlkwd{scale_color_viridis}\hlstd{(}\hlstr{"alpha 1"}\hlstd{,} \hlkwc{option}\hlstd{=}\hlstr{"A"}\hlstd{,} \hlkwc{direction} \hlstd{=} \hlopt{-}\hlnum{1}\hlstd{)} \hlopt{+}
\hlstd{+ }    \hlkwd{ggtitle}\hlstd{(}\hlstr{"Posterior standard deviation"}\hlstd{)} \hlopt{+}
\hlstd{+ }    \hlkwd{theme_bw}\hlstd{()} \hlopt{+} \hlkwd{theme}\hlstd{(}\hlkwc{axis.text.x} \hlstd{=} \hlkwd{element_text}\hlstd{(}\hlkwc{angle} \hlstd{=} \hlnum{45}\hlstd{,} \hlkwc{hjust} \hlstd{=} \hlnum{1}\hlstd{))}
\hlstd{R> }
\hlstd{R> }\hlstd{season.length.median.plt} \hlkwb{<-} \hlkwd{ggplot}\hlstd{(season.length.median)} \hlopt{+}
\hlstd{+ }    \hlkwd{geom_sf}\hlstd{(}\hlkwc{data} \hlstd{= season.length.median,} \hlkwd{aes}\hlstd{(}\hlkwc{color} \hlstd{= season.length.median),}
\hlstd{+ }            \hlkwc{size}\hlstd{=size,} \hlkwc{shape}\hlstd{=}\hlnum{15}\hlstd{)} \hlopt{+}
\hlstd{+ }    \hlkwd{scale_color_viridis}\hlstd{(}\hlstr{"season length\textbackslash{}ndays"}\hlstd{,} \hlkwc{direction} \hlstd{=} \hlopt{-}\hlnum{1}\hlstd{)} \hlopt{+}
\hlstd{+ }    \hlkwd{ggtitle}\hlstd{(}\hlstr{"Posterior median"}\hlstd{)} \hlopt{+}
\hlstd{+ }    \hlkwd{theme_bw}\hlstd{()} \hlopt{+} \hlkwd{theme}\hlstd{(}\hlkwc{axis.text.x} \hlstd{=} \hlkwd{element_text}\hlstd{(}\hlkwc{angle} \hlstd{=} \hlnum{45}\hlstd{,} \hlkwc{hjust} \hlstd{=} \hlnum{1}\hlstd{))}
\hlstd{R> }
\hlstd{R> }\hlstd{season.length.sd.plt} \hlkwb{<-} \hlkwd{ggplot}\hlstd{(season.length.sd)} \hlopt{+}
\hlstd{+ }    \hlkwd{geom_sf}\hlstd{(}\hlkwc{data} \hlstd{= season.length.sd,} \hlkwd{aes}\hlstd{(}\hlkwc{color} \hlstd{= season.length.sd),}
\hlstd{+ }            \hlkwc{size}\hlstd{=size,} \hlkwc{shape}\hlstd{=}\hlnum{15}\hlstd{)} \hlopt{+}
\hlstd{+ }    \hlkwd{scale_color_viridis}\hlstd{(}\hlstr{"season length\textbackslash{}ndays"}\hlstd{,} \hlkwc{option}\hlstd{=}\hlstr{"A"}\hlstd{,} \hlkwc{direction} \hlstd{=} \hlopt{-}\hlnum{1}\hlstd{)} \hlopt{+}
\hlstd{+ }    \hlkwd{ggtitle}\hlstd{(}\hlstr{"Posterior standard deviation"}\hlstd{)} \hlopt{+}
\hlstd{+ }    \hlkwd{theme_bw}\hlstd{()} \hlopt{+} \hlkwd{theme}\hlstd{(}\hlkwc{axis.text.x} \hlstd{=} \hlkwd{element_text}\hlstd{(}\hlkwc{angle} \hlstd{=} \hlnum{45}\hlstd{,} \hlkwc{hjust} \hlstd{=} \hlnum{1}\hlstd{))}
\hlstd{R> }
\hlstd{R> }\hlkwd{plot_grid}\hlstd{(alpha1.median.plt, alpha1.sd.plt,}
\hlstd{+ }          \hlstd{season.length.median.plt, season.length.sd.plt,}\hlkwc{align} \hlstd{=} \hlstr{"hv"}\hlstd{)}
\end{alltt}
\end{kframe}

{\centering \includegraphics[width=\maxwidth]{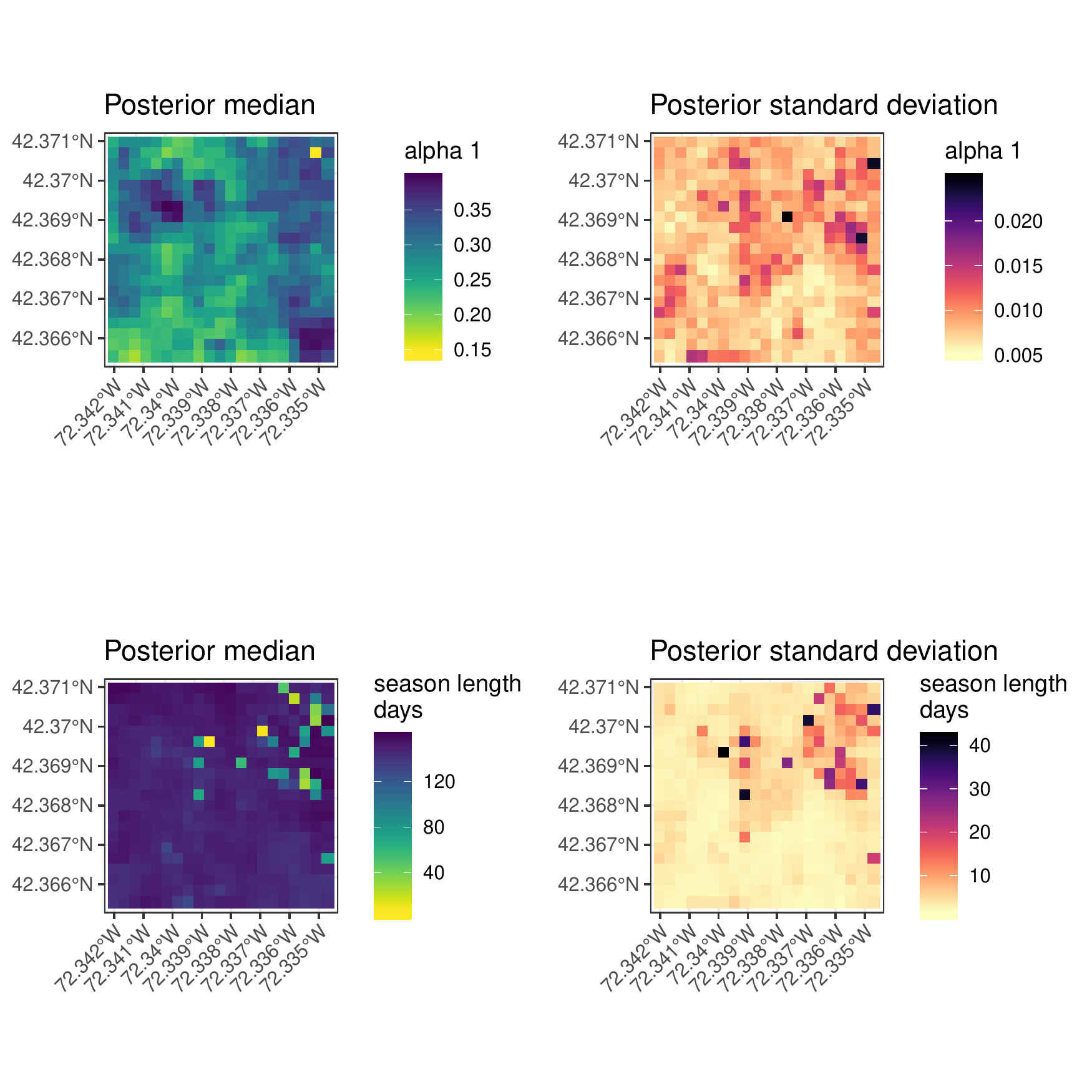} 

}

\end{knitrout}

\pagebreak
Results of a time test of \code{phenoBrick} are shown in the following figure. This test was carried out on a Linux workstation with 128 GBs of RAM and an AMD 3970X 32 core processor. The filled in points are at 1, 2, 4, 8, 16, and 32 cores. It is evident from the figure that the run time is halved when the number of cores used is doubled.

\begin{knitrout}
\definecolor{shadecolor}{rgb}{0.969, 0.969, 0.969}\color{fgcolor}\begin{kframe}
\begin{alltt}
\hlstd{R> }\hlstd{time.test} \hlkwb{<-} \hlkwd{cbind}\hlstd{(}\hlnum{1}\hlopt{:}\hlnum{32}\hlstd{,}\hlkwd{c}\hlstd{(}\hlnum{486}\hlstd{,}\hlnum{244}\hlstd{,}\hlnum{167}\hlstd{,}\hlnum{124}\hlstd{,}\hlnum{99}\hlstd{,}\hlnum{84}\hlstd{,}\hlnum{72}\hlstd{,}\hlnum{65}\hlstd{,}\hlnum{57}\hlstd{,}\hlnum{52}\hlstd{,}\hlnum{48}\hlstd{,}\hlnum{44}\hlstd{,}\hlnum{41}\hlstd{,}\hlnum{38}\hlstd{,}\hlnum{36}\hlstd{,}\hlnum{34}\hlstd{,}\hlnum{32}\hlstd{,}\hlnum{30}\hlstd{,}
\hlstd{+ }                          \hlnum{30}\hlstd{,}\hlnum{28}\hlstd{,}\hlnum{27}\hlstd{,}\hlnum{26}\hlstd{,}\hlnum{25}\hlstd{,}\hlnum{23}\hlstd{,}\hlnum{23}\hlstd{,}\hlnum{22}\hlstd{,}\hlnum{21}\hlstd{,}\hlnum{21}\hlstd{,}\hlnum{20}\hlstd{,}\hlnum{20}\hlstd{,}\hlnum{19}\hlstd{,}\hlnum{18}\hlstd{))}
\hlstd{R> }
\hlstd{R> }\hlkwd{plot}\hlstd{(time.test,} \hlkwc{bty} \hlstd{=} \hlstr{"n"}\hlstd{,} \hlkwc{ylab} \hlstd{=} \hlstr{"time (seconds)"}\hlstd{,} \hlkwc{xlab} \hlstd{=} \hlstr{"Number of CPU cores"}\hlstd{,}
\hlstd{+ }     \hlkwc{axes} \hlstd{= F,} \hlkwc{ylim} \hlstd{=} \hlkwd{c}\hlstd{(}\hlnum{0}\hlstd{,}\hlnum{500}\hlstd{))}
\hlstd{R> }\hlkwd{axis}\hlstd{(}\hlnum{1}\hlstd{,} \hlkwc{at} \hlstd{=} \hlkwd{c}\hlstd{(}\hlnum{1}\hlstd{,}\hlnum{4}\hlstd{,}\hlnum{8}\hlstd{,}\hlnum{12}\hlstd{,}\hlnum{16}\hlstd{,}\hlnum{20}\hlstd{,}\hlnum{24}\hlstd{,}\hlnum{28}\hlstd{,}\hlnum{32}\hlstd{))}
\hlstd{R> }\hlkwd{axis}\hlstd{(}\hlnum{2}\hlstd{,} \hlkwc{at} \hlstd{=} \hlkwd{c}\hlstd{(}\hlnum{0}\hlstd{,}\hlnum{100}\hlstd{,}\hlnum{200}\hlstd{,}\hlnum{300}\hlstd{,}\hlnum{400}\hlstd{,}\hlnum{500}\hlstd{),}\hlkwc{las} \hlstd{=} \hlnum{2}\hlstd{)}
\hlstd{R> }\hlkwd{abline}\hlstd{(}\hlkwc{h} \hlstd{= time.test[}\hlkwd{c}\hlstd{(}\hlnum{1}\hlstd{,}\hlnum{2}\hlstd{,}\hlnum{4}\hlstd{,}\hlnum{8}\hlstd{,}\hlnum{16}\hlstd{,}\hlnum{32}\hlstd{),}\hlnum{2}\hlstd{],} \hlkwc{col} \hlstd{=} \hlstr{"gainsboro"}\hlstd{)}
\hlstd{R> }\hlkwd{points}\hlstd{(time.test[}\hlkwd{c}\hlstd{(}\hlnum{1}\hlstd{,}\hlnum{2}\hlstd{,}\hlnum{4}\hlstd{,}\hlnum{8}\hlstd{,}\hlnum{16}\hlstd{,}\hlnum{32}\hlstd{),],} \hlkwc{pch} \hlstd{=} \hlnum{19}\hlstd{)}
\end{alltt}
\end{kframe}

{\centering \includegraphics[width=\maxwidth]{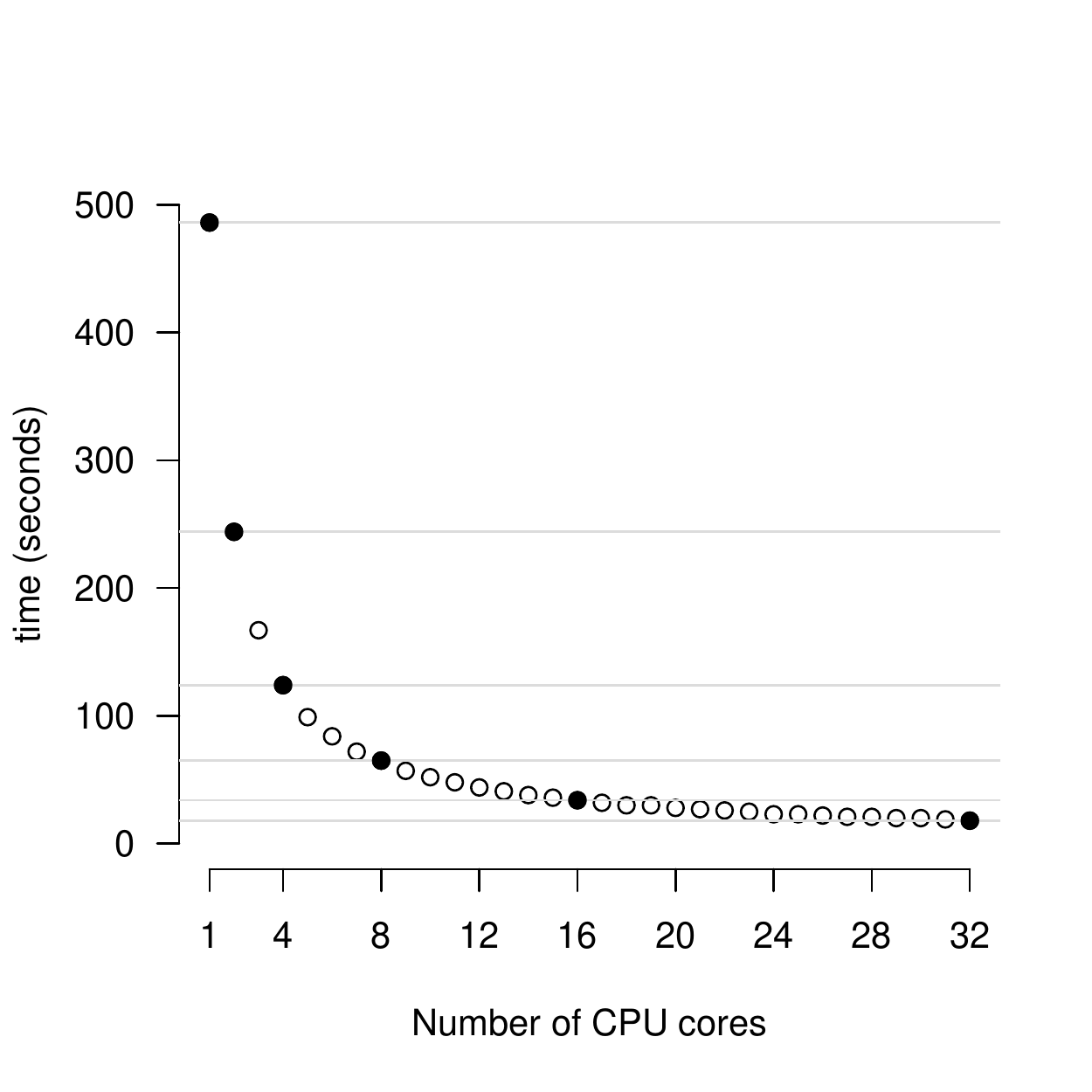} 

}

\end{knitrout}

\pagebreak

\subsection{\R~code to create the \code{phenoBrick} function}\label{phenoBrick}
\begin{knitrout}\scriptsize
\definecolor{shadecolor}{rgb}{0.969, 0.969, 0.969}\color{fgcolor}\begin{kframe}
\begin{alltt}
\hlstd{R> }\hlstd{phenoBrick}
\end{alltt}
\begin{verbatim}
## function(VI.rast, doy, gamma = c(0, 1), family = "normal", starting, tuning, priors, n.samples, sub.sample, n.cores = 1, 
##                        eval.rast, t.normal.bounds = c(0,1)){
##     
##     if(missing(sub.sample)){sub.sample <- list("start" = 1, "thin" = 1, "end" = n.samples)}
##     
##     if(!("end" %in% names(sub.sample))){sub.sample$end <- n.samples}
##     
##     n.samples.keep <- length(seq(as.integer(sub.sample$start), as.integer(sub.sample$end), by=as.integer(sub.sample$thin)))
##     
##     rast.dim <- dim(VI.rast)
##     
##     cl <- makeCluster(n.cores)
##     registerDoSNOW(cl)
##     
##     alpha1.samps <- brick(extent(VI.rast),nrows=rast.dim[1],ncols=rast.dim[2],nl=n.samples.keep,crs=proj4string(VI.rast))
##     alpha2.samps <- brick(extent(VI.rast),nrows=rast.dim[1],ncols=rast.dim[2],nl=n.samples.keep,crs=proj4string(VI.rast))
##     alpha3.samps <- brick(extent(VI.rast),nrows=rast.dim[1],ncols=rast.dim[2],nl=n.samples.keep,crs=proj4string(VI.rast))
##     alpha4.samps <- brick(extent(VI.rast),nrows=rast.dim[1],ncols=rast.dim[2],nl=n.samples.keep,crs=proj4string(VI.rast))
##     alpha5.samps <- brick(extent(VI.rast),nrows=rast.dim[1],ncols=rast.dim[2],nl=n.samples.keep,crs=proj4string(VI.rast))
##     alpha6.samps <- brick(extent(VI.rast),nrows=rast.dim[1],ncols=rast.dim[2],nl=n.samples.keep,crs=proj4string(VI.rast))
##     alpha7.samps <- brick(extent(VI.rast),nrows=rast.dim[1],ncols=rast.dim[2],nl=n.samples.keep,crs=proj4string(VI.rast))
##     sigma2.samps <- brick(extent(VI.rast),nrows=rast.dim[1],ncols=rast.dim[2],nl=n.samples.keep,crs=proj4string(VI.rast))
##     acc.rate     <- brick(extent(VI.rast),nrows=rast.dim[1],ncols=rast.dim[2],nl=2,crs=proj4string(VI.rast))
##     
##     run.idx <- 1:(rast.dim[1]*rast.dim[2])
##     if(!missing(eval.rast)){run.idx <- run.idx[eval.rast[,,1]]}
##     
##     pb <- progress_bar$new(format = "[:bar] :percent | elapsed: :elapsedfull", total = length(run.idx), clear = FALSE)
##     progress <- function(n){pb$tick()}
##     opts <- list(progress = progress)
##     
##     res <-
##         foreach(i=1:length(run.idx),.packages = c("raster","coda","rsBayes"),.combine="rbind",.options.snow=opts)%dopar%{
##             
##             idx <- run.idx[i]
##             y <- as.vector(VI.rast[idx])
##             t <- doy[!is.na(y)]
##             y <- y[!is.na(y)]
##             
##             out <- pheno(y ~ t, priors=priors, n.samples=n.samples, sub.sample = sub.sample, starting = starting, 
##                          tuning=tuning, family=family, verbose = F, t.normal.bounds = t.normal.bounds, gamma = gamma)
##             
##             rbind(t(out$p.theta.samples),out$MH.acceptance)
##             
##         }
##     stopCluster(cl)
##     nrow.res <- nrow(res)
##     
##     alpha1.samps[run.idx] <- res[seq(1,nrow.res,9),]
##     alpha2.samps[run.idx] <- res[seq(2,nrow.res,9),]
##     alpha3.samps[run.idx] <- res[seq(3,nrow.res,9),]
##     alpha4.samps[run.idx] <- res[seq(4,nrow.res,9),]
##     alpha5.samps[run.idx] <- res[seq(5,nrow.res,9),]
##     alpha6.samps[run.idx] <- res[seq(6,nrow.res,9),]
##     alpha7.samps[run.idx] <- res[seq(7,nrow.res,9),]
##     sigma2.samps[run.idx] <- res[seq(8,nrow.res,9),]
##     acc.rate[run.idx]     <- res[seq(9,nrow.res,9),1:2]
##     
##     return(list("alpha1.samps" = alpha1.samps, "alpha2.samps" = alpha2.samps, "alpha3.samps" = alpha3.samps,
##                 "alpha4.samps" = alpha4.samps, "alpha5.samps" = alpha5.samps, "alpha6.samps" = alpha6.samps,
##                 "alpha7.samps" = alpha7.samps, "sigma2.samps" = sigma2.samps, "acc.rate" = acc.rate))
## }
\end{verbatim}
\end{kframe}
\end{knitrout}

\pagebreak
\subsection{Documentation for \code{phenoBrick}}\label{documentation}
\HeaderA{phenoBrick}{phenoBrick}{phenoBrick}
\begin{Description}\relax
Function to run \code{rsBayes::pheno} for a vegetation index (VI) time series raster using multiple CPU cores.
\end{Description}
\begin{Usage}
\begin{verbatim}
phenoBrick(VI.rast, doy, gamma = c(0, 1), family = "normal", starting, tuning, 
           priors, n.samples, sub.sample, n.cores = 1, eval.rast, 
           t.normal.bounds = c(0,1))
\end{verbatim}
\end{Usage}
\begin{Arguments}
\begin{ldescription}
\item[\code{VI.rast}] a \code{RasterBrick} object holding the observed VIs. Each band holds VI observations for a specific day of the year (DOY). There should be as many bands as there are observation DOYs. NAs can be included where VI observations are missing.

\item[\code{doy}] a numeric vector of length \code{nlayers(VI.rast)} holding DOYs that correspond to the DOYs for each band in \code{VI.rast}.

\item[\code{gamma}] a vector of length two that holds the VI variable's theoretical bounds. Lower and upper bounds are given in element 1 and 2, respectively. These must be finite values.

\item[\code{family}] a quoted string that indicates which likelihood to use for modeling the VI variable. Options are Beta, Normal, and truncated Normal, specified using quoted argument values \code{beta}, \code{normal}, and \code{t.normal}, respectively. The \eqn{\sigma^2}{} parameter is the variance for the Beta, Normal, and truncated Normal likelihoods. See Details for VI variable's support for each likelihood.

\item[\code{starting}] a list with each tag corresponding to a parameter name. Valid tags are \code{alpha.1}, \code{alpha.2}, \ldots, \code{alpha.7}, and \code{sigma.sq}. The value portion of each tag is the parameter's starting value. See the Details for additional guidance on selecting starting values. 

\item[\code{tuning}] a list with each tag corresponding to a parameter name. Valid tags are \code{alpha.1}, \code{alpha.2}, \ldots, \code{alpha.7}, and \code{sigma.sq}.  The value portion of each tag defines the variance of the Metropolis sampler Normal proposal distribution. Tuning values should be selected to keep the acceptance rate between approximately 20 and 50 percent.

\item[\code{priors}] a list with tags \code{alpha} and \code{sigma.sq.IG}. \code{alpha} should be a list comprising \code{alpha.1.Unif}, \code{alpha.2.Unif}, \ldots, \code{alpha.7.Unif} with each of these tags set equal to a vector of length two that holds the lower and upper bound for the Uniform  priors on the given \code{alpha}. The \code{sigma.sq.IG} is a vector of length two that holds the shape and scale parameters for the \eqn{\sigma^2}{}'s inverse-Gamma prior distribution. At minimum the \code{priors} list  must include a prior for \code{sigma.sq.IG}. If \eqn{\alpha}{} priors are not specified then they are assumed to follow their default values, see the Details section below.

\item[\code{n.samples}] the number of MCMC samples to collect.

\item[\code{sub.sample}] an optional list that specifies the subset of MCMC samples should be returned. Valid tags are \code{start}, \code{end}, and \code{thin}. The default values are \code{start=1}, \code{end=n.samples} and \code{thin=1}, i.e., no burn in or thinning.

\item[\code{n.cores}] number of CPU cores to use for model fitting.

\item[\code{eval.rast}] an optional TRUE/FALSE \code{RasterLayer} object indicating which pixels in \code{VI.rast} should be fit. TRUE signifies the model should be fit for the pixel and FALSE signifies it should not.
\end{ldescription}
\end{Arguments}

\begin{Details}
Selection of the likelihood via the \code{family} argument should be respective of possible VI value range. The Beta likelihood assumes support for VI between 0 and 1. The Normal likelihood assumes support for VI on the whole real line. The default for the truncated Normal likelihood is support between 0 and 1; however, specifying the optional argument \code{t.normal.bounds} allows for user defined support bounds.

The default priors for the \eqn{\alpha}{}'s are:

\begin{center}
\begin{tabular}{lll}
$\alpha_1 \sim Unif(\gamma_1, \gamma_2)$,&$\alpha_2 \sim Unif(0, \gamma_2-\alpha_1)$,&$\alpha_3 \sim Unif(0, 1)$,\\
$\alpha_4 \sim Unif(0, \alpha_7)$,&$\alpha_5 \sim Unif(0, 0.001)$,&$\alpha_6 \sim Unif(0, 1)$,\\
$\alpha_7 \sim Unif(1, 365)$&&\\
\end{tabular}
\end{center}
The hyperparameters of these prior distributions can be changed using the \code{priors} argument.

An error will be thrown if an \eqn{\alpha}{} starting value provided in the \code{starting} list is outside the default or supplied priors. Note, \eqn{\alpha_2}{} and \eqn{\alpha_4}{} have upper bounds determined by starting values of \eqn{\alpha_1}{} and \eqn{\alpha_7}{}, respectively.
\end{Details}

\begin{Value}
The output is a list of 9 \code{RasterBrick} objects. List components include:
\begin{ldescription}
  \item[\code{alpha1.samps}] is a \code{RasterBrick} holding the thinned post-burnin MCMC samples for $\alpha_1$ at each evaluated pixel.
  \item[\code{alpha2.samps}] is a \code{RasterBrick} holding the thinned post-burnin MCMC samples for $\alpha_2$ at each evaluated pixel.
  \item[\code{alpha3.samps}] is a \code{RasterBrick} holding the thinned post-burnin MCMC samples for $\alpha_3$ at each evaluated pixel.
  \item[\code{alpha4.samps}] is a \code{RasterBrick} holding the thinned post-burnin MCMC samples for $\alpha_4$ at each evaluated pixel.
  \item[\code{alpha5.samps}] is a \code{RasterBrick} holding the thinned post-burnin MCMC samples for $\alpha_5$ at each evaluated pixel.
  \item[\code{alpha6.samps}] is a \code{RasterBrick} holding the thinned post-burnin MCMC samples for $\alpha_6$ at each evaluated pixel.
  \item[\code{alpha7.samps}] is a \code{RasterBrick} holding the thinned post-burnin MCMC samples for $\alpha_7$ at each evaluated pixel.
  \item[\code{sigma2.samps}] is a \code{RasterBrick} holding the thinned post-burnin MCMC samples for $\sigma^2$ at each evaluated pixel.
  \item[\code{acc.rate}] is a 2 band \code{RasterBrick} holding the overall Metropolis acceptance rate in the first band and the last batch acceptance rate in the second band (MH.acceptance output for \code{pheno}).
\end{ldescription}
\end{Value}

\begin{Author}
Chad Babcock \email{cbabcock@umn.edu},
Andrew O. Finley \email{finleya@msu.edu}
\end{Author}

\end{document}